\documentclass{aastex631}

\begin{document}

\title{Mass distribution for single-lined hot subdwarf stars in LAMOST}

\author[0000-0003-2362-6637]{Zhenxin Lei}
\affiliation{Key Laboratory of Stars and Interstellar Medium, 
Xiangtan University, Xiangtan 411105, People's Republic of China}
\affiliation{Physics Department, Xiangtan University, Xiangtan, 411105, People’s Republic of China}
\email{leizhenxin2060@163.com}

\author{Ruijie He}
\affiliation{Key Laboratory of Stars and Interstellar Medium, 
Xiangtan University, Xiangtan 411105, People's Republic of China}
\affiliation{Physics Department, Xiangtan University, Xiangtan, 411105, People’s Republic of China}

\author[0000-0003-0963-0239]{P\'eter N\'emeth}
\affiliation{Astronomical Institute of the Czech Academy of Sciences, CZ-251\,65, Ond\v{r}ejov, Czech Republic}
\affiliation{Astroserver.org, F\H{o} t\'er 1, 8533 Malomsok, Hungary}

\author{Xuan Zou}
\affiliation{Key Laboratory of Stars and Interstellar Medium, 
Xiangtan University, Xiangtan 411105, People's Republic of China}
\affiliation{Physics Department, Xiangtan University, Xiangtan, 411105, People’s Republic of China}

\author{Huaping Xiao}
\affiliation{Key Laboratory of Stars and Interstellar Medium, 
Xiangtan University, Xiangtan 411105, People's Republic of China}
\affiliation{Physics Department, Xiangtan University, Xiangtan, 411105, People’s Republic of China}

\author{Yong Yang}
\affiliation{Key Laboratory of Optical Astronomy, National Astronomical Observatories, Chinese Academy of Sciences, Beijing 100012, China}

\author[0000-0003-2868-8276]{Jingkun Zhao}
\affiliation{Key Laboratory of Optical Astronomy, National Astronomical Observatories, Chinese Academy of Sciences, Beijing 100012, China}



\begin{abstract}
Masses for 664 single-lined hot subdwarf stars identified in LAMOST were calculated by 
comparing synthetic fluxes from spectral energy distribution (SED) with observed fluxes from virtual observatory service.
Three groups of hot subdwarf stars were selected from the whole sample according to their parallax precision to study the mass distributions. 
We found, that He-poor sdB/sdOB stars present a wide mass distribution from 0.1 to 1.0 $\mathrm{M}_{\odot}$ with a sharp mass peak around at 0.46 $\mathrm{M}_{\odot}$, which is consistent with canonical binary model prediction. 
He-rich sdB/sdOB/sdO stars present a much flatter mass distribution than He-poor sdB/sdOB stars and with a mass peak around 0.42 $\mathrm{M}_{\odot}$. 
By comparing the observed mass distributions to the predictions of different formation scenarios, we concluded that the binary merger channel, including two helium white dwarfs (He-WDs) and He-WD + main sequence (MS) merger, cannot be the only main formation channel for He-rich hot subdwarfs, and other formation channels such as the surviving companions from type Ia supernovae (SNe Ia) could also make impacts on producing this special population, especially for He-rich hot subdwarfs with masses less than 0.44 $\mathrm{M}_{\odot}$. 
He-poor sdO stars also present a flatter mass distribution with an inconspicuous peak mass at 0.18 $\mathrm{M}_{\odot}$. 
The similar mass - $\Delta RV_\mathrm{max}$ distribution between He-poor sdB/sdOB and sdO stars supports the scenario that He-poor sdO stars could be the subsequent evolution stage of He-poor sdB/sdOB stars. 

\end{abstract}

\keywords{}


\section{Introduction} \label{sec:intro}
Hot subdwarf stars are located at the region between the main sequence (MS) and the white dwarf (WD) cooling sequence in the Hertzsprung-Russell (HR) diagram \citep{2009ARA&A..47..211H, 2016PASP..128h2001H}. These blue stars have small stellar masses around 0.5 $\mathrm{M}_{\odot}$,  high effective temperatures (e.g., roughly 20000 K $\leq$ $T_\mathrm{eff}$  $\leq$ 70000 K) and gravity (e.g., roughly 5.0 $\leq$ $\mathrm{log}\ g$ $\leq$ 6.5 ), and they can be roughly classified into B and O type (sdB/O) based on their spectral line features. Many hot subdwarf stars are burning Helium (He) in their cores and occupy the bluest positions of horizontal branch (HB), thus known as extreme HB (EHB) stars. Other types of stars could also cross the hot subdwarf region in the HR diagram, such as post-EHB stars, blue hook stars in globular clusters \citep{2010ApJ...718.1332B, 2016ApJ...822...44B}, post asymptotic giant branch (post-AGB) stars, post red giant branch (post-RGB) stars, low and extremely low mass (pre-)white dwarfs (pre-ELM WDs), etc \citep{2016PASP..128h2001H}. 

The bulk of the currently known hot subdwarf stars were discovered only recently in the data releases of large surveys, such as the Sloan Digital Sky Survey (\citealt{2015A&A...577A..26G,2015MNRAS.446.4078K, 2016MNRAS.455.3413K, 2019MNRAS.486.2169K}), ESO Supernova Ia Progenitor Survey 
(SPY, \citealt{ 2005A&A...430..223L,2007A&A...462..269S}),  The Hamburg Quasar Survey (HQS, \citealt{2003A&A...400..939E}), The Arizona-Montr\'eal Spectroscopic Program (\citealt{2014ASPC..481...83F}),
The Large Sky Area Multi-Object Fiber Spectroscopic Telescope spectra survey (LAMOST, \citealt{2016ApJ...818..202L, 2019ApJ...881....7L, 2021ApJS..256...28L, 2018ApJ...868...70L,2019ApJ...881..135L, 2020ApJ...889..117L, 2023ApJ...942..109L}), Galaxy Evolution Explorer survey (GALEX, \citealt{ 2011MNRAS.410.2095V, 2012MNRAS.427.2180N}) and Transiting Exoplanet Survey
Satellite (TESS, \citealt{2022MNRAS.516.1509K}). Reliable atmospheric parameters (e.g., $T_{\rm eff}$, $\log{g}$ and $\log{(n{\rm He}/n{\rm H})}$ were obtained by spectral analysis in these studies, and it provides great convenience on studying these special blue stars in our Galaxy. 

A catalog of known hot subdwarf stars was compiled from literature and reported by \citet{2017A&A...600A..50G}, which contains 5613 objects and a lot of useful information, including multi-band photometry, ground-based proper motions, classifications, published atmospheric parameters. 
Thanks to the continuously running large surveys mentioned above, the number of confirmed hot subdwarfs was updated successively to 5874 and 6616 by \citet{2020A&A...635A.193G} and \citet{2022A&A...662A..40C}, respectively. 

Hot subdwarf stars can be further classified into more sub-types according to the different strength of hydrogen (H) and He lines in their spectra, e.g., He-poor sdB/sdOB/sdO stars and their He-rich counterparts, He-rich sdB/sdOB/sdO stars \citep{1990A&AS...86...53M, 2017A&A...600A..50G, 2018ApJ...868...70L}. \citet{2013A&A...551A..31D} designed an MK (Morgan–Keenan)-like spectral classification scheme for hot subdwarf stars, in which a spectral class, a luminosity class, and a helium class are needed for spectral classification. With this detailed classification scheme, \citet{2019PASJ...71...41L} classified 56 hot subdwarf stars found in LAMOST DR1, and \citet{2021MNRAS.501..623J} classified 107 hot subdwarf stars observed by the Southern African Large Telescope (SALT).

Since most of the sdB type hot subdwarfs were found in close binary systems \citep{2001MNRAS.326.1391M,2004Ap&SS.291..321N, 2011MNRAS.415.1381C}, binary evolution is considered to be the main formation channel for these stars (see \citealt{1976ApJ...204..488M} for early discussions on this topic). Based on the results from detailed binary population synthesis, \citet{2002MNRAS.336..449H, 2003MNRAS.341..669H} found that stable Roche lobe overflow (RLOF), common envelope (CE) ejection and the merger of two He-WDs due to binary evolution can produce sdB stars, and most of the properties of these stars, such as the period and mass distribution, positions in the $T_{\rm eff}$ - $\log{g}$ plane, etc (also see \citealt{2013MNRAS.434..186C}) are consistent with observations. 
These models predicted that the mass range of sdB stars extends from 0.3 to 0.8 $\mathrm{M}_{\odot}$.

The two He-WD merger channel is considered to be the main formation channel for single He-rich hot subdwarf stars (\citealt{1984ApJ...277..355W}). \citet{2012MNRAS.419..452Z} studied the merger of two He-WDs and found that a composite model of slow accretion from a debris disk and fast accretion into a corona can reproduce the observed properties of $T_{\rm eff}$, $\log{g}$, nitrogen (N) and carbon (C) abundances for He-rich hot subdwarf stars. Moreover, \citet{2017ApJ...835..242Z} also studied the post-merger models of a He-WD with a MS star. They found that the merger of He-WD+MS channel could produce intermediate He-rich hot subdwarfs (iHe-rich, e.g., -1.0 $<\mathrm{log}(n\mathrm{He}/n\mathrm{H}<$ 1.0), and a mass range of 0.48 - 0.52 ${\rm M}_{\odot}$ was predicted for these stars in their model. Recently, companions surviving from Type Ia supernovae (SNe Ia) explosions were also predicted to evolve into single hot subdwarfs.  \citet{2021MNRAS.507.4603M} studied the surviving companions from SNe Ia explosions of WD + MS binaries and found that this channel could produce single iHe-rich hot subdwarfs. Based on the results of detailed binary population synthesis, they obtained a Galactic birth rate of 2.3 - 6 $\times 10^{-4}\ {\rm yr}^{-1}$ for their spin-up/down model.   

On the other hand, \citet{2008A&A...491..253M} proposed that He-rich hot subdwarf stars could be formed through single stellar evolution. In their study, after losing nearly the whole envelope at the tip of RGB, stars could experience a delayed core He flash while on the way to the WD cooling curve or already on it (also see \citealt{1993ApJ...407..649C} for late He flash scenarios on WD cooling curves). This delayed core He flash could drive a deep mixing in the envelope thus leading to He and C enhancement at the stellar surface. However, based on the observed distribution of rotation rates for the companions in known wide hot subdwarf binaries, \citet{2020A&A...642A.180P} found that binary interaction is always required for the formation of hot subdwarfs. This result seems to contradict the single formation channel of He-rich hot subdwarfs. Furthermore, \citet{2022A&A...661A.113G} studied the radial velocities (RVs) for 646 single-lined hot subdwarfs with multi-epoch observed spectra in SDSS and LAMOST. No significant RV variations were detected for nearly all He-rich hot subdwarfs, but the exact opposite is true for He-poor ones. Their results support the merger formation channel for He-rich hot subdwarfs.   

Mass is a very important parameter to test and constrain formation channels for hot subdwarf stars. Unfortunately, based on the observational data available until now, obtaining exact masses for a large number of hot subdwarfs is difficult. To study the empirical mass distribution for sdB stars, \citet{2012A&A...539A..12F} collected 16 sdB stars with known masses on the basis of asteroseismology and 11 sdB stars in binary systems with masses determined by light curve modeling and spectroscopy. They found a relatively sharp mass distribution with a mean mass around 0.47 ${\rm M}_{\odot}$, but with a wide mass range of 0.28-0.63 ${\rm M}_{\odot}$. Their results were consistent with model predictions from \cite{2002MNRAS.336..449H,2003MNRAS.341..669H} when selection effects were considered. 
Furthermore, combining with the distance from \textit{Gaia} parallax \citep{2021A&A...649A...1G} and atmospheric parameters, \citet{2022A&A...666A.182S} obtained angular diameters thus radius and masses for 68 sdB stars in binary systems by comparing observed fluxes with synthetic fluxes calculated from spectral energy distributions (SED). 
They found that the mass distribution of sdB stars with cool, low-mass stellar, and sub-stellar companions differ from those with WD companions, which demonstrated that they most likely come from different populations. 

Although more than 6000 hot subdwarf stars have become known \citep{2022A&A...662A..40C} until now, among which more than 3000 stars have reliable atmospheric parameters (e.g., $T_{\rm eff}$, $\log{g}$ and $\log{(n{\rm He}/n{\rm H})}$) from spectral fitting, there is still a lack of masses for a large number of hot subdwarf stars to study the mass distribution statistically, thus to investigate the formation and evolution of these stars. 
In this study, we build on the atmospheric parameters obtained in our previous studies \citep{2018ApJ...868...70L, 2019ApJ...881..135L,2020ApJ...889..117L}.
From those models, we calculated synthetic SEDs for 664 single-lined hot subdwarfs identified in LAMOST and obtained their radii, masses, and luminosity by comparing the synthetic fluxes with observed fluxes from the virtual observatory (VO) service \citep{2008A&A...492..277B}. Based on these results, we studied the possible formation channels for hot subdwarf stars with different spectral classifications. Section 2 described the method used to obtain masses for hot subdwarf stars. Our results are given in Section 3, and a discussion is presented in Section 4. Finally, a summary of our study is given in Section 5. 

\section{Methodology}
\subsection{Basic principle}
Once the synthetic spectral energy distribution (SED) of a star is known, we can obtain the model flux density at the stellar surface, $\it{F(\lambda)}$. 
On the other hand, the observed flux density at the Earth, $\it{f(\lambda)}$, can be obtained by converting photometric magnitudes into flux density (see \citealt{2018OAst...27...35H}, \citealt{2021MNRAS.503.2157B} and \citealt{2022A&A...666A.182S} for more information). 
With known $\it{F(\lambda)}$, $\it{f(\lambda)}$ and distance from the Earth, $d$, one can easily obtain the stellar angular diameter, $\Theta$, and stellar radius, $R$, by the following equation:
\begin{equation}
    \frac{\it{f(\lambda)}}{\it{F(\lambda)}}=\frac{\Theta^{2}}{4}=\frac{R^{2}}{d^{2}}
\label{eq:1}
\end{equation} 
Then, the stellar mass of a hot subdwarf star can be obtained by:
\begin{equation}
    M=\frac{gR^2}{G}
\label{eq:2}
\end{equation}
where $g$ is the surface gravity and $G$ is the gravitational constant. 

\subsection{Sample selection,  distances estimation, magnitudes conversion into photometric fluxes}

With the useful information from Gaia DR2 \citep{2018A&A...616A...1G}, e.g., magnitudes, parallaxes, colors, etc, 
\citet{2018ApJ...868...70L, 2019ApJ...881..135L, 2020ApJ...889..117L}  
identified 864 single-lined hot subdwarf stars in LAMOST. Since all of these hot subdwarf stars have good quality LAMOST spectra (e.g., signal-to-noise ratio, SNR, greater than 10.0 in the u band), they obtained reliable atmospheric parameters (e.g., $T_{\rm eff}$, $\log{g}$ and $\mathrm{log}(n\mathrm{He}/n\mathrm{H})$) by fitting H and He profiles with synthetic spectra, which were calculated with {\sc Synspec} (version 49, \citealt{2007ApJS..169...83L}) from non-Local Thermodynamic Equilibrium (NLTE) {\sc Tlusty} model atmospheres (version 204; \citealt{2017arXiv170601859H}). 
Based on the obtained atmospheric parameters, synthetic SEDs were also calculated for each star in the same way as the synthetic spectra.

Gaia EDR3 \citep{2021A&A...649A...1G} provided astrometry and photometry for 1.8 billion objects collected in the first 34 months of the satellite mission, among which 1.5 billion sources have parallaxes, proper motions, and $G$, $G_{\rm BP}$, $G_{\rm RP}$ magnitudes. With precise parallaxes, we could calculate distances for a huge number of objects which are bright and relatively nearby to the Sun. 
However, distances cannot be obtained simply by directly reversing parallaxes for faint and distant objects, since the parallax uncertainties are large, even negative parallaxes can occur. 
To provide more reliable distances for as many stars in Gaia EDR3 as possible, \citet{2021AJ....161..147B} used a probabilistic approach, which employed a prior constructed from a three-dimensional model of our Galaxy, and estimated stellar distances based on Gaia parallaxes and photometry. 

In this study, 864 single-lined hot subdwarf stars identified in \citet{2018ApJ...868...70L, 2019ApJ...881..135L,2020ApJ...889..117L} were cross-matched  with \textit{Gaia} EDR3 data \citep{2021A&A...649A...1G} to obtain their parallaxes. 
By inspecting quasar data in \textit{Gaia} EDR3 \cite{2021A&A...649A...2L} found that parallaxes can be biased. 
\citet{2021A&A...649A...4L} investigated the variation of the parallax bias for \textit{Gaia} EDR3 data with magnitude, colour, and ecliptic latitude. They provided a python implementation\footnote{\url{https://www.cosmos.esa.int/web/gaia/edr3-code}} to correct the bias for \textit{Gaia} EDR3 parallaxes, which was also adopted for our selected sample in this study. 
Since the uncertainty of parallax can affect the precision of distance directly, thus affecting the determination of stellar radius and mass \citep{2021A&A...650A.102I}, only objects with positive parallax and a relative parallax uncertainty of less than 20\% (e.g., $\varpi>0$ and $\sigma_{\varpi}/\varpi\leq0.2$, see \citealt{2022A&A...658A..22R}) were reserved for the following analysis. 
After this step, 727 hot subdwarf stars with \textit{Gaia} EDR3 parallax remained in our selection, and their distances were obtained directly by using the inverse of parallaxes after applying a zero-point correction.  

We used the VO Sed Analyzer\footnote{\url{http://svo2.cab.inta-csic.es/svo/theory/vosa/}} (VOSA, \citealt{2008A&A...492..277B}) of the Spanish Virtual Observatory (SVO) to search for photometric data for our sample and convert observed magnitudes to fluxes. 
VOSA is a public web tool designed to help users search for observed photometric data from Virtual Observatory (VO) services and build SEDs. It can compare observed photometry with synthetic photometry from theoretical models to obtain important physical parameters for stars, such as $T_{\rm eff}$, $\log{g}$, metallicity, radius, luminosity, etc. 
We obtained interstellar extinctions in the visual band, \textit{$A_{\rm V}$}, for the selected sample through Galactic dust reddening maps\footnote{\url{https://irsa.ipac.caltech.edu/applications/DUST/}} \citep{1998ApJ...500..525S, 2011ApJ...737..103S} with extinction parameter \textit{$R_{\rm V}$} = 3.1. 
All this information together with coordinates (e.g., RA, DEC) of the sample was uploaded into VOSA, then the conversion of photometric fluxes from  different filters could be obtained for each sample stars. 

\begin{figure}
    \centering    
    \includegraphics[width=160mm]{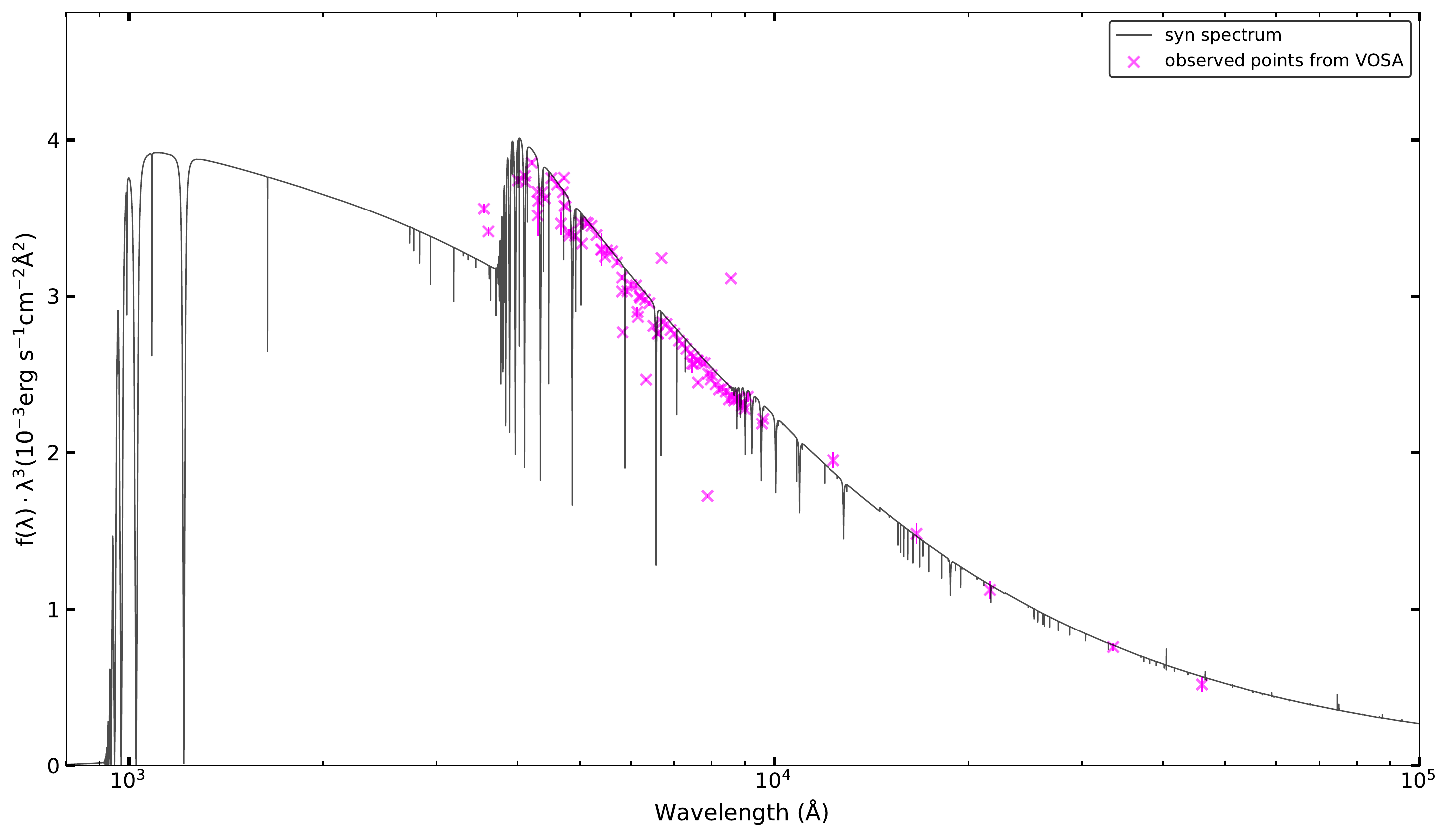}
    \caption{Example of comparison synthetic SED with observed photometric fluxes. Flux density is multiplied  by wavelength to the power of three to eliminate the steep slope of SED.  Black solid curve denotes the synthetic SED calculated based on the atmospheric parameters obtained from LAMOST spectra fitting, while magenta crosses denote the observed photometric fluxes with different filters retrieved from VOSA. }
    \label{fig1}
\end{figure}

\subsection{Dilution parameter, radius and luminosity}
To compare synthetic SEDs with observed photometric fluxes, one needs to convolve synthetic SEDs with the transmission curve of the specific filter. Thanks to the Filter Profile Service\footnote{\url{http://svo2.cab.inta-csic.es/theory/fps3/}} in SVO \citep{2012ivoa.rept.1015R, 2020sea..confE.182R}, which provided information for more than 2300 astronomical filters, including transmission curves and zero points in the Vega, AB, and ST photometric systems, we obtained all the transmission curves and zero point information of the astronomical filters listed in the Filter Profile Service. Then, we convolved the synthetic SEDs with the filter transmission curves and compared them with the observed photometric fluxes obtained from VOSA. By doing this, we obtained the dilution factor $M_{\rm d}=R^2/d^2$ and stellar angular diameter $\Theta$ (see Equation \ref{eq:1}) for our selected sample. 

Fig \ref{fig1} gives an example of a comparison between synthetic SED and observed photometric fluxes for one of the selected stars, LAMOST\_obsid = 580907209. 
The observed photometric data were collected from about 100 filters within different passbands equipped on various photometric systems, such as the Galaxy Evolution Explorer (GALEX, \citealt{2005ApJ...619L...1M}), Gaia mission \citep{2016A&A...595A...1G}, Two Micron All Sky Survey (2MASS, \citealt{2006AJ....131.1163S}), Wide-field Infrared Survey Explorer (WISE, \citealt{2010AJ....140.1868W}), etc. 
Fig \ref{fig1} shows that most of the observed photometric fluxes (e.g., magenta crosses) are consistent with the synthetic SED (e.g., black solid curve). 
                       
Once the stellar radii and effective temperatures are known, one can calculate the luminosity of a star by comparing the two parameters with the solar standard model using the following equation:
\begin{equation}
    \frac{L}{L_\odot}=\left(\frac{R}{R_\odot}\right)^{2}\left(\frac{T_\mathrm{eff}}{T_{\odot}}\right)^{4}
\end{equation}
where $R_{\odot}$ and $T_{\odot}$ are radius and surface effective temperature of the Sun, which were set to $6.959\times10^{8}$ m  and 5777 K respectively.


\subsection{Mass calculation and uncertainty estimation}

With known radius and $\log{g}$, we can calculate the stellar masses for the selected sample through Equation \ref{eq:2}. To estimate mass uncertainties, a Monte Carlo (MC) approach was adopted in the calculations. 
Based on the values and uncertainties for the parameters involved in mass determination (i.e., parallaxes, gravity, etc), a set of values was generated for each parameter by using a Gaussian distribution. 
With these values, the same number of masses was calculated for each sample star. 
In the calculation, with the symmetric uncertainty, we assumed a Gaussian distribution of parallax for each star. 
However, the transformation from parallax to distance is nonlinear, thus the distances obtained for each star are not a Gaussian-like distribution and their uncertainties are asymmetric (for a detailed discussion see \citealt{2015PASP..127..994B, 2016ApJ...832..137A, 2021AJ....161..147B}). 
This is also valid for the obtained mass values. 
Therefore, for each sample star, we calculated the median (e.g., 50th percentile) mass value as the final mass. 
Moreover, the 16th and 84th percentile mass values form a 68\% conﬁdence interval around the median mass, and the differences between them and the median mass are considered as the asymmetric uncertainties of the mass (see Column 14 in Table 1).

Atmospheric parameters for hot subdwarf stars in \citet{2018ApJ...868...70L, 2019ApJ...881..135L, 2020ApJ...889..117L} were obtained with NLTE model atmospheres and only statistical uncertainties were presented, while no systematic errors were considered in their studies. 
A realistic assessment of systematics must be based on independent studies, therefore, we estimated systematic errors for $T_{\rm eff}$ and $\log{g}$ by comparing atmospheric parameters in \citet{2018ApJ...868...70L, 2019ApJ...881..135L, 2020ApJ...889..117L} with the results from \citet{2022A&A...666A.182S}. 
Finally, we found systematic differences of about 560 K and 0.013 dex for the two parameters, respectively. Moreover, different atmospheric models could also lead to systematic errors in parameter determination. 
\citet{2000A&A...363..198H} determined atmospheric parameters for three pulsating sdB stars by using blanketed NLTE and LTE model atmospheres respectively. 
They found that there is no obvious difference in $T_{\rm eff}$ determination between NLTE and LTE models, but NLTE $\log{g}$ is slightly lower than LTE ones by at least  0.05 dex (see Section 4.1 in their study for detailed discussion). 
In this study, the systematic errors for $T_{\rm eff}$ and $\log{g}$ discussed above were added to the statistical uncertainties in quadrature for the selected hot subdwarf stars to estimate their mass uncertainties.

Based on the method described above, we obtained masses for 727 hot subdwarf stars from \citet{2018ApJ...868...70L, 2019ApJ...881..135L, 2020ApJ...889..117L}.
For some of these hot subdwarf stars, we got excessively large mass values (e.g., larger than 1.0 ${\rm M}_{\odot}$), which are not likely real hot subdwarfs, or their mass values are unreliable due to large uncertainties of distances and/or gravity. 
On the other hand, a few stars turned out to have very low mass values (e.g., less than 0.1 ${\rm M}_{\odot}$). 
These stars may be extremely low-mass white dwarfs or the mass values are unreliable due to large uncertainties from related parameters as well. 
Such extreme cases are not reported here. 
In total, 664 hot subdwarf stars having mass values between 0.1 and 1.0 $\mathrm{M}_{\odot}$, are reported and analyzed in the following sections. 

\section{Results}

Table \ref{tab:main} presents the calculated mass values and some useful information for 664 selected hot subdwarf stars. 
From left to right, columns 1-3 give right ascension (RA), declination (DEC) and LAMOST\_obsid, respectively. 
Columns 4-6 give Gaia EDR3 source\_id, $G$ magnitudes, and parallaxes after zero-point correction. 
Columns 7-10 show $T_{\rm eff}$, $\log{g}$\footnote{The systematic errors discussed in Section 2.4 have been added to the uncertainties of  $T_{\rm eff}$ and $\log{g}$.}, and spectral classification from \citet{2018ApJ...868...70L,2019ApJ...881..135L, 2020ApJ...889..117L}. 
Column 10 gives E(B-V) values, while columns 11-14 give angular diameters, stellar radius, stellar luminosity, and masses, respectively, which were calculated in this study. 
For each sample star, we calculated the 16th, 50th (median), and 84th percentile values for the last four parameters, and reported the median values as the final values listed in Table \ref{tab:main}, while the differences between the 16th, 84th percentile values and the median value were reported as asymmetric uncertainties (see Section 2.4 for more information).   

\begin{longrotatetable}
\begin{deluxetable*}{@{\extracolsep{1pt}}llllllllcccccl@{}}
\tablecaption{Main parameters for the 664 hot subdwarfs selected in this study.}
\label{tab:main}
\tablewidth{100pt}
\tabletypesize{\tiny}
\tablehead{
\colhead{RA} & \colhead{DEC} &
\colhead{obs\_id}  &\colhead{source\_id} & \colhead{${G}$}  & \colhead{parallax}  & \colhead{$T_\mathrm{eff}$} & \colhead{ $\mathrm{log}\ g$} &  \colhead{spclass} & \colhead{E(B-V)} & 
\colhead{Angular diameter} & \colhead{Radius}  & \colhead{Luminosity} & 
\colhead{Mass}  \\
\colhead{LAMOST} & \colhead{LAMOST} & \colhead{LAMOST} &  \colhead{\textit{Gaia} EDR3}  & \colhead{\textit{Gaia} EDR3 (mag)} & \colhead{\textit{Gaia} EDR3 (mas)} & \colhead{(\textit{K})} & \colhead{(cm\ s$^{-2}$)}& &  &\colhead{log(rad)} &  \colhead{$R/R_{\odot}$}  &  \colhead{$L/L_{\odot}$} &  \colhead{$M/M_{\odot}$}
} 
\colnumbers
\startdata
0.0767 & 22.30079 & 385816105 & 2847977322031768832 & 14.219558$\pm$0.002796 & 1.2039$\pm$0.0359 & 28810$\pm$630 & 5.57$\pm$0.06 & sdB & 0.058$\pm$0.004 & -11.013$^{+0.004}_{-0.006}$ & 0.176$^{+0.006}_{-0.004}$ & 19.2$^{+2.3}_{-1.8}$ & 0.42$^{+0.08}_{-0.06}$\\
0.278039 & 11.010084 & 66605009 & 2765454164004531328 & 13.589941$\pm$0.002805 & 1.6454$\pm$0.0352 & 27030$\pm$602 & 5.48$\pm$0.05 & sdB & 0.071$\pm$0.002 & -10.856$^{+0.003}_{-0.003}$ & 0.191$^{+0.004}_{-0.004}$ & 17.4$^{+1.9}_{-1.6}$ & 0.40$^{+0.06}_{-0.05}$\\
0.535283 & 19.987016 & 385805170 & 2846319155418023424 & 15.581263$\pm$0.002853 & 0.4198$\pm$0.0496 & 31560$\pm$722 & 5.58$\pm$0.08 & sdB & 0.035$\pm$0.002 & -11.348$^{+0.008}_{-0.008}$ & 0.233$^{+0.035}_{-0.024}$ & 48.5$^{+16.3}_{-10.4}$ & 0.77$^{+0.29}_{-0.20}$\\
0.981681 & 27.810404 & 492112073 & 2854091981071865344 & 13.308936$\pm$0.002796 & 1.7692$\pm$0.0329 & 28000$\pm$572 & 5.56$\pm$0.06 & sdB & 0.048$\pm$0.002 & -10.826$^{+0.041}_{-0.010}$ & 0.188$^{+0.003}_{-0.004}$ & 19.4$^{+1.7}_{-1.7}$ & 0.47$^{+0.08}_{-0.06}$\\
1.8907183 & 13.5993244 & 619614193 & 2767874292175410560 & 13.047603$\pm$0.002814 & 2.0899$\pm$0.0285 & 29560$\pm$577 & 5.41$\pm$0.05 & sdB & 0.11$\pm$0.005 & -10.746$^{+0.015}_{-0.029}$ & 0.188$^{+0.003}_{-0.003}$ & 24.0$^{+2.4}_{-1.7}$ & 0.33$^{+0.05}_{-0.04}$\\
2.1021972 & 49.083822 & 593009050 & 393589879591384576 & 15.888843$\pm$0.002816 & 0.6389$\pm$0.0432 & 26640$\pm$899 & 5.53$\pm$0.09 & sdB & 0.108$\pm$0.002 & -11.293$^{+0.010}_{-0.006}$ & 0.178$^{+0.014}_{-0.013}$ & 14.3$^{+3.5}_{-2.4}$ & 0.40$^{+0.14}_{-0.09}$\\
2.188996 & 12.288674 & 66613068 & 2766032919436757504 & 14.60169$\pm$0.002839 & 1.183$\pm$0.0489 & 32230$\pm$654 & 6.00$\pm$0.09 & sdB & 0.071$\pm$0.003 & -11.127$^{+0.012}_{-0.017}$ & 0.139$^{+0.007}_{-0.005}$ & 18.8$^{+2.8}_{-2.0}$ & 0.70$^{+0.22}_{-0.15}$\\
\enddata
\end{deluxetable*}
\end{longrotatetable}

\subsection{Mass vs. atmospheric parameters}

\begin{figure}

    \includegraphics[width=90mm]{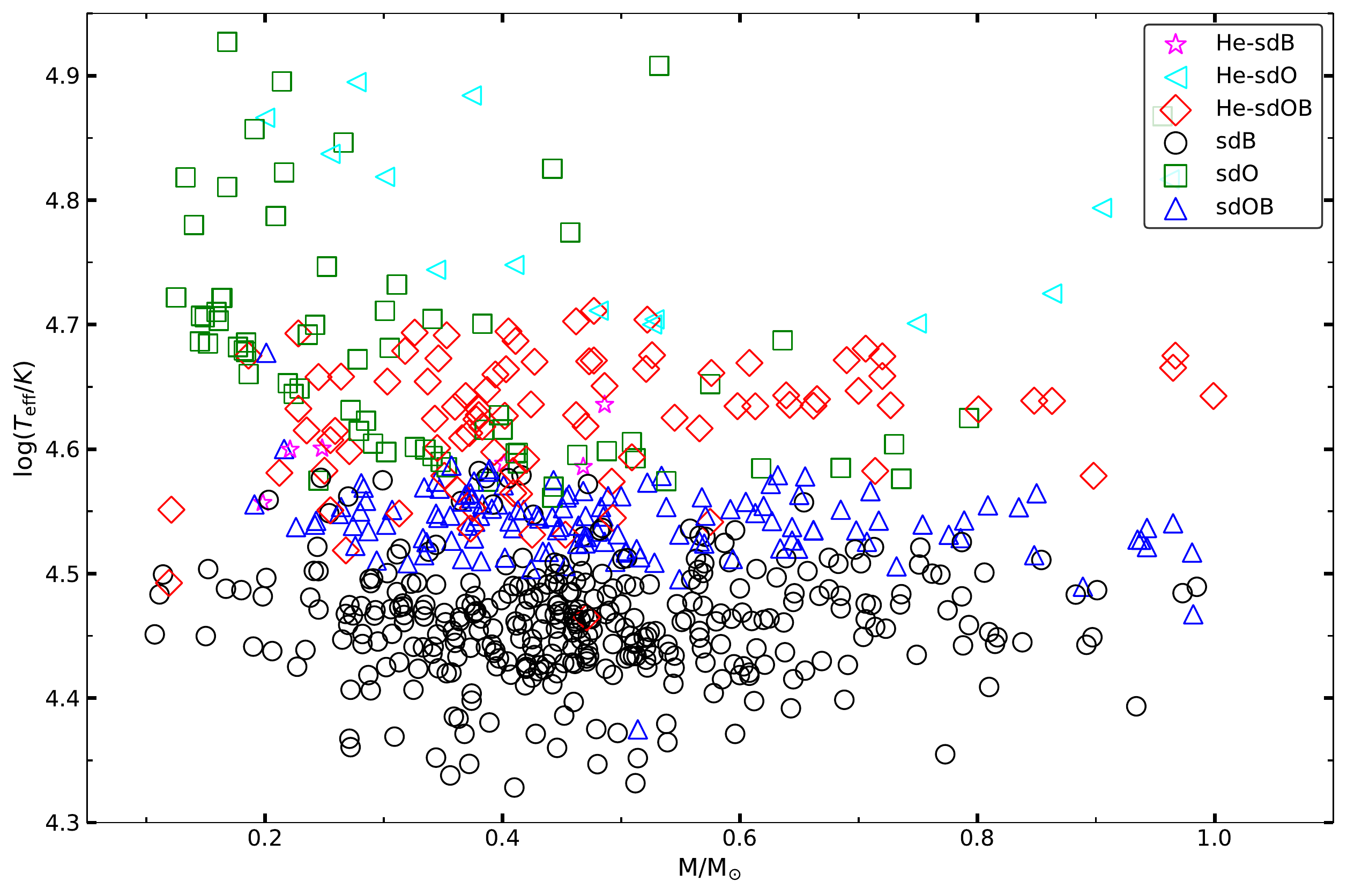}
    \includegraphics[width=89mm]{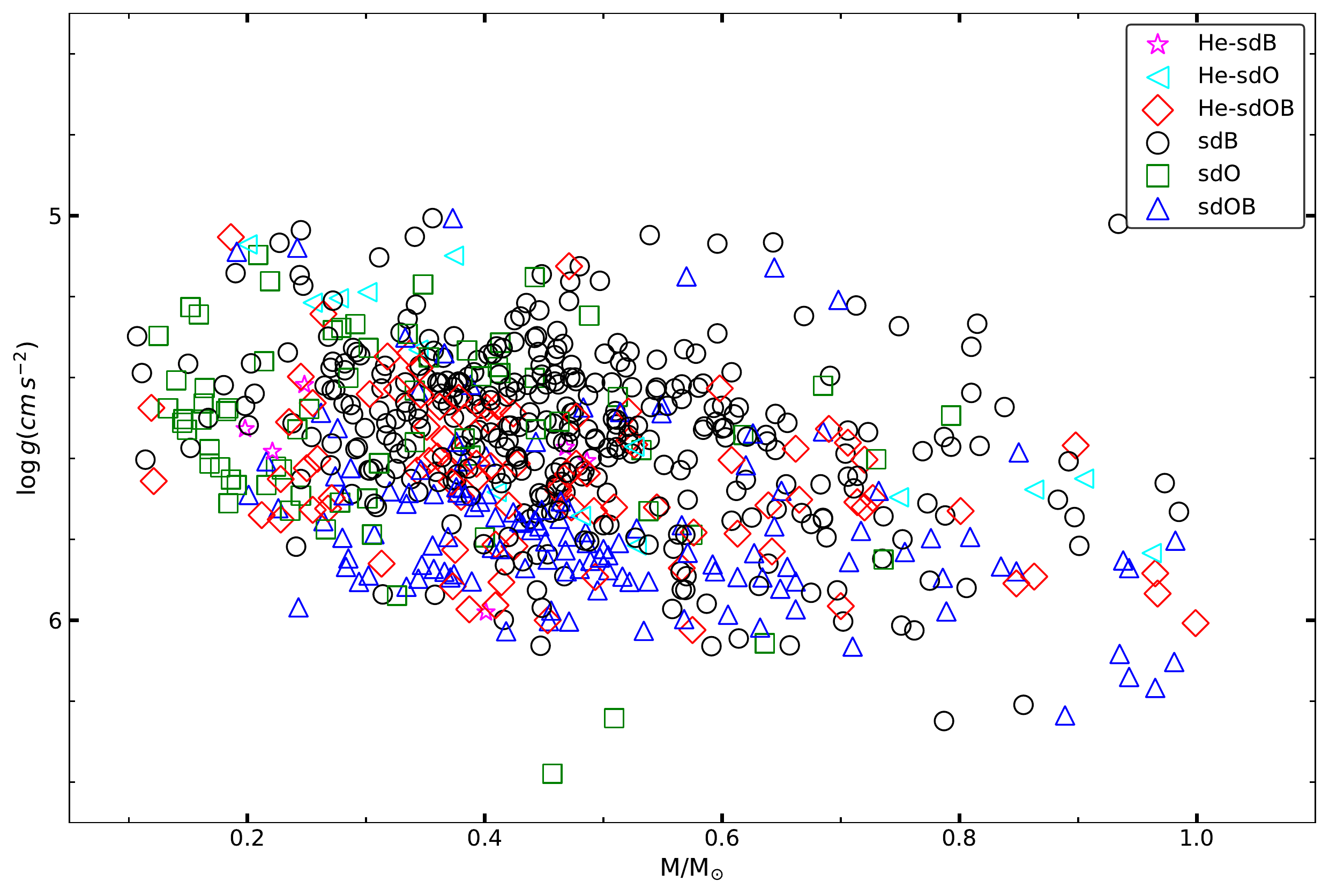}\\
    \includegraphics[width=90mm]{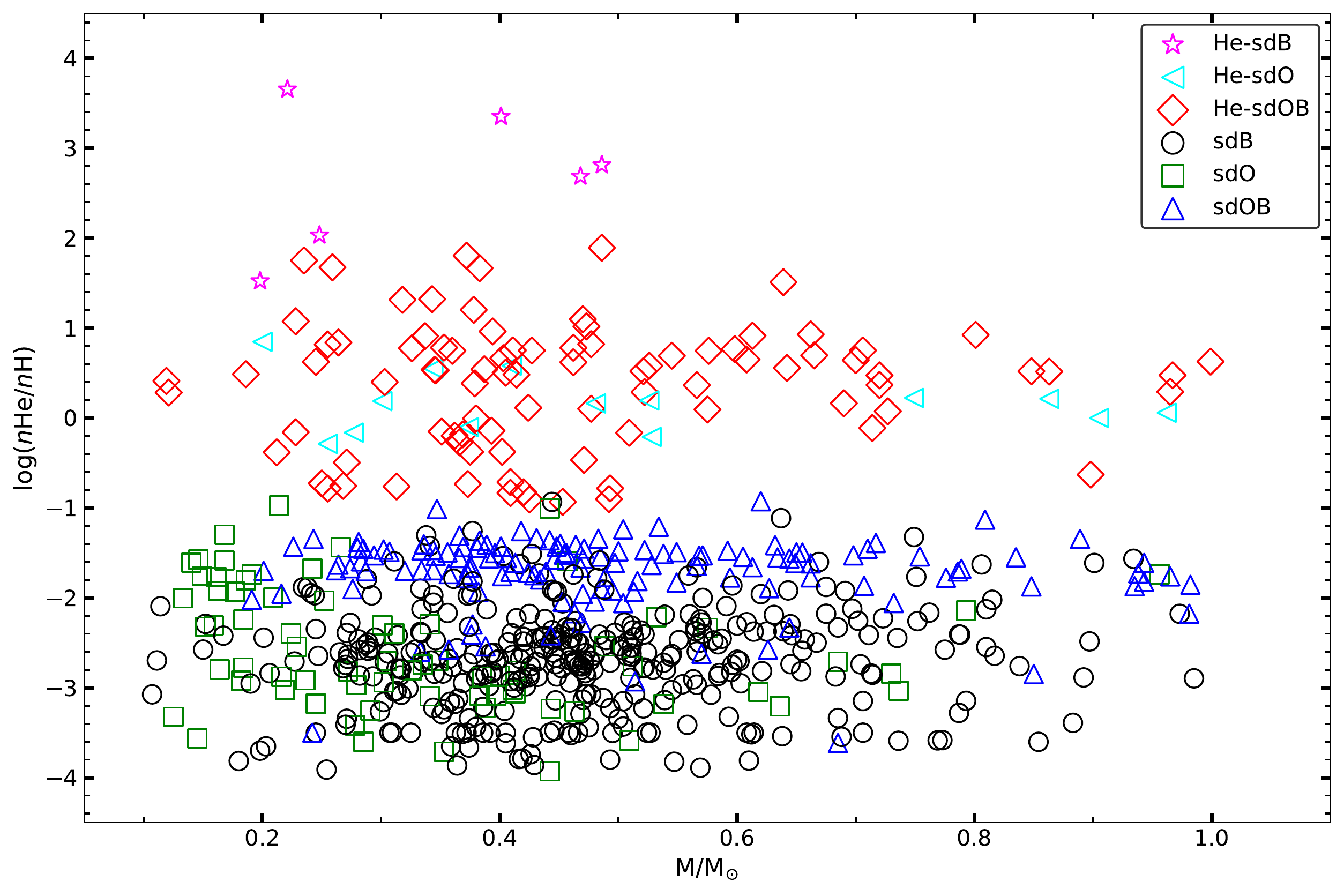}
    \includegraphics[width=89mm]{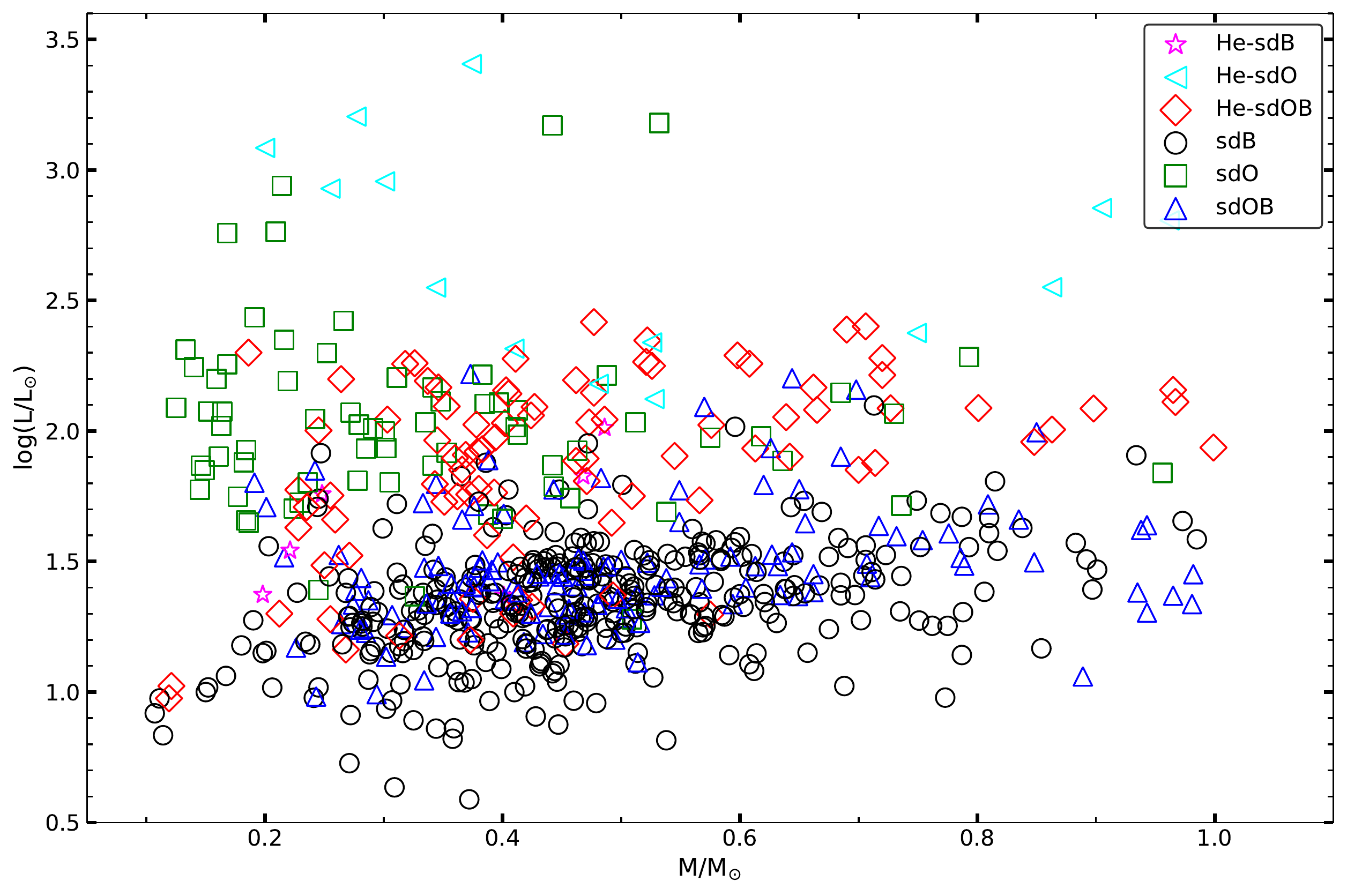}
    
    \caption{Relationship between mass and atmospheric parameters for the selected hot subdwarfs. 
    From upper left to bottom right, it shows mass - $T_{\rm eff}$ plane, mass -- $\log{g}$ plane, mass -- $\log{(n{\rm He}/n{\rm H})}$ plane and mass -- luminosity plane, respectively. 
    The labels with different colors denote the spectral classification of hot subdwarfs from \citet{2018ApJ...868...70L, 2019ApJ...881..135L, 2020ApJ...889..117L}. Parameter errors are not shown for clarity.}
    \label{fig2}
\end{figure}

The four panels in Fig \ref{fig2} show the relationships we found between masses and atmospheric parameters (e.g., $T_{\rm eff}$, $\log{g}$, $\log{(n{\rm He}/n{\rm H})}$ and luminosity) for the selected hot subdwarfs. 
As one can see that, no obvious characteristics are contained in mass -- $T_{\rm eff}$ plane (upper left panel), mass -- $\log{g}$ plane (upper right panel) and mass -- $\log{(n{\rm He}/n{\rm H})}$ plane (bottom left panel). Note that the detailed discussion of the relationships among atmospheric parameters for the selected hot subdwarfs can be found in Section 3.2 of \citet{2018ApJ...868...70L} and Section 4.1 of \citet{2019ApJ...881..135L}. 

 \begin{figure}
    \centering
    \includegraphics[width=120mm]{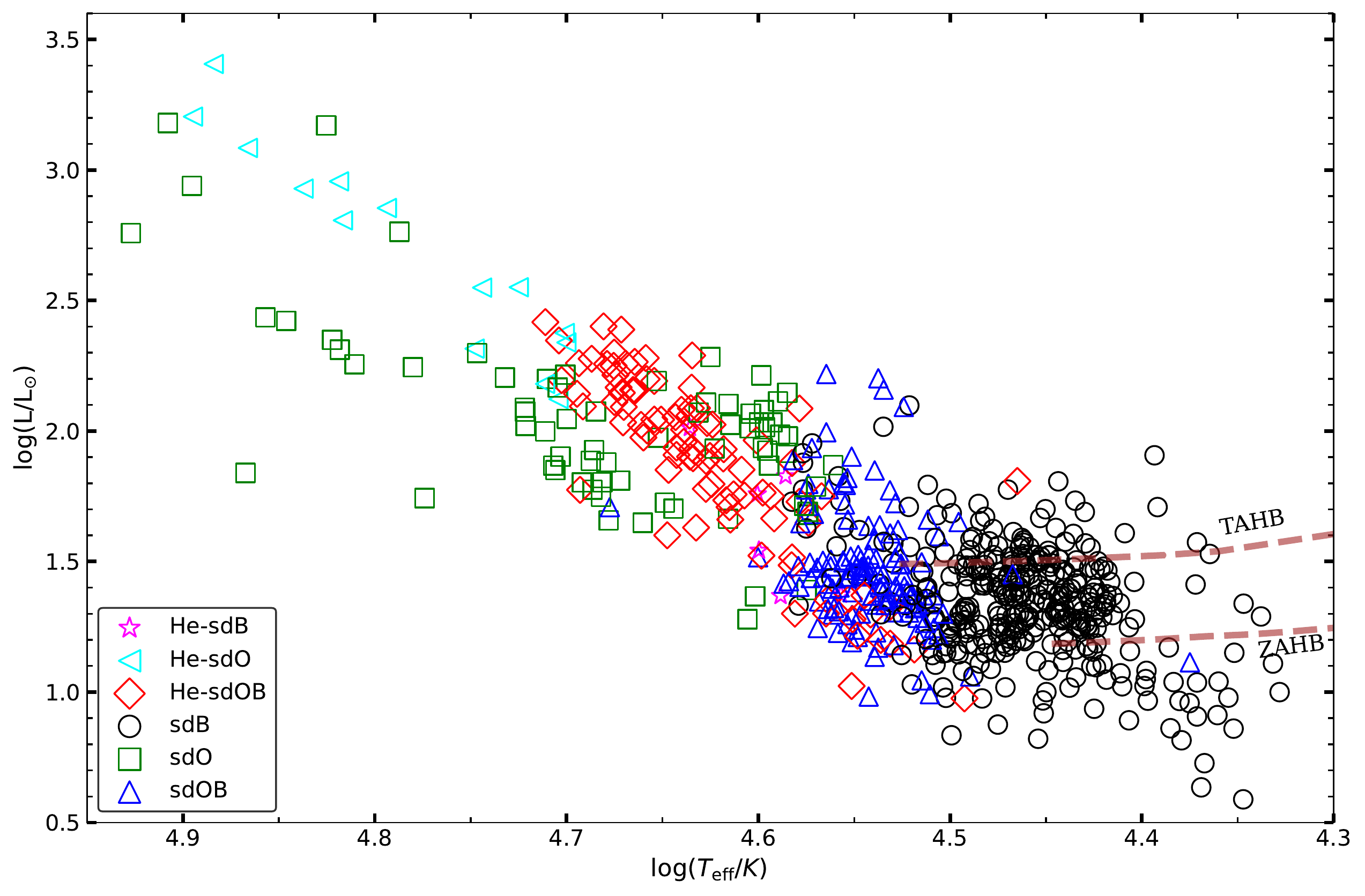}
    \caption{$T_{\rm eff}$ -- luminosity plane for the hot subdwarf stars selected in this study. 
    The labels with different colors denote the same spectral classifications as in Fig \ref{fig2}. 
    The two brown dashed lines denote the ZAHB and TAHB positions for HB stars with $[{\rm Fe}/{\rm H}]=-1.48$ from \citet{1993ApJ...419..596D}.}
    \label{fig3}
\end{figure}

On the other hand, as shown in the mass -- luminosity plane (bottom right panel), though a wide mass distribution (e.g., roughly from 0.1 to 1.0 ${\rm M}_{\odot}$, see Section 3.3 for detailed discussion) is presented, most of sdB (black circles) and sdOB (blue triangles) stars settle in a narrow range at a relatively low luminosity (e.g., roughly from $\log{(L/L_{\odot})}=0.5$ to 1.5), which demonstrates that these stars are still at the core He burning stage. 
 
However, most sdO (green squares) and He-rich stars, i.e., He-sdO (aqua right triangles), He-sdOB (red diamonds), and He-sdB (magenta stars), show a much higher and wider range of luminosities than sdB and sdOB stars. 
Some of them even present luminosity near 3000 $L_{\odot}$ (e.g., $\log{(L/L_{\odot})}=3.5$). 
These results demonstrate that they could either have evolved off the core He-burning stage and now are on the way to the WD cooling curve, or could be pre-EHB stars that just cross the hot subdwarf region. 
This result is also clearly shown in Fig \ref{fig3}, in which the relationships between $T_{\rm eff}$ and luminosity for the selected hot subdwarfs are presented. As shown in the figure, most sdB and sdOB stars show a much narrower $T_{\rm eff}$ and luminosity range than that of sdO and He-rich hot subdwarfs (also see Fig 10 in \citealt{2022A&A...666A.182S}), and located well in the region defined by the zero-age horizontal branch (ZAHB) and the terminal-age horizontal branch (TAHB) lines.  

 \begin{figure}
    \centering
    \includegraphics[width=130mm]{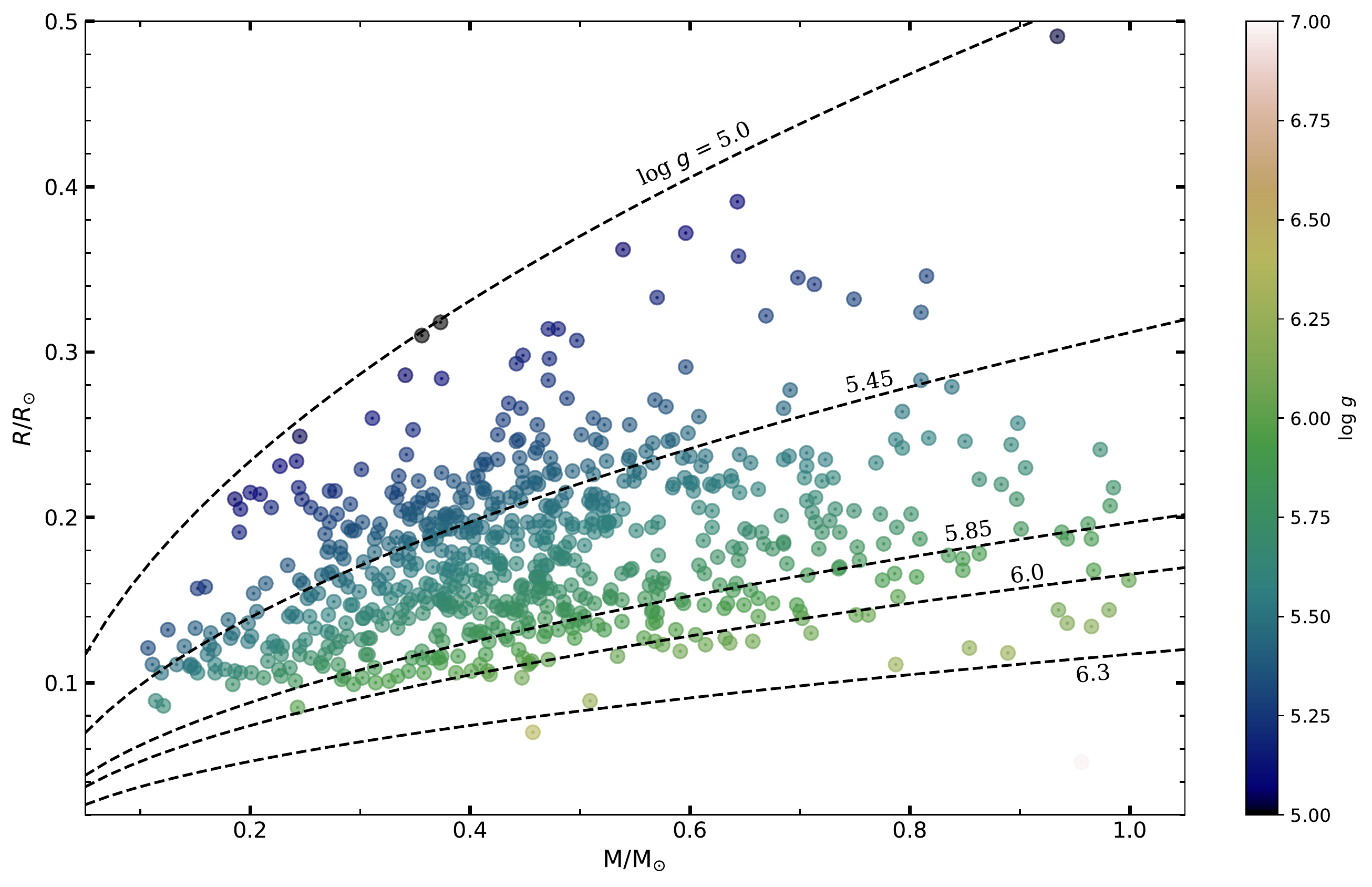}
    \caption{Mass -- radius plane for the hot subdwarf stars selected in this study. Color bar in the right denotes different $\log{g}$ values, while black dashed curves from top to bottom denote constant $\log{g}$ values of 5.0, 5.45, 5.85, 6.0 and 6.3 dex, respectively.}
    \label{fig4}
\end{figure}

Fig \ref{fig4} shows the relationship between mass and radius for our selected sample. 
$\log{g}$ values for each object were denoted by the color gradient showed in the right. Most of hot subdwarf stars in our sample have radius between 0.1 and 0.4 R$_{\odot}$, while most of the objects have $\log{g}$ values between 5.0 and 6.3 dex. The black dashed curves reveal the relationship between mass and radius that square of radius has positive correlation with stellar mass when $\log{g}$ keen constant (see Equation \ref{eq:2}). Since all spectral type of hot subdwarfs present relative wide distributions of $\log{g}$ (see the upper right panel in Fig \ref{fig2}), it is not easy to distinguish them in this figure. Therefore, stars with different spectral classifications were not denoted by different labels as in Fig \ref{fig2} and \ref{fig3}. 

\subsection{Mass values comparison with other studies} 

\begin{figure}
    \centering
    \includegraphics[width=140mm]{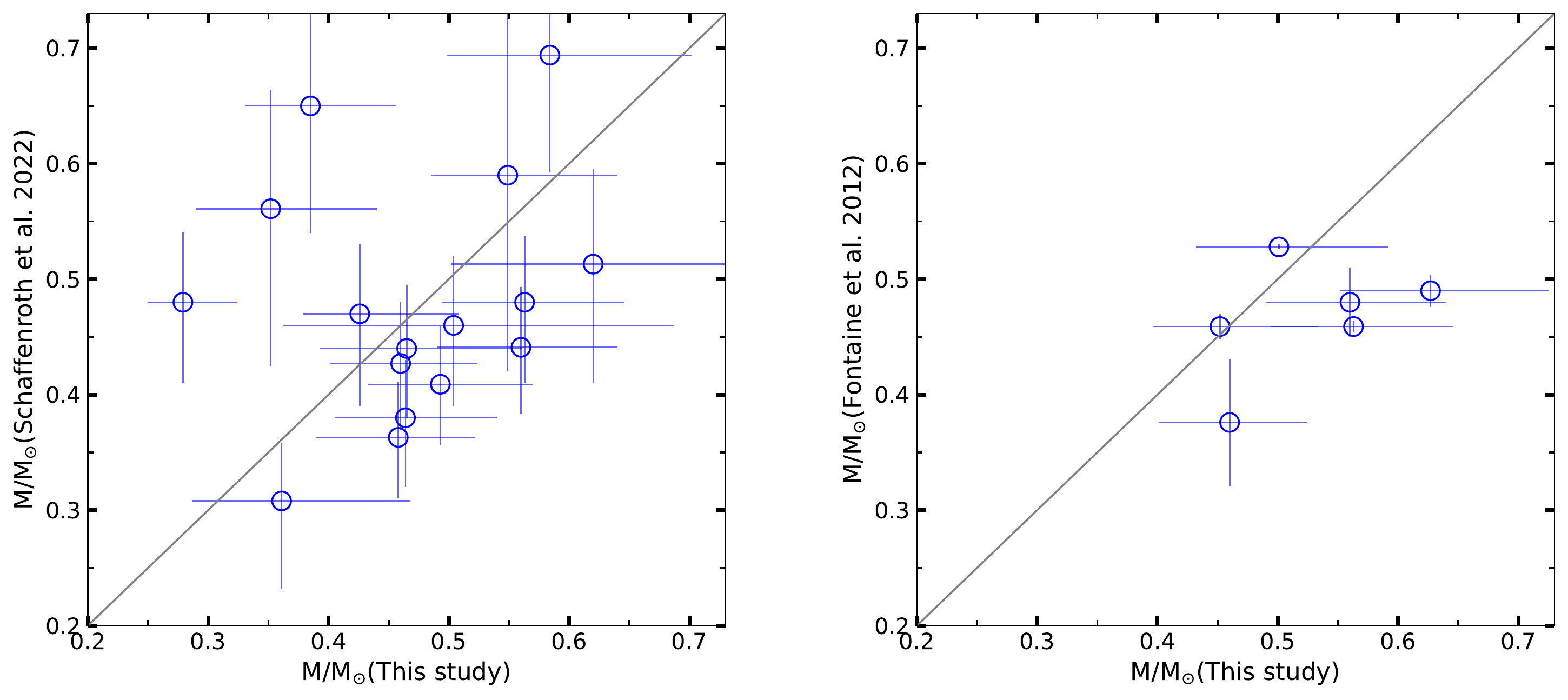}
    \caption{Left panel: Mass comparison between this study and \citet{2022A&A...666A.182S}. 
    Right panel: Mass comparison between this study and \cite{2012A&A...539A..12F}.}
    \label{fig5}
\end{figure}

To check the reliability of the mass obtained in this study, we cross-matched the selected stars with the sample analyzed by \citet{2022A&A...666A.182S} and \citet{2012A&A...539A..12F} (see Section 1), and found 18 and 6 common objects, respectively. 
In Fig \ref{fig5}, we compared masses obtained in this study with the values of common objects in \citet{2022A&A...666A.182S} (left panel) and \citet{2012A&A...539A..12F} (right panel). 
Although considerable dispersion is present in both panels, the masses obtained in this study are consistent with the two former studies when the uncertainties are also considered. 
Note that, the uncertainties for our mass values are mainly from the large uncertainties of $\log{g}$ and distances. 
Considering that we have a large number of samples (e.g., 664 hot subdwarf stars), it will not influence much the analysis and interpretation of the mass distribution in this study. 

\begin{figure}
    \centering
    \includegraphics[width=140mm]{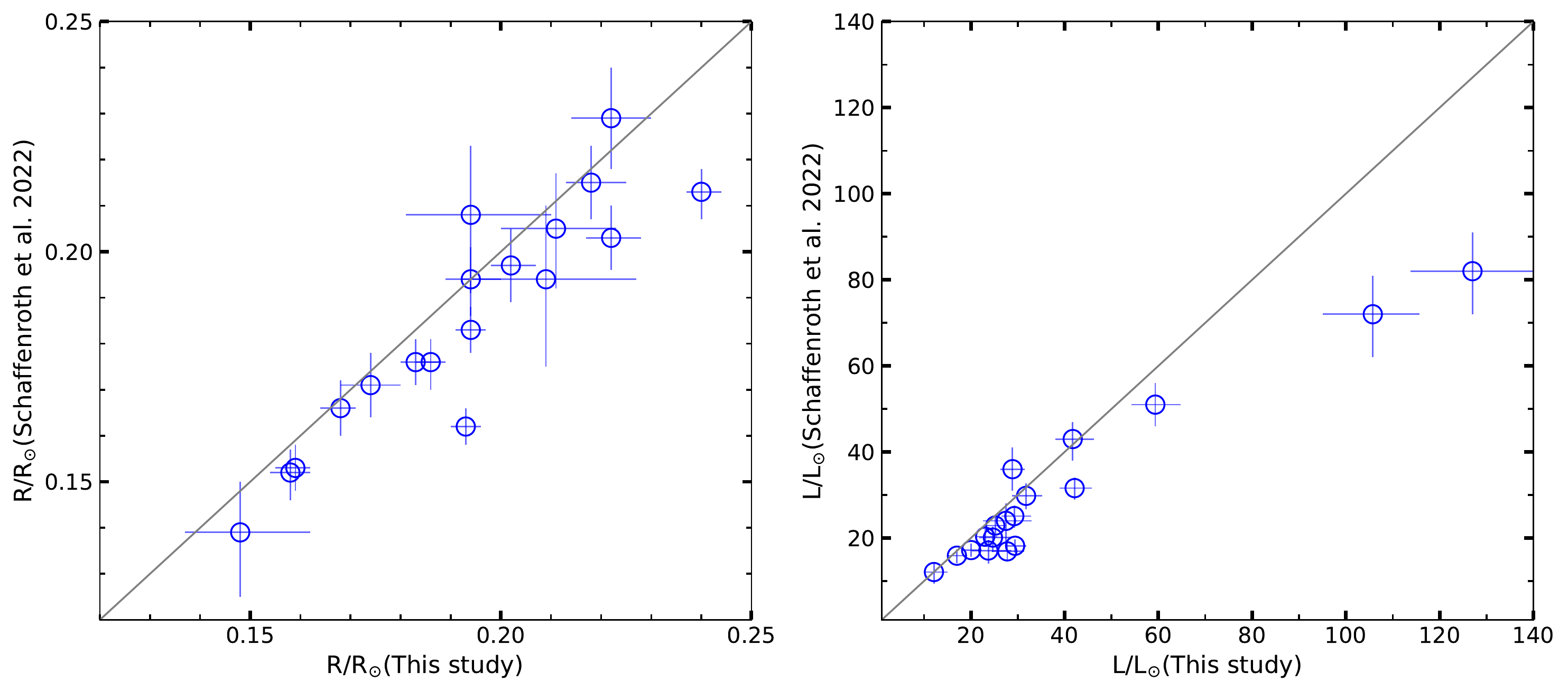}
    \caption{Radius and luminosity comparison between this study and \citet{2022A&A...666A.182S}}
    \label{fig6}
\end{figure}

In Fig \ref{fig6}, we also compared the radius and luminosity obtained in this study with the values by \citet{2022A&A...666A.182S}. 
As shown in the two panels, since $\log{g}$ values were not used when calculating radius and luminosity, both the radius and luminosity values show much better consistency than the direct mass comparison between the two studies.  

\begin{deluxetable*}{c|cccccc}
\tablenum{2}
\tablecaption{Statistical properties for hot subdwarf stars in three groups.}
\tablewidth{120pt}
\tabletypesize{\normalsize}
\tablehead{
\colhead{Group} &\colhead{Subset} & \colhead{Total} & 
\colhead{Peak mass} &  
\colhead{Mean mass} &\colhead{Massive sample }  & \colhead{Massive sample} \\
\colhead{Name} &\colhead{Name} &  Number  & \colhead{$\mathrm{M}_{\odot}$} & \colhead{$\mathrm{M}_{\odot}$} & \colhead{Number} & \colhead{Fraction(\%)}
}
\startdata
 &All sample & 664 & 0.46 & 0.46 & 136 & 20  \\
\textbf{Group 1} & sdB/sdOB & 483 & 0.46 & 0.48 & 105 & 22\\
(\textbf{$\sigma_{\varpi}/\varpi \leq 0.2$}) & He-rich sample & 110 & 0.40 & 0.46 & 24 & 22\\
 &sdO & 71 & 0.18 & 0.34 & 7 & 10 \\
 \hline 
&All sample & 503 & 0.46 & 0.47 & 100 & 20  \\
\textbf{Group 2} & sdB/sdOB & 376 & 0.46 & 0.48 & 80 & 21\\
(\textbf{$\sigma_{\varpi}/\varpi \leq 0.1$}) & He-rich sample & 75 & 0.36 & 0.46 & 15 & 20\\
 &sdO & 52 & 0.18 & 0.33 & 5 & 10 \\
  \hline 
&All sample & 289 & 0.46 & 0.48 & 58 & 20  \\
\textbf{Group 3} & sdB/sdOB & 220 & 0.46 & 0.48 & 44 & 20\\
(\textbf{$\sigma_{\varpi}/\varpi \leq 0.05$}) & He-rich sample & 39 & 0.42 & 0.50 & 9 & 23\\
 &sdO & 30 & 0.18 & 0.36 & 5 & 16 \\
\enddata
\tablecomments{We define massive samples as objects with a mass larger than 0.6 $\mathrm{M}_{\odot}$.}
\end{deluxetable*}

\subsection{Mass distributions for the selected hot subdwarf stars}

To study the mass distribution at different parallax precisions, we selected three groups of hot subdwarf stars from our sample for a mass distribution analysis, i.e., objects with $\sigma_{\varpi}/\varpi \leq 0.2$ were selected as Group 1, objects with $\sigma_{\varpi}/\varpi\leq 0.1$ were selected as Group 2, and objects with $\sigma_{\varpi}/\varpi\leq0.05$ were selected as Group 3. 
The statistical information for subsets with different spectral types in the three groups is listed in Table 2. 
From left to right, it gives the group name, subset name, number count, peak mass of the distribution, mean mass, number of stars in the massive sample, and the relative fraction of the massive sample to the total, respectively.

With an increasing parallax precision the number of stars within each group decreases. 
Group 1 contains all the 664 selected hot subdwarfs, Group 2 has 503 hot subdwarfs, and Group 3 consists of 289 hot subdwarfs. 
The number fractions of each subset with different spectral types are consistent among the three groups, e.g., sdB/sdOB stars are the largest population in each group which represents about 73\% of the total, while He-rich stars (He-rich sdB/sdOB/sdO) and sdO stars in each group present about 16\% and 10\%, respectively. 
Furthermore, the peak of the mass distribution for each subset in the three groups is nearly the same, e.g., sdB/sdOB stars present a peak mass of about 0.46 ${\rm M}_{\odot}$ (also see Fig \ref{fig8}), He-rich stars present a peak mass at 0.4 ${\rm M}_{\odot}$, while sdO stars present an inconspicuous peak mass of about 0.18 ${\rm M}_{\odot}$.

 \begin{figure}
    \includegraphics[width=0.5\textwidth]{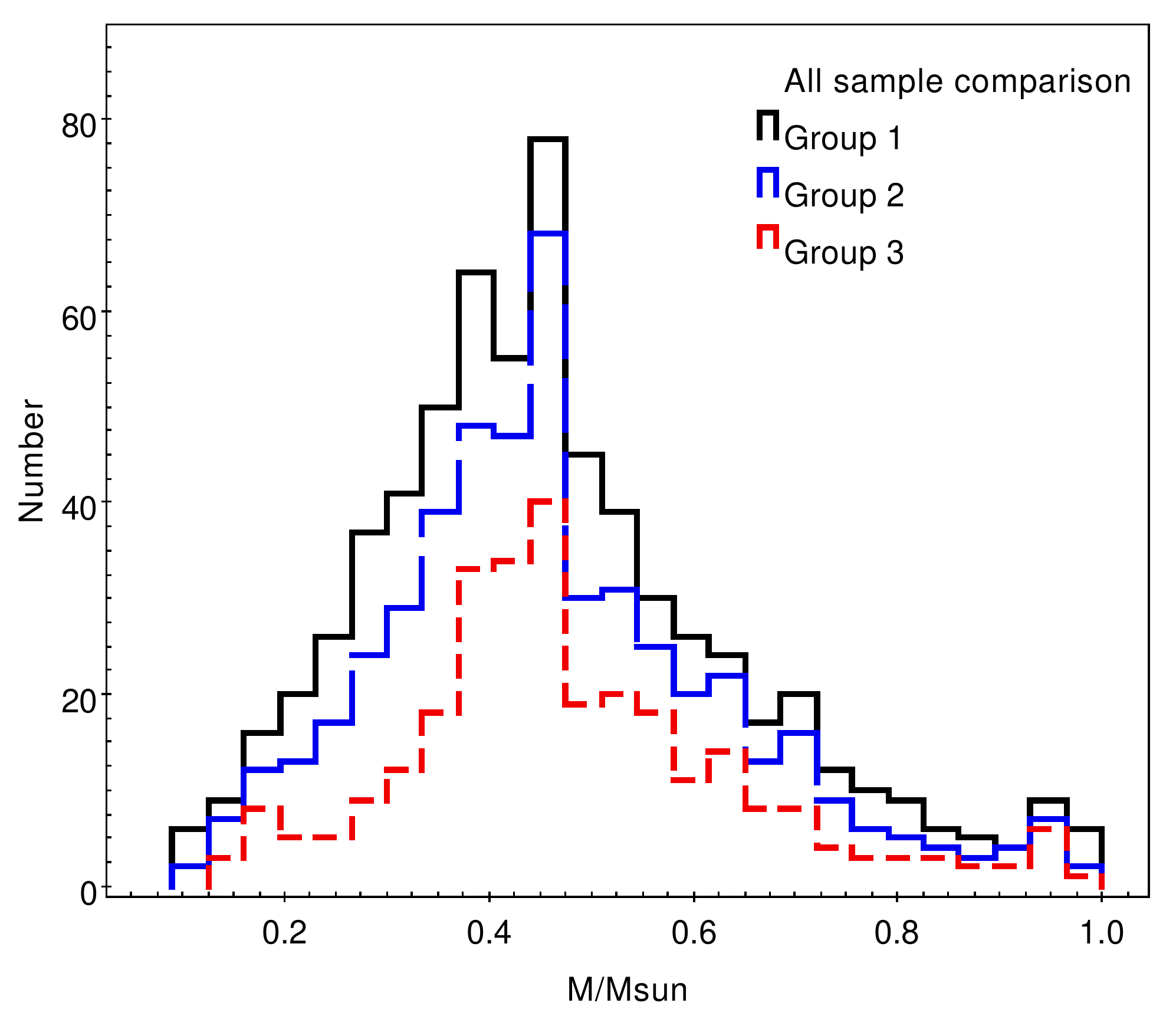}
    \includegraphics[width=0.5\textwidth]{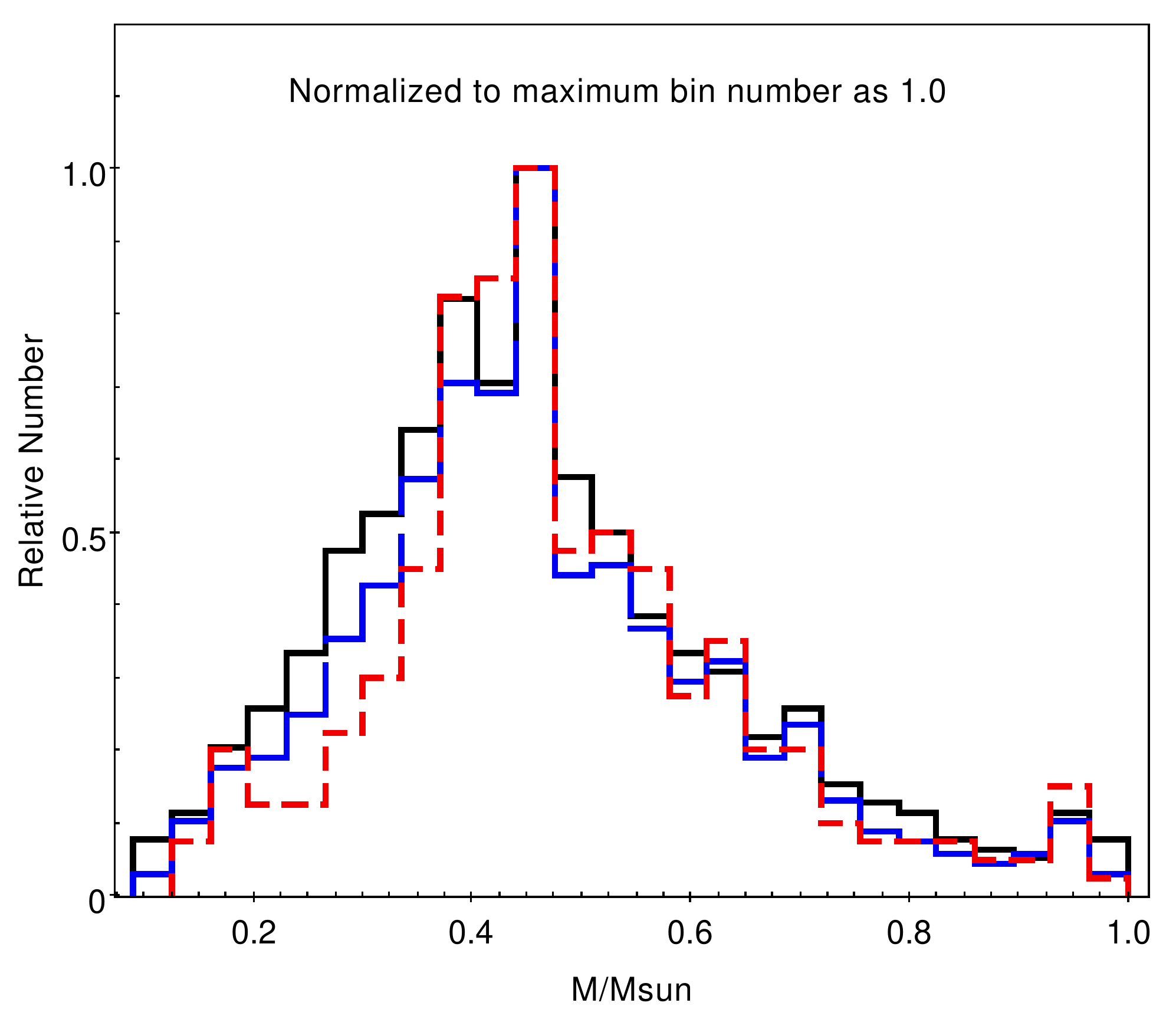}
    \caption{Mass distributions for all types of hot subdwarf stars in three groups. Left panel: Absolute numbers in each bin. Right panel: Numbers in each group were normalized to the maximum. Black solid, blue long-dashed, and red short-dashed histograms denote the mass distribution for hot subdwarf stars in Groups 1, 2, and 3, respectively. The bin size is set to 0.035 ${\rm M}_{\odot}$, and 26 bins are present in the mass range of 0.1 to 1.0 ${\rm M}_{\odot}$.}
    \label{fig7}
\end{figure}

The left panel in Fig \ref{fig7} shows the mass distributions for the whole sample in each group. 
The mass distributions are very similar among the three groups, most of the hot subdwarf stars have masses between 0.37 and 0.5 ${\rm M}_{\odot}$, and a sharp peak at 0.46 ${\rm M}_{\odot}$ is present.
However, Group 3 (i.e.,red-dotted histogram) seems to have much fewer stars with a mass between 0.2 and 0.37 ${\rm M}_{\odot}$ than Group 1 (i.e., black-solid histogram) and Group 2 (i.e., blue-dashed histogram). 
This feature is shown more clearly in the right panel of Fig \ref{fig7}, in which the number counts in each bin were normalized. 
One can see that with the parallax (distance) precision increasing, the relative number of less massive objects (e.g., mass between 0.2 and 0.37 ${\rm M}_{\odot}$) decreases, which leads to a sharper peak at the low mass side. 
It means that with larger uncertainties in parallax (distance) the relative contribution of low-mass objects show an increase. 
On the other hand, massive stars (e.g., M $>0.6$) present nearly the same mass distribution among three groups, independent of the parallax precision.

The left panel in Fig \ref{fig8} shows the mass distributions of hot subdwarfs with different spectral types in Group 1.
One can see that the whole sample presents a very wide mass distribution (black solid histogram) from 0.1 ${\rm M}_{\odot}$ to 1.0 ${\rm M}_{\odot}$, with a sharp peak (mean mass) at 0.46 ${\rm M}_{\odot}$. 
Since sdB and sdOB stars have the largest fraction in the group (e.g., 483 objects in a total of 664, or about 73\% in percentage), they show very similar mass distributions (magenta dash-dotted histogram) as the whole sample. 
They also present a sharp mass peak around 0.46 ${\rm M}_{\odot}$ (see Table 2). 

 \begin{figure}
    \includegraphics[width=0.333\textwidth]{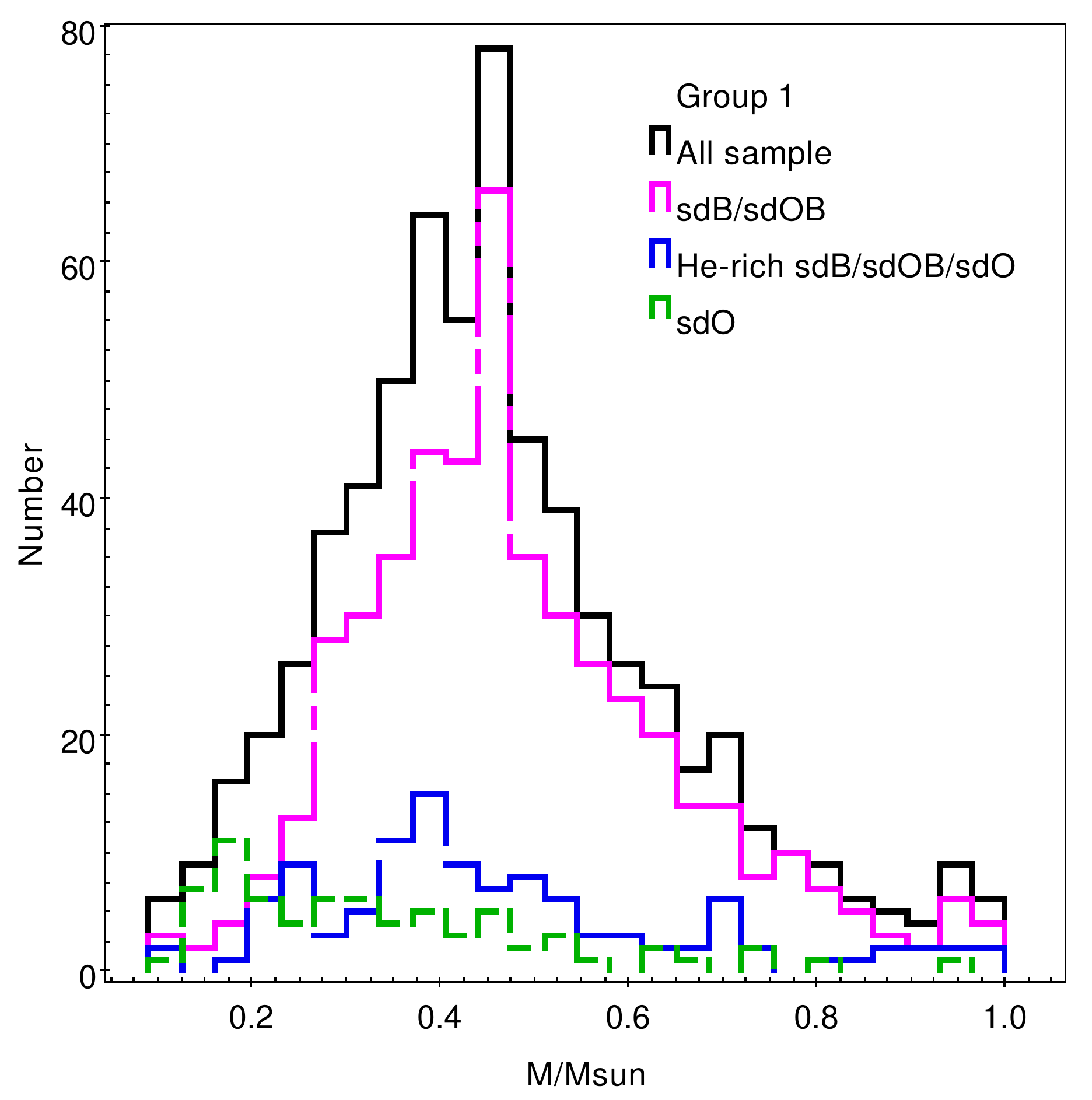}
    \includegraphics[width=0.333\textwidth]{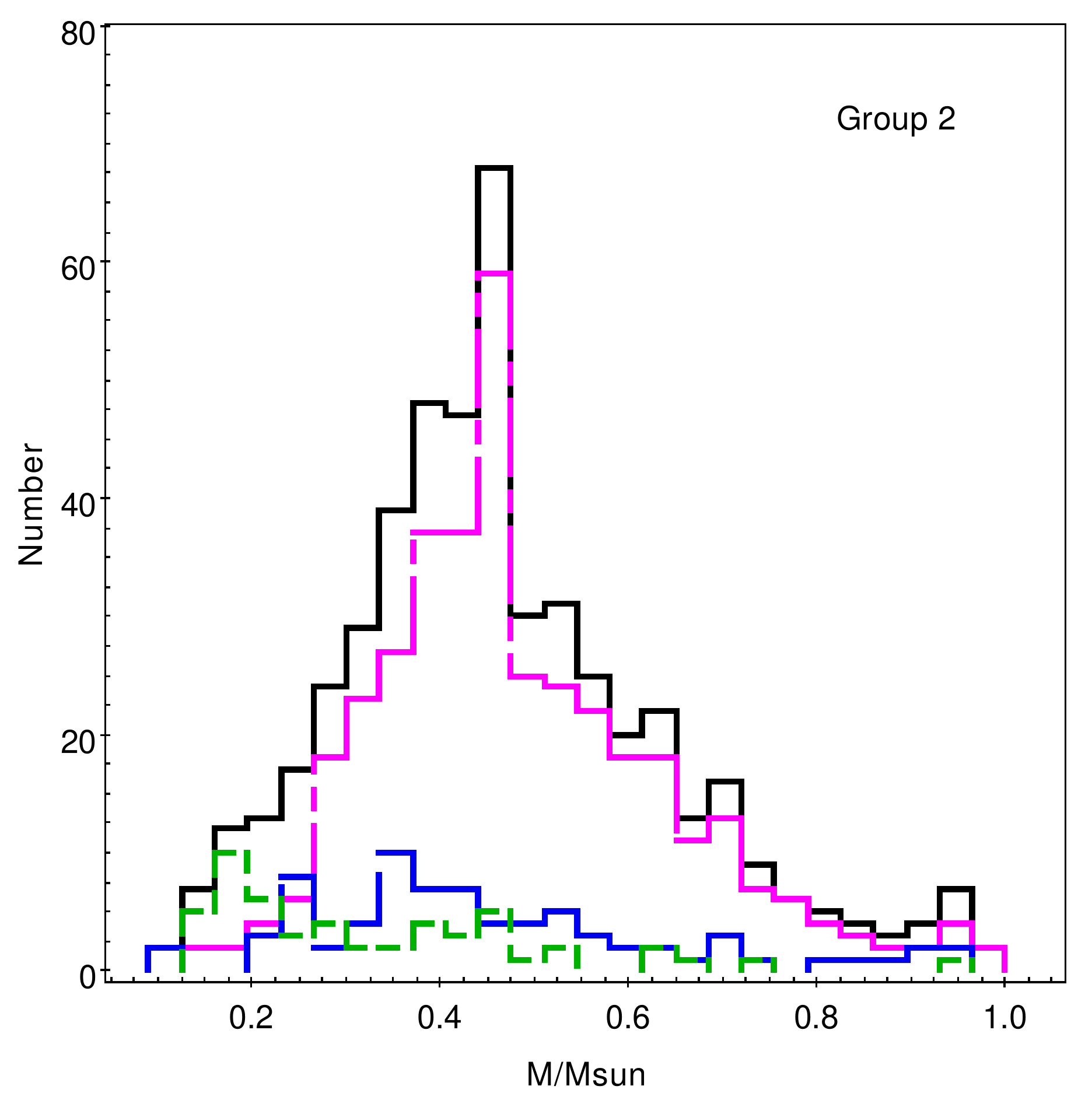}
    \includegraphics[width=0.333\textwidth]{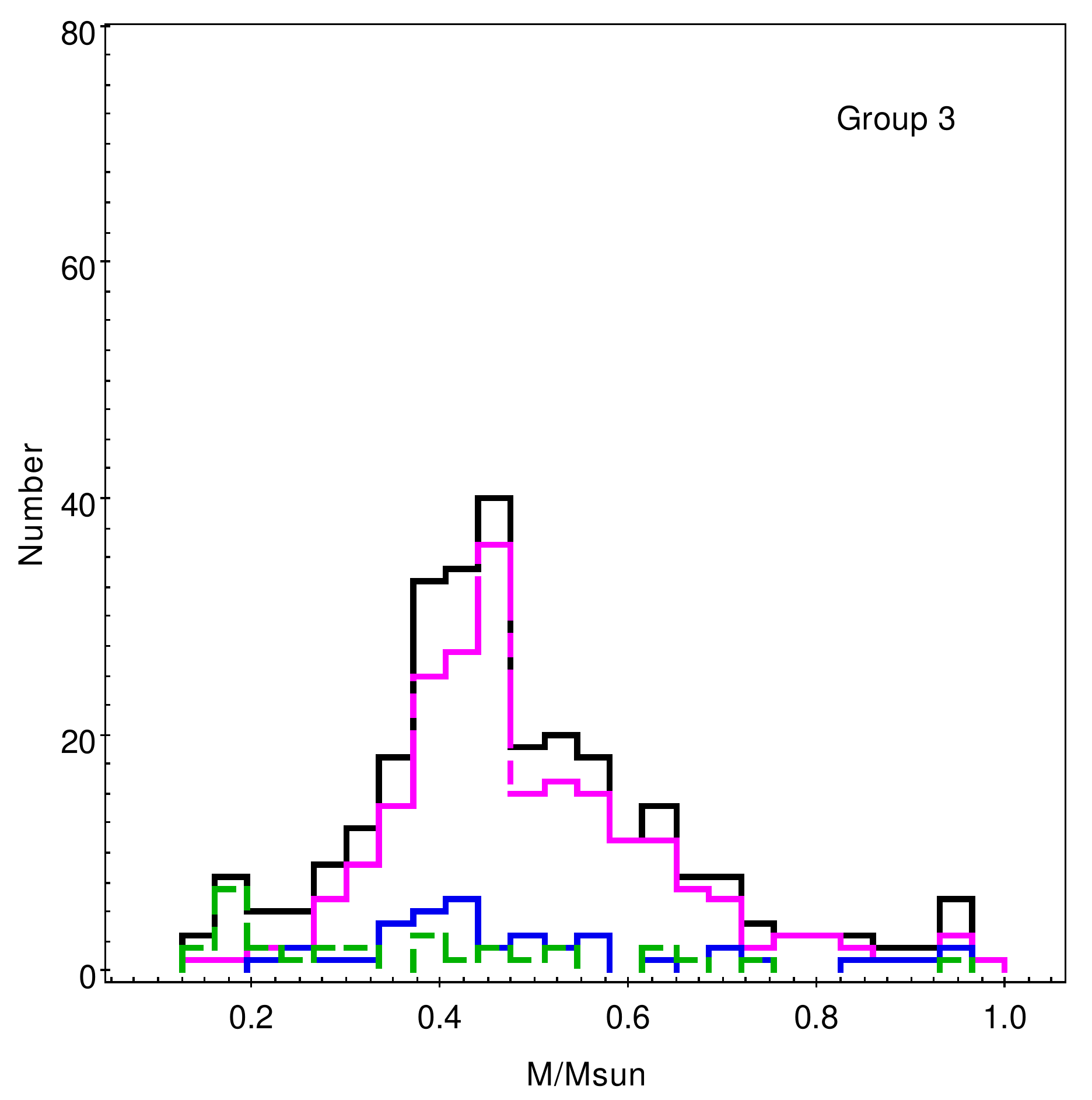}    
    \caption{Mass distribution for the selected hot subdwarfs with different spectral classifications in three groups. In each group, the mass  distributions for all samples, sdB/sdOB stars, He-rich stars, and sdO stars are denoted by black solid, magenta dash-dotted, blue dashed, and green dotted histograms, respectively. The bin size is the same as in Fig \ref{fig7}.}
    \label{fig8}
\end{figure}  

On the other hand, both He-rich stars (e.g., He-sdB, He-sdO, He-sdOB) and sdO stars present much flatter mass distributions (denoted by blue dashed and green dotted histograms, respectively) when compared with sdB/sdOB stars. 
Moreover, He-rich stars show a small mass peak near 0.40 ${\rm M}_{\odot}$ and have a mean mass of about 0.46 ${\rm M}_{\odot}$.
While sdO stars present an inconspicuous peak mass around at 0.18 ${\rm M}_{\odot}$ and a mean mass of about 0.34 ${\rm M}_{\odot}$, which is obviously smaller than the values of sdB/sdO and He-rich stars. 
If we consider stars with a mass greater than 0.6 ${\rm M}_{\odot}$ as massive hot subdwarfs, the fraction of massive stars in the sdO sample is about 10\% (e.g., 7 out of 71, see Table 2), while this fraction among other types of hot subdwarfs (e.g., sdB/sdOB, He-rich sample) are both around at 20\%. 
Note that, according to the results of \citet{2002MNRAS.336..449H}, the minimum core mass for He ignition in a star is about 0.3 ${\rm M}_{\odot}$ (see Table 1 in their study). It means that the low-mass sdO stars found in our sample probably did not undergo a core He burning phase (see detailed discussion in Section 4.4). 

The middle and right panel of Fig \ref{fig8} show the mass distribution of hot subdwarfs with different spectral types in Group 2 and 3 respectively. 
Mass distributions for the various subsets in the two groups are very similar to the distributions of the corresponding subsets in Group 1, except that less massive stars become fewer with an increasing parallax precision (distance) from Group 1 to Group 3, as shown in Fig \ref{fig7}. 
Considering that higher parallax precision (distance) means more reliable mass determination, we used hot subdwarf stars in Group 3 to compare the mass distribution with other studies in the following sections.

\section{Discussion}
\subsection{Mass distribution comparison with other studies}

Combining high-quality light curves from the Transiting Exoplanet Survey Satellite (TESS, \citealt{2015JATIS...1a4003R}) and the \emph{K}2 space mission \citep{2014PASP..126..398H} with the fit of SED, distance from \emph{Gaia} and atmospheric parameters from literature, \citet{2022A&A...666A.182S} not only derived the properties of primary and secondary stars in about 200 sdB binaries but also obtained absolute masses, radii, and luminosities for 68 sdB stars with different types of companions. 
They found a broad mass distribution for sdB stars with cool, low-mass stellar or sub-stellar companions (i.e., sdB+dM/BD systems), which present a peak mass at 0.46 ${\rm M}_{\odot}$. 
Though the mass distribution of sdB stars with WD companions (i.e., sdB+WD systems) also has a similar width as sdB+dM/BD systems, they have a much smaller peak mass around 0.38 ${\rm M}_{\odot}$ (see Section 3.2 and Fig 12 in their study). 
Based on this result, they concluded that sdB+dM/BD and sdB+WD systems originate from different populations.

\begin{figure}
    \centering
    \includegraphics[width=86mm]{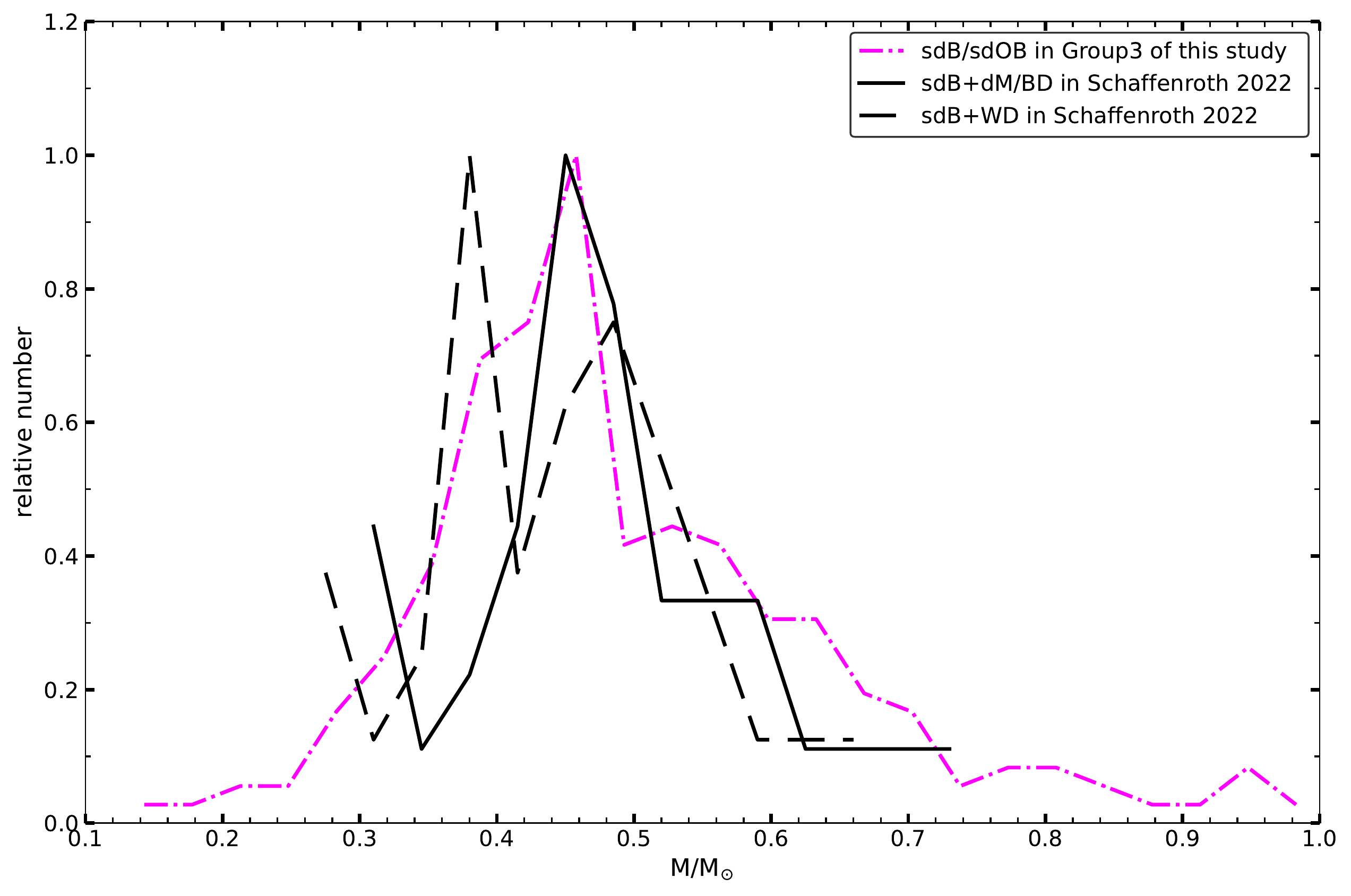}
    \includegraphics[width=86mm] {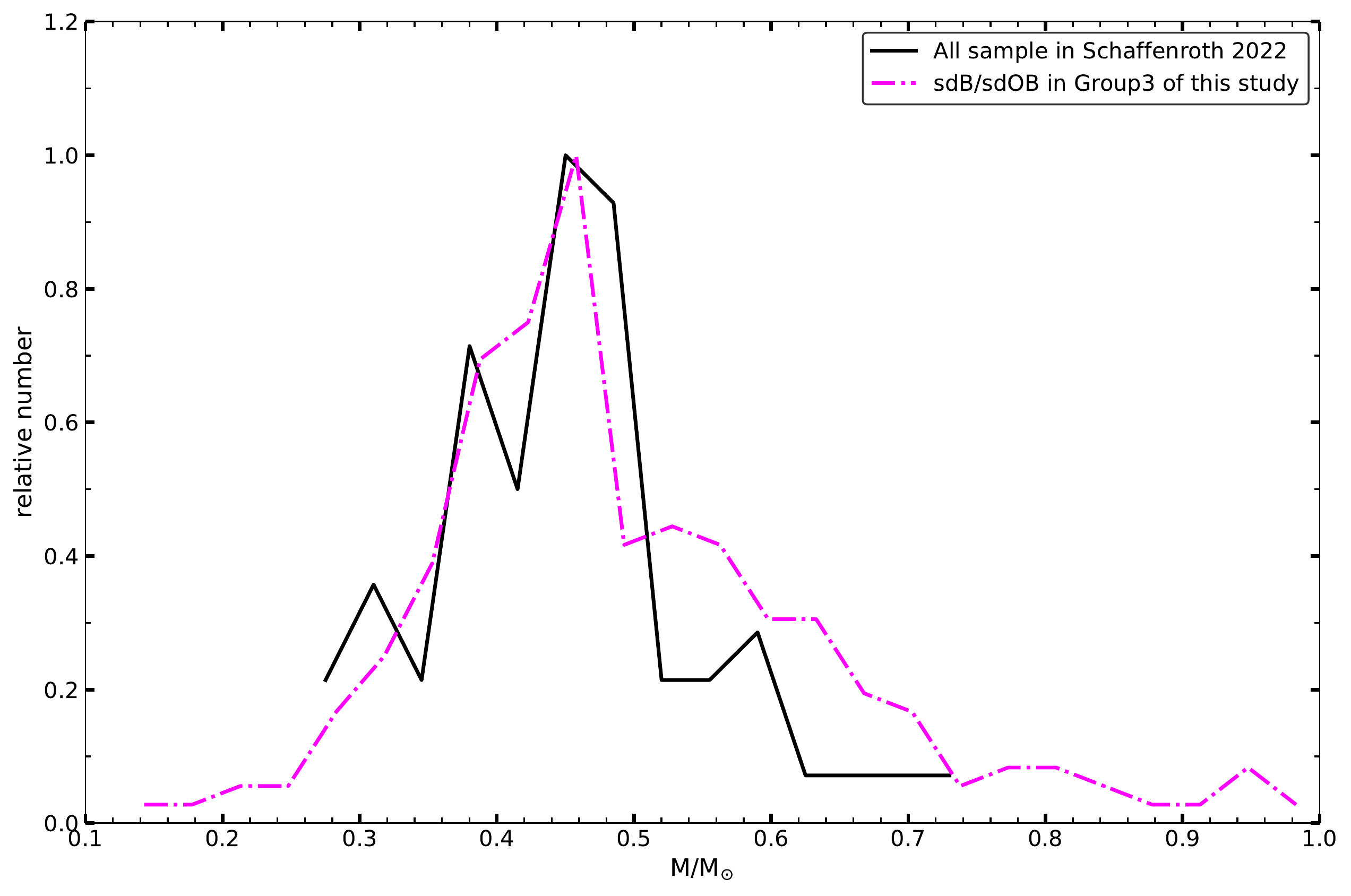}
    \caption{Mass distribution comparison for sdB/sdOB stars in Group 3 with sdB stars in \citet{2022A&A...666A.182S}. 
    Left panel: Mass distribution for sdB+dM/BD (black solid curve) and sdB+WD systems (black dashed curve) in \citet{2022A&A...666A.182S} are plotted separately. 
    Right panel: Both sdB+dM/BD and sdB+WD systems in \citet{2022A&A...666A.182S} were put together to plot the mass distribution (black solid curve). 
    In each panel, the magenta dash-dotted curve denotes the mass distribution for sdB/sdOB stars in Group 3. 
    All distributions were normalized to the maximum values for a better comparison.}
    \label{fig9}
\end{figure}

In the left panel of Fig \ref{fig9}, we compared the mass distributions for sdB+dM/BD and sdB+WD systems obtained in \citet{2022A&A...666A.182S} separately with the mass distribution for sdB/sdOB stars of Group 3 in this study. 
All the sample numbers in each bin are normalized to the maximum bin number which has been set to 1. 
One can see that the mass distribution for sdB/sdOB stars in Group 3 (magenta dash-dotted curve, e.g., roughly from 0.1 to 1.0 ${\rm M}_{\odot}$) is generally wider than the mass distribution for both sdB+dM/BD (black solid curve, e.g., roughly from 0.3 to 0.74 $\mathrm{M}_{\odot}$) and sdB+WD systems (black dashed curve, e.g., from 0.28 to 0.68 ${\rm M}_{\odot}$). 
Note that, the atmospheric parameters for our sample are obtained by fitting H and He profiles in \citet{2018ApJ...868...70L, 2019ApJ...881..135L,2020ApJ...889..117L}, with no metal opacities included. 
The inclusion of metals can potentially change the $T_{\rm eff}$ and $\log{g}$ values, thus resulting in slightly different masses.
Moreover, the number of sdB stars analyzed in this study is much bigger than that of in \citet{2022A&A...666A.182S}, e.g., 220 vs 68. 
Though these factors would act towards a wider mass distribution, they will not influence the observed mass distribution tendencies as significantly. 
The peak of the mass distribution for our sdB/sdOB stars is nearly the same as the one for sdB+dM/BD systems (at 0.46 ${ \rm M}_{\odot}$). 
Furthermore, there is also a second mass peak around 0.38 ${\rm M}_{\odot}$ for our sdB/sdOB stars, which is corresponding to the peak mass of sdB+WD systems in \citet{2022A&A...666A.182S}. 

Note that, we cannot separate sdB+dM/BD and sdB+WD systems in our sample without additional information, e.g., light curves. 
Therefore, we compared the total mass distribution in the right panel of Fig \ref{fig9} for all sdB binaries in \citet{2022A&A...666A.182S} (black solid curve) with the mass distribution for our sdB/sdOB stars in Group 3 (magenta dash-dotted curve). 
We find the two mass distributions to be consistent with each other, especially on the low mass side. 
However, the mass distribution in \citet{2022A&A...666A.182S} seems to present a little wider peak and have more stars with masses around 0.5 ${\rm M}_{\odot}$ than ours. 
It could be due to differences in the atmospheric parameters and their uncertainties adopted in the two studies, especially for $\log{g}$.

\subsection{Mass distribution comparison with theoretical models}

 \citet{2003MNRAS.341..669H} carried out a detailed binary population synthesis to study the formation of sdB stars. 
 Three main formation channels were investigated in their study, i.e., Roche lobe overflow (RLOF), common-envelope (CE) ejection, and merger of two He-WDs. 
 sdB stars produced from the best-fitting model of these channels could satisfy most observational characteristics, such as $P-M_{\rm comp}$ plane, $T_{\rm eff}-\log{g}$ plane, orbital period distribution, mass function distribution, birth rates, etc. 
 To study the effects of input parameters (e.g., metallicity, initial mass-ratio distribution, the critical mass ratio for stable RLOF $q_{\rm crit}$, CE ejection efficiency $\alpha_{\rm CE}$ and thermal contribution to the binding energy of the envelope $\alpha_{\rm th}$, etc) on the formation of sdB stars, they conducted 12 sets of Monte Carlo simulations by varying the parameters in a reasonable range (see Table 1 and 2 in their study). 
 \citet{2003MNRAS.341..669H} give mass distributions for the 12 sets of simulations and chose set 2 as their best-fitting models after comparing their results with observations (see Fig 11 in their study). 
 These results provide a great convenience to study the mass distribution of sdB stars and thus their formation channels.

  \begin{figure}
    \centering
    \includegraphics[width=0.49\textwidth]{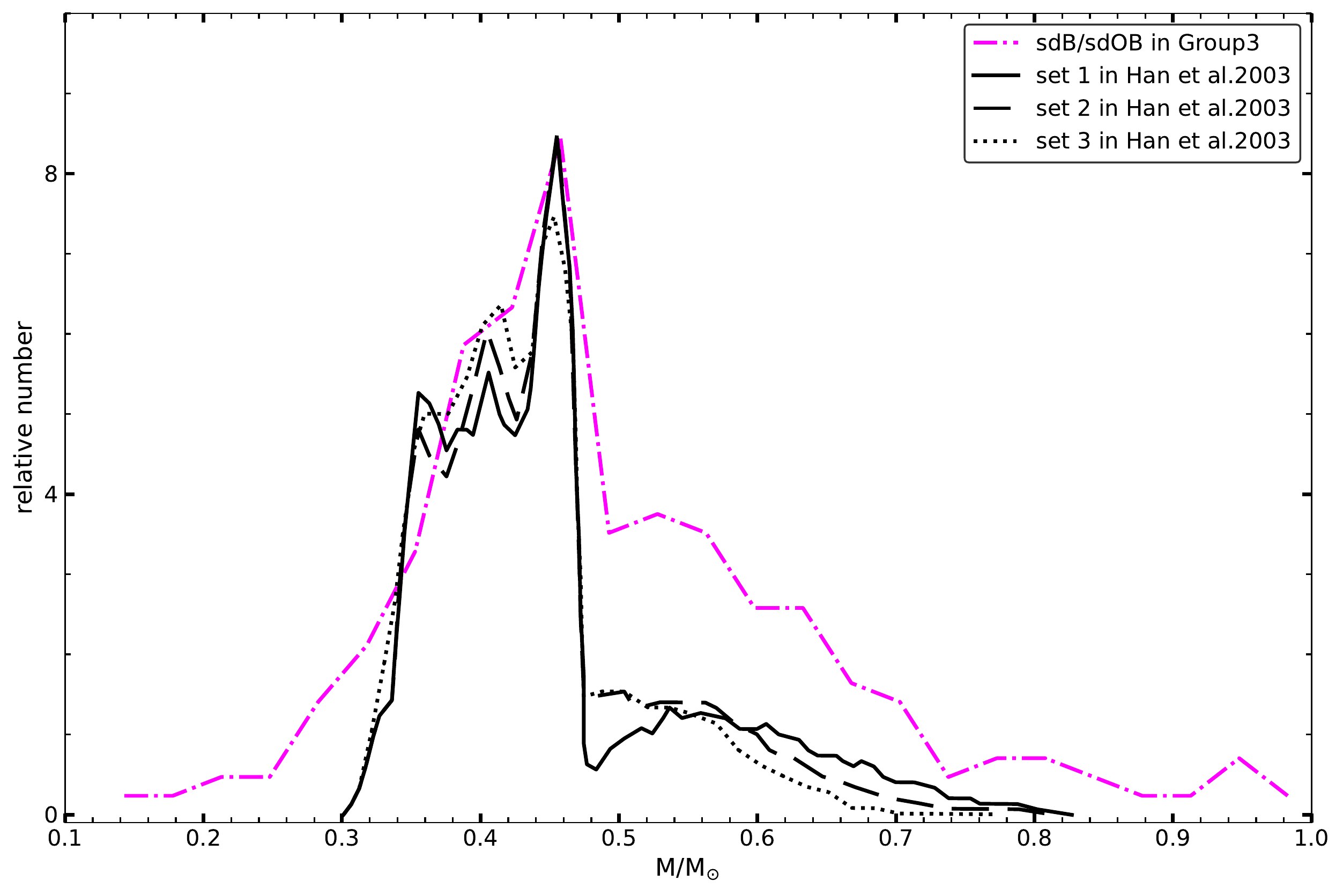}
    \includegraphics[width=0.49\textwidth]{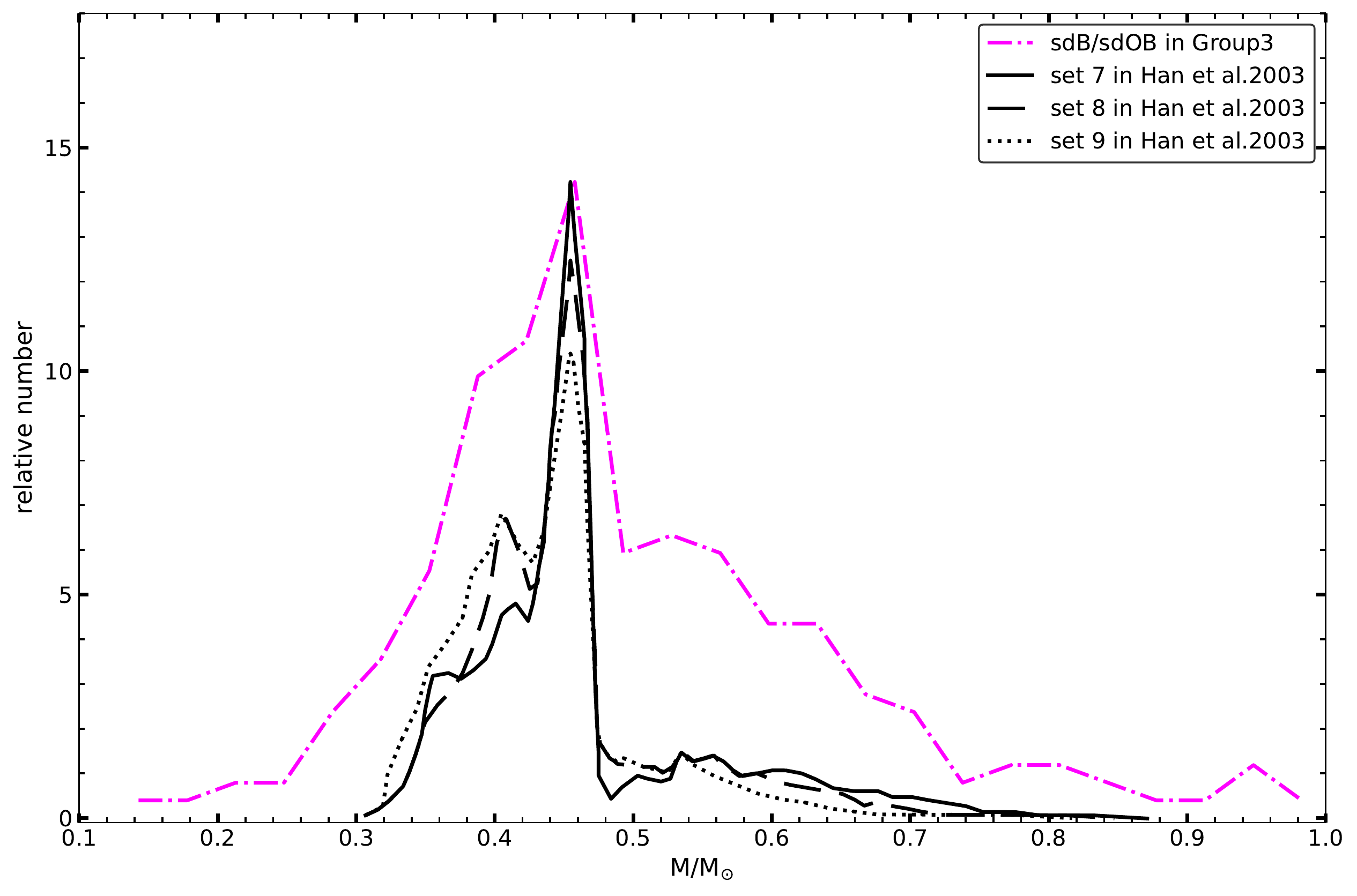}\\
    \includegraphics[width=0.49\textwidth]{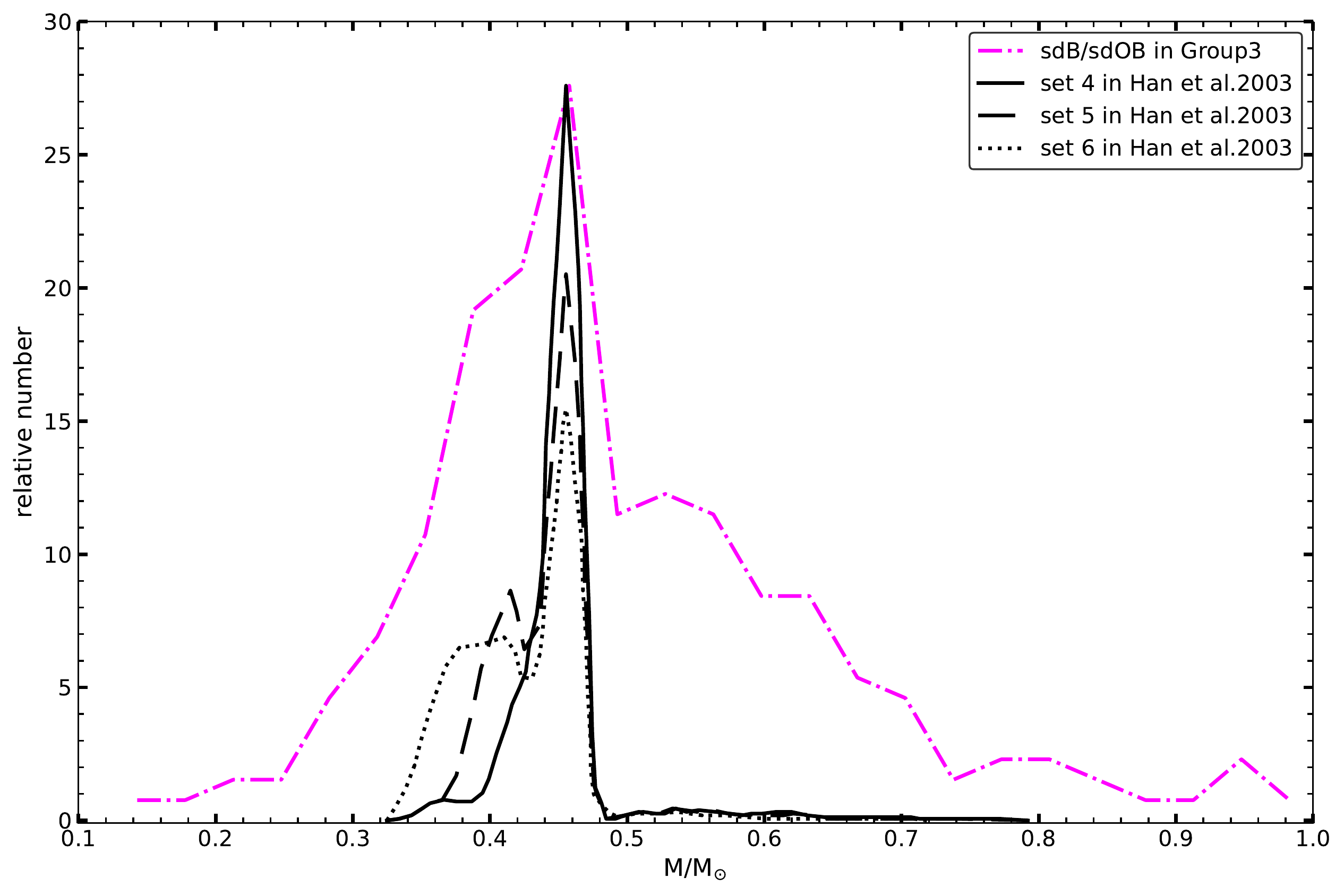}
      \includegraphics[width=0.49\textwidth]{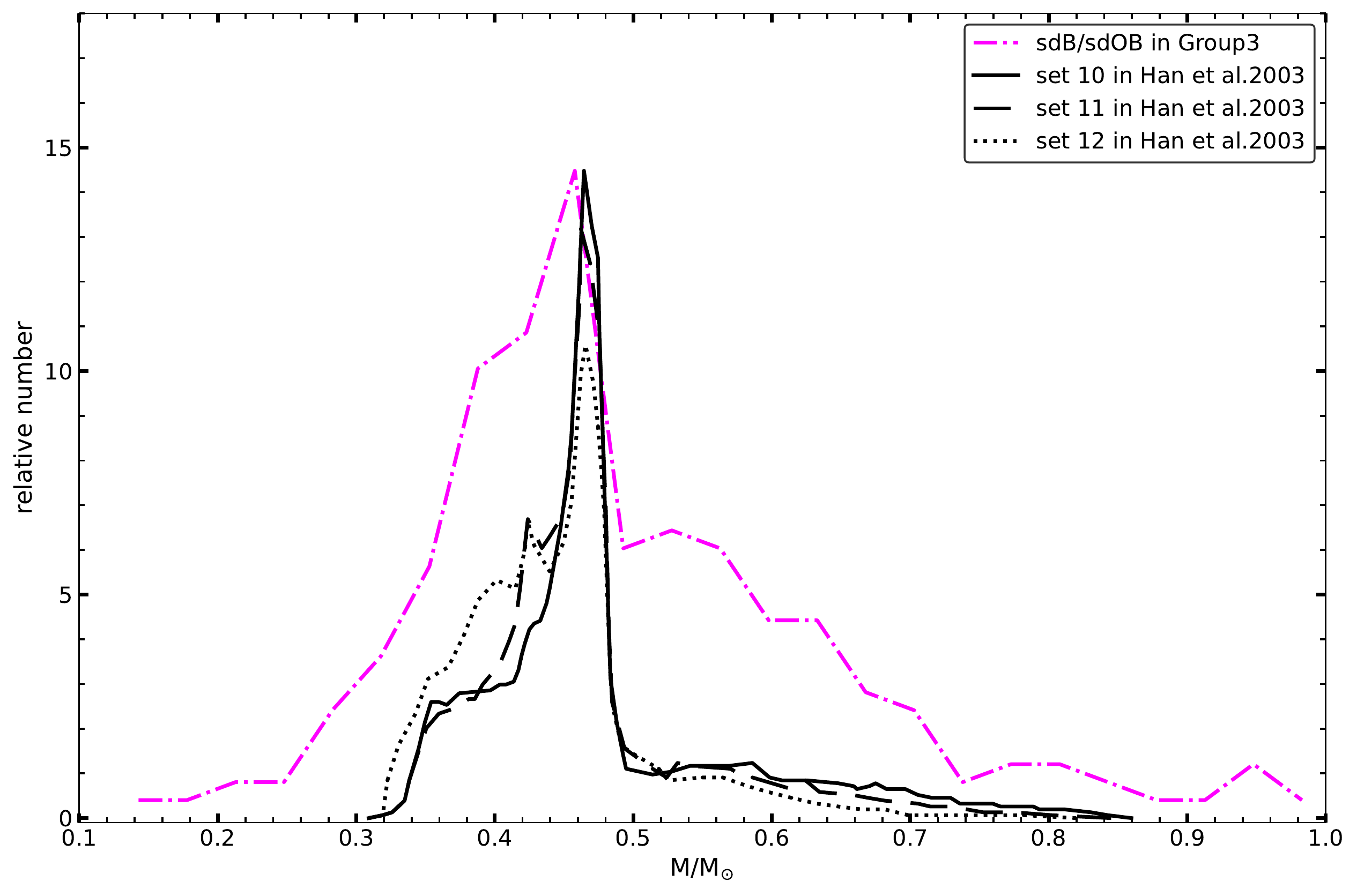}  
    \caption{Mass distribution comparison between sdB/sdOB stars in Group 3 and the results predicted by \citet{2003MNRAS.341..669H}. From the upper left to bottom right panel, the mass distribution predicted by 12 sets of simulation in \citet{2003MNRAS.341..669H} are denoted separately by black solid, dashed, and dotted curves (see legend box) in each panel, while the mass distribution for sdB/sdOB stars in Group 3 of this study is present by the magenta dash-dotted curve. 
    Bin numbers for sdB/sdOB stars in Group 3 were normalized to the maximum bin number of \citet{2003MNRAS.341..669H} in each panel. }
    \label{fig10}
\end{figure} 

In Fig \ref{fig10}, we compared the mass distribution of our sdB/sdOB stars from Group 3 with mass distributions for the 12 sets of simulation in \citet{2003MNRAS.341..669H}. 
 In the upper left panel, the mass distribution for our sdB/sdOB stars from Group 3 (e.g., magenta dash-dotted curve) is generally consistent with the mass distribution predicted by set 1, 2 and 3 simulations in \citet{2003MNRAS.341..669H}, which are denoted by black solid, dashed and dotted curves respectively. 
 Furthermore, the mass distribution peaks predicted by \citet{2003MNRAS.341..669H}, i.e.,  at 0.46 ${\rm M}_{\odot}$, agree quite well with the one presented in this study. 
 However, the sdB mass distribution in \citet{2003MNRAS.341..669H} displays a much steeper fall for masses larger than 0.48 ${\rm M}_{\odot}$. 
 It seems to demonstrate that the main channels in \citet{2003MNRAS.341..669H} predicted fewer sdB stars above 0.48 ${\rm M}_{\odot}$ than the results from this study. 
 Note that, to compare with the observational results of \citet{2001MNRAS.326.1391M}, which is biased against sdB binaries with bright F/G type companions, the GK selection in \citet{2003MNRAS.341..669H} removed sdB binary systems with companions having $T_{\rm eff}$ higher than 4000 K or brighter than the sdB primaries. 
 Due to the strict selection criterion, we did not compare the results with corrections for the GK selection effects in \citet{2003MNRAS.341..669H} with the results obtained in this study directly. 
 Since the minimum core mass for igniting He in a non-degenerate core is about 0.3 ${\rm M}_{\odot}$, while the minimum core mass to ignite He in degenerate stars could be larger (see Table 1 and Table 2 in \citealt{2002MNRAS.336..449H}), sdB stars have a low mass cutoff around 0.3 ${\rm M}_{\odot}$ in \citet{2003MNRAS.341..669H}. 
 There are still some stars less massive than 0.3 ${\rm M}_{\odot}$ in our sdB/sdOB sample, which can be low-mass WDs or pre-ELM WDs.  

The three other (e.g., bottom left, upper right, and bottom right) panels in Fig \ref{fig10} present the mass distribution comparison between our sdB/sdOB stars from Group 3 and the simulations of set 4 to 12 in \citet{2003MNRAS.341..669H},  respectively. 
Though the peaks of mass distributions are roughly consistent between the two studies, models from set 4 to 12 in \citet{2003MNRAS.341..669H} predicted obviously fewer stars at the low mass side (e.g., 0.3 - 0.42 ${\rm M}_{\odot}$) than this study. 
Therefore, the mass distribution for sdB/sdOB stars from Group 3 in this study prefers the model predictions of set 1 to 3 in \citet{2003MNRAS.341..669H}, especially for set 2 (e.g., the black dashed curve in the upper left panel, with $Z=.02$, a flat initial mass ratio distribution,  $q_{\rm crit}$=1.5, $\alpha_{\rm CE}$=$\alpha_{\rm th}$=0.75) even by visual inspections.

To study how parameter variations in binary evolution impact the production of sdB stars, \citet{2012ApJ...746..186C} also conducted a grid of binary population synthesis models with different parameter assumptions, such as minimum core mass for He ignition, envelope binding energy, CE ejection efficiency, amount of mass and angular momentum loss during stable mass transfer, and the criteria for a stable mass transfer on the RGB and in the Hertzsprung gap, etc. (see Table 1 in their study). 
They found that the variations of these parameters separately or together can significantly change the production of sdB binaries. 
They present mass distributions of sdB stars for 14 sets of simulations in Fig 14 of their study, most of which predicted similar mass distributions as in \citet{2003MNRAS.341..669H}, especially for their Run 6 which used very similar parameters as the best-fitting model (set 2) in \citet{2003MNRAS.341..669H}. 
However, most of the models in \citet{2012ApJ...746..186C} predicted a most probable mass for sdB stars around 0.48 ${\rm M}_{\odot}$, which is higher by about 0.02 ${\rm M}_{\odot}$ than the value obtained in this study and by \citet{2003MNRAS.341..669H}. 
They also predicted a population of post-RGB stars in their models, which has a lower mean stellar mass (e.g., less than 0.22 ${\rm M}_{\odot}$, see Fig 14 in their study) than sdB stars and failed to ignite He burning in the cores due to an excessive mass transfer on the RGB, but still crossed the sdB region in $T_{\rm eff}$ - $\log{g}$ plane. This population also appears in our results as we mentioned above, e.g., stars less massive than 0.3 ${\rm M}_{\odot}$ in Group 3 (see Fig \ref{fig8}).  
Furthermore, the models by \citet{2012ApJ...746..186C} predicted an obvious mass gap between the sdB and post-RGB population between 0.22-0.26 ${\rm M}_{\odot}$ (see Fig 14 in their study), but this feature is not present in our sdB/sdOB stars (see Fig \ref{fig8}). It could be averaged out due to a different bin size  adopted in our sample, and the overlap of different types of hot subdwarf stars would also mask this feature in our mass distributions. We will continue to discuss the nature of this population in Section 4.4.

\subsection{Mass distribution for He-rich hot subdwarf stars and their binary nature }

\begin{deluxetable*}{c|cccc}
\tablenum{3}
\tablecaption{Statistical properties of hot subdwarfs in the three groups also having detected RV variations by \textcolor{blue}{Geier et al. (2022)}.}
\tablewidth{140pt}
\tabletypesize{\normalsize}
\tablehead{
\colhead{Group} & \colhead{Subset} & \colhead{Total} & 
\colhead{Close binary} &  
\colhead{Close binary} \\
\colhead{Name}& \colhead{Name} &  Number  & \colhead{Number} & \colhead{Fraction(\%)}
}
\startdata
& All sample & 216 & 86 & 40  \\
\textbf{Group 1}& sdB/sdOB & 173 & 72 & 42 \\
& He-rich sample & 16 & 1 & 6 \\
& sdO & 27 & 13 & 48 \\
  \hline
& All sample & 159 & 66 & 42  \\
\textbf{Group 2}& sdB/sdOB & 130 & 54 & 42 \\
& He-rich sample & 10 & 1 & 10 \\
& sdO & 19 & 11 & 58 \\
  \hline
& All sample & 96 & 42 & 44  \\
\textbf{Group 3}& sdB/sdOB & 78 & 34 & 44 \\
& He-rich sample & 7 & 1 & 14 \\
& sdO & 11 & 7 & 64 \\
\enddata
\tablecomments {Close binaries are defined by $\log{p < -4.0}$ (see text for details).} 
\end{deluxetable*}

\citet{2022A&A...661A.113G} performed a radial velocity (RV) variability study for 646 single-lined hot subdwarfs having multi-epoch observed spectra in SDSS and LAMOST. 
The distribution of the maximum RV variations, (i.e., $\Delta RV_{\rm max}$, defined as the difference between the maximum and minimum RV values of a star, see Section 2.3 in their study), was used as diagnostics to study the RV variability thus the binarity of hot subdwarfs. 
If a false-detection probability calculated in their study, $p$ is smaller than 0.01\% ($\log{p < -4.0}$), the detection of RV variability was considered to be significant, thus a close hot subdwarf binary should be expected. 
Employing this method, they confirmed 164 hot subdwarf stars in the  sample showing significant RV variations. 
They also found a distinctive difference between He-poor and He-rich hot subdwarfs, that the former present a high fraction of close binaries, while nearly no significant RV variations were detected for the latter. 
This result made them conclude that there is no evolutionary connection between He-poor and He-rich hot subdwarfs. 
Moreover, their results also indicated that He-rich hot subdwarfs should be formed from the binary merger channels when considering the results from \citet{2020A&A...642A.180P} that binary interaction is always required to form hot subdwarf stars. 

To investigate the binarity of hot subdwarf stars in our study, 
we cross-matched the selected stars in the three groups with the hot subdwarfs studied by \citet{2022A&A...661A.113G} respectively. 
The statistical properties of the common objects are shown in Table 3. 
From left to right, it gives the group name, subset name, number counts, close binary number counts, and close binary fractions, respectively.
There are in total 216 hot subdwarfs in Group 1 also having RV variation values detected by \citet{2022A&A...661A.113G}, among which 86 stars present significant RV variations (e.g., $\log{p < -4.0}$). 
It means that about 40\% of hot subdwarfs are expected to be in close binaries in this group. 
Specifically, sdB/sdOB stars present a binary fraction of 42\% and sdO stars present a binary fraction of 48\%. 
He-rich (e.g., He-sdB, He-sdOB and He-sdO) stars in our sample present a much lower close binary fraction of 6\%, and only one star has a significant RV variation among the 16 He-rich stars. 
With an increasing parallax precision from Group 1 to 3, the fractions of close binaries in each subset for three groups are roughly consistent. 
Furthermore, though both He-poor and He-rich stars in this study  present a little higher close binary fraction than the counterparts in \citet{2022A&A...661A.113G}, e.g., 30 $\pm$ 2\% for He-poor stars and about 3\% for He-rich stars (see Section 3.1 in their study for details), these results are still consistent with each other when considering the randomness of cross-matching between the two sets of samples. 

\begin{figure}
    \centering
    \includegraphics[width=140mm]{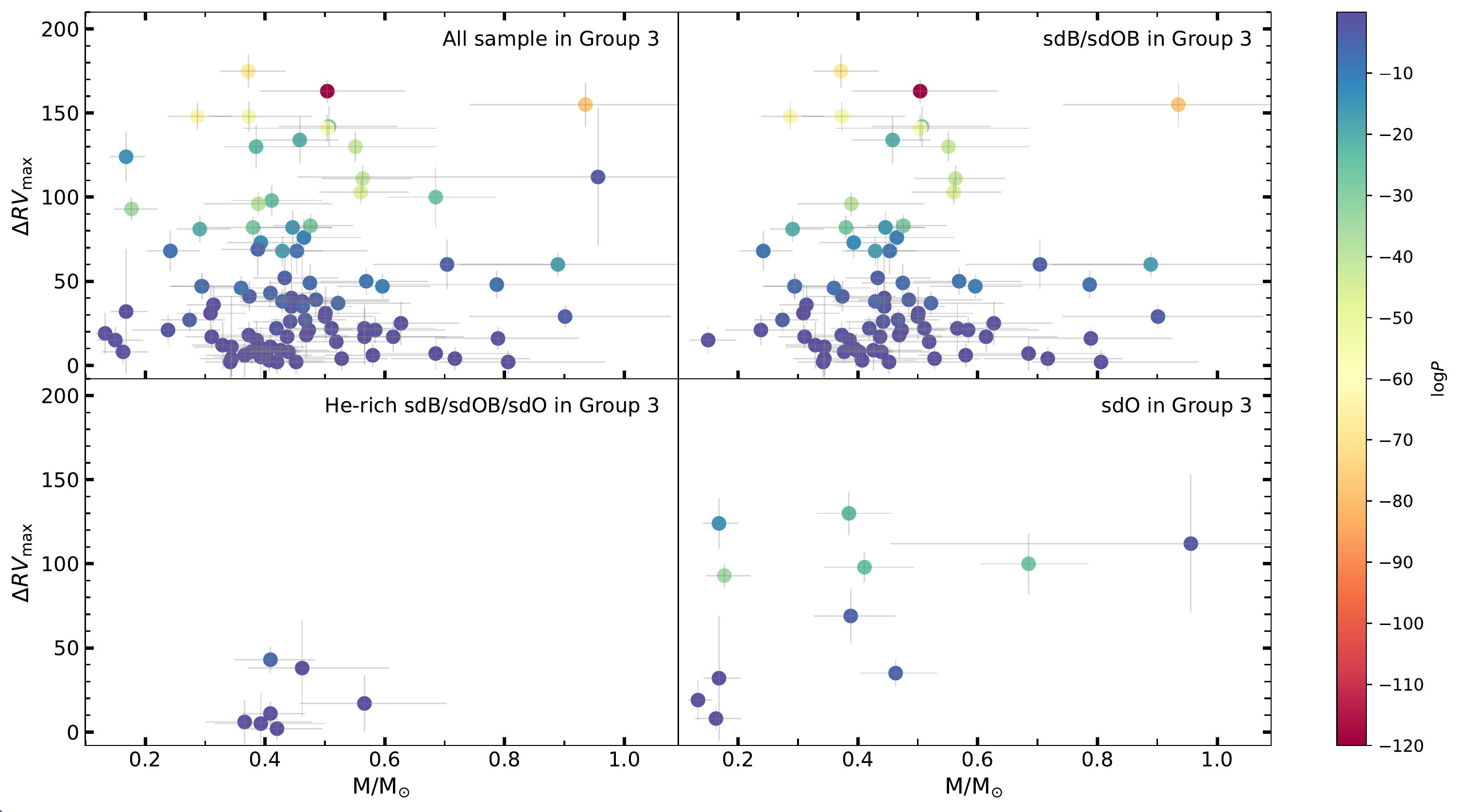}
    \caption{Mass - $\Delta RV_{\rm max}$ plane for hot subdwarf stars in Group 3 also having detected RV variations in \citet{2022A&A...661A.113G}. 
    The four panels in the figure show the relationship between mass and binarity for the different populations, i.e., all sample (upper left), sdB/sdO stars (upper right), He-rich stars (bottom left), and sdO stars (bottom right), respectively. 
    The color bar on the right is scaled to the values of false-detection probability ($\log{p}$). 
    Note that stars with $\log{p < -4.0}$ are considered to have significant RV variations and thus expected to be in close binaries, see Section 2.3 in \citet{2022A&A...661A.113G} for more details.}
    \label{fig11}
\end{figure}

Fig \ref{fig11} shows the relationship between mass distribution and the maximum RV variations for the 96 common objects between Group 3 in this study and \citet{2022A&A...661A.113G}. 
As clearly shown in the figure, there are a considerable number of objects that present large maximum RV variations and have very small false-detection probabilities (denoted by the color bar on the right of the figure. e.g., $\log{p < -4.0}$ means significant RV variability) for all sample (upper left panel), sdB/sdOB stars (upper right panel) and sdO stars (bottom right panel), which indicates a moderate close binary fraction among these types of hot subdwarfs (e.g., higher than 40\%, see Table 3). 
While in the bottom left panel, He-rich stars present nearly no significant RV variations, which is very different from sdB/sdOB and sdO stars. 
Based on this feature, \citet{2022A&A...661A.113G} came to the conclusion that there is no evolutionary connection between He-rich and He-poor hot subdwarf stars, and He-rich stars are formed by the binary merger channels.
If this is the case, He-rich hot subdwarfs should present a higher fraction of massive stars than He-poor hot subdwarfs, since hot subdwarf stars formed from merger channel shall have larger masses on average according to the model prediction of \citet{2002MNRAS.336..449H, 2003MNRAS.341..669H}, also see the results from \citet{2012MNRAS.419..452Z} and \citet{2017ApJ...835..242Z}. 
However, based on our results presented in Table 2, He-rich stars have nearly the same fraction of massive stars (e.g., about 20\%) as He-poor sdB/sdOB stars, which does not support binary mergers as the only formation channel for He-rich hot subdwarfs. 

\begin{figure}
    \centering
    \includegraphics[width=120mm]{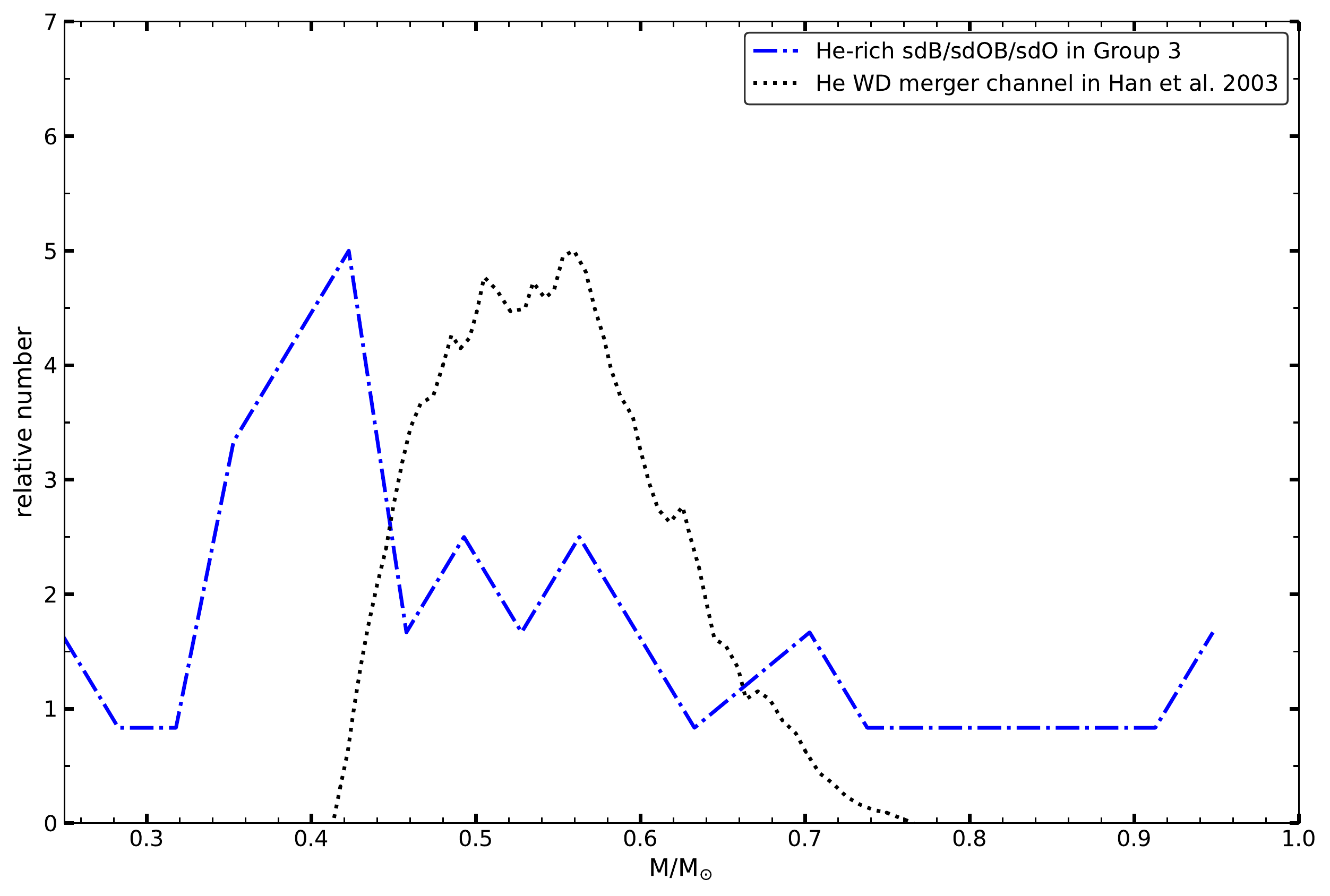}
    \caption{Mass distribution comparison for He-rich hot subdwarfs in Group 3 of this study (blue dash-dotted curve) with the results predicted by two He-WD merger channel in \citet{2003MNRAS.341..669H} (black dotted curve). Bin numbers in Group 3 were normalized to the maximum bin number of \citet{2003MNRAS.341..669H}. }
    \label{fig12}
\end{figure}

To understand this issue more clearly, in Fig \ref{fig12}, we compared the mass distribution of He-rich stars from Group 3 in this study with the results predicted by two He-WD merger channels from set 2 simulation in \citet{2003MNRAS.341..669H}. 
As seen in the figure, the two He-WD merger channel (black dotted curve) predicts a wide mass distribution of hot subdwarfs, e.g., roughly from 0.42 to 0.76 ${\rm M}_{\odot}$, with a fairly wide peak between 0.5 and 0.6 ${\rm M}_{\odot}$. 
On the other hand, He-rich stars in this study (blue dash-dotted curve) present a wider mass distribution, e.g., from 0.3 to nearly 1.0 ${\rm M}_{\odot}$, with a sharp peak around 0.42 ${\rm M}_{\odot}$\footnote{We also studied He-rich stars in Group 1 and 2, and similar mass distributions were obtained as in Group 3.}, which is much lower than the wide peak predicted by binary merger channel. 
Furthermore, our He-rich sample presents a much lower relative contribution than the model predicts in the broad mass peak area (e.g., 0.5-0.6 ${\rm M}_{\odot}$). 
\citet{2017ApJ...835..242Z} also studied the merger of a He-WD with an MS star, which can produce intermediate helium-rich (iHe-rich, e.g, $-1.0<\log{(n{\rm He}/n{\rm H})}<1.0$) hot subdwarfs. 
However, most of the iHe-rich hot subdwarfs formed through the He-WD+MS merger channel have masses in the range of 0.48 – 0.50  ${\rm M}_{\odot}$, and a few of them have masses up to 0.52 ${\rm M}_{\odot}$, which is also more massive than the mass peak in our He-rich hot subdwarf sample. 

As discussed above, these comparisons of results indicated that the two He-WD or the He-WD+MS merger channels cannot be the only formation channels for He-rich hot subdwarfs, some other mechanisms must exist and make contributions to form this kind of stars. 
\citet{2021MNRAS.507.4603M} proposed that the MS companions of massive WDs can survive a Type Ia supernova (SN Ia) explosions and evolve into hot subdwarfs. The hot subdwarfs formed in this channel are also single and He enriched. Moreover, the results based on binary population synthesis can explain some observed features of iHe-rich hot subdwarfs, especially when spin-up/spin-down models were considered. 
Surprisingly, the spin-up/spin-down models predict the production of hot subdwarfs in the mass range from 0.35 to 1.0 ${\rm M}_{\odot}$, and a distinct mass peak around 0.4 ${\rm M}_{\odot}$ (see Fig 9 in their study), which is consistent with the mass distribution of He-rich stars obtained in this study. 
However, the same model predicted a low Galactic birth rate for He-rich hot subdwarf stars (e.g., 2.3 - 6 $\times 10^{-4} {\rm yr}^{-1}$, see Section 4.1 in their study), which demonstrates that some other channels still contribute to the formation of He-rich hot subdwarfs. 

On the other hand, \citet{2022MNRAS.511L..66W} discovered two He-sdO stars with unusually strong carbon and oxygen lines (named CO-sdO stars in their study). This new type of hot subdwarf star can neither be explained by a late hot He-core flash \citep{2018A&A...614A.136B} nor by the merger of two He-WDs, which are considered as the formation channels for canonical He-sdO stars. \citet{2022MNRAS.511L..60M} proposed a new formation channel for the two CO-sdO stars discovered by \citet{2022MNRAS.511L..66W}. 
In this scenario, the merger of a more massive He-WD and a less massive CO-WD could ignite He core or shell burning and become a CO-sdO star. 
The material of the CO-WD could be accreted on the top of He-WD and leads to the CO-enriched envelope. 
This scenario explains the surface parameters and composition of CO-sdO stars. 
However, \citet{2022MNRAS.511L..60M} did not give the mass distribution for this formation channel, thus we cannot compare our results with it directly. 
As discussed in \citet{2022MNRAS.511L..60M}, the scenario proposed in their study can be a small sub-channel, since the parameter space of the progenitor system needs to be well defined (see Section 2 in their study for detailed discussion) and CO-sdO stars are just a minority compared to the hot subdwarf population. 
Therefore, this scenario plays a limited role in the formation of He-sdO stars.

Based on the results mentioned above, one can conclude that in addition to the two He-WD merger channels, there should be some other channels that still make contributions to the formation of He-rich hot subdwarf stars, e.g., the SNe Ia explosion channel could make some contributions on the formation of less massive He-rich hot subdwarfs (e.g., less than 0.44 ${\rm M}_{\odot}$, see Fig \ref{fig12}), while the binary merger channel could dominate the formation of massive He-rich hot subdwarfs (e.g., larger than 0.44 ${\rm M}_{\odot}$). 

\subsection{The nature of low-mass hot subdwarf stars}

\begin{figure}
    \centering
    \includegraphics[width=0.50\textwidth]{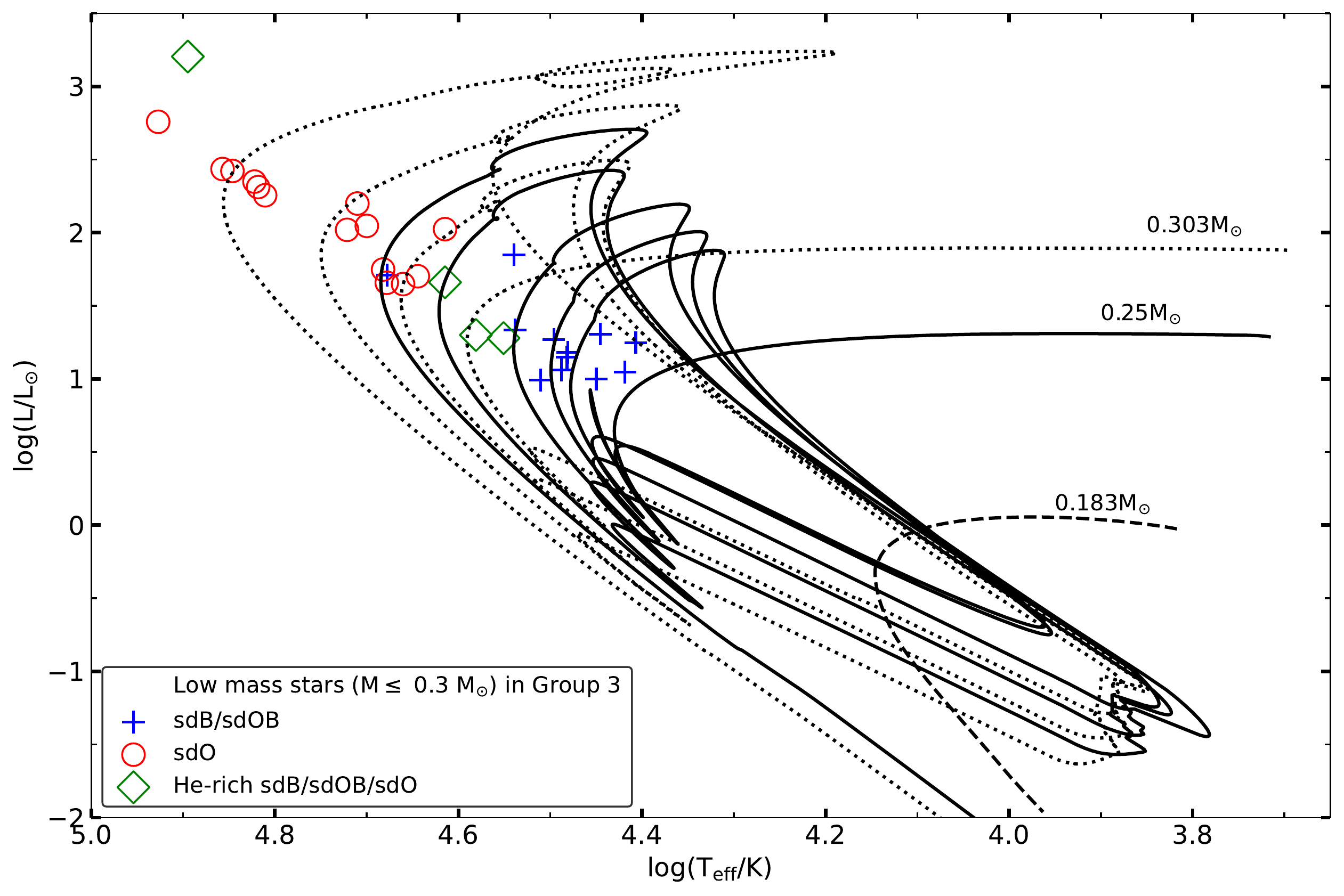}
    \includegraphics[width=0.49\textwidth]{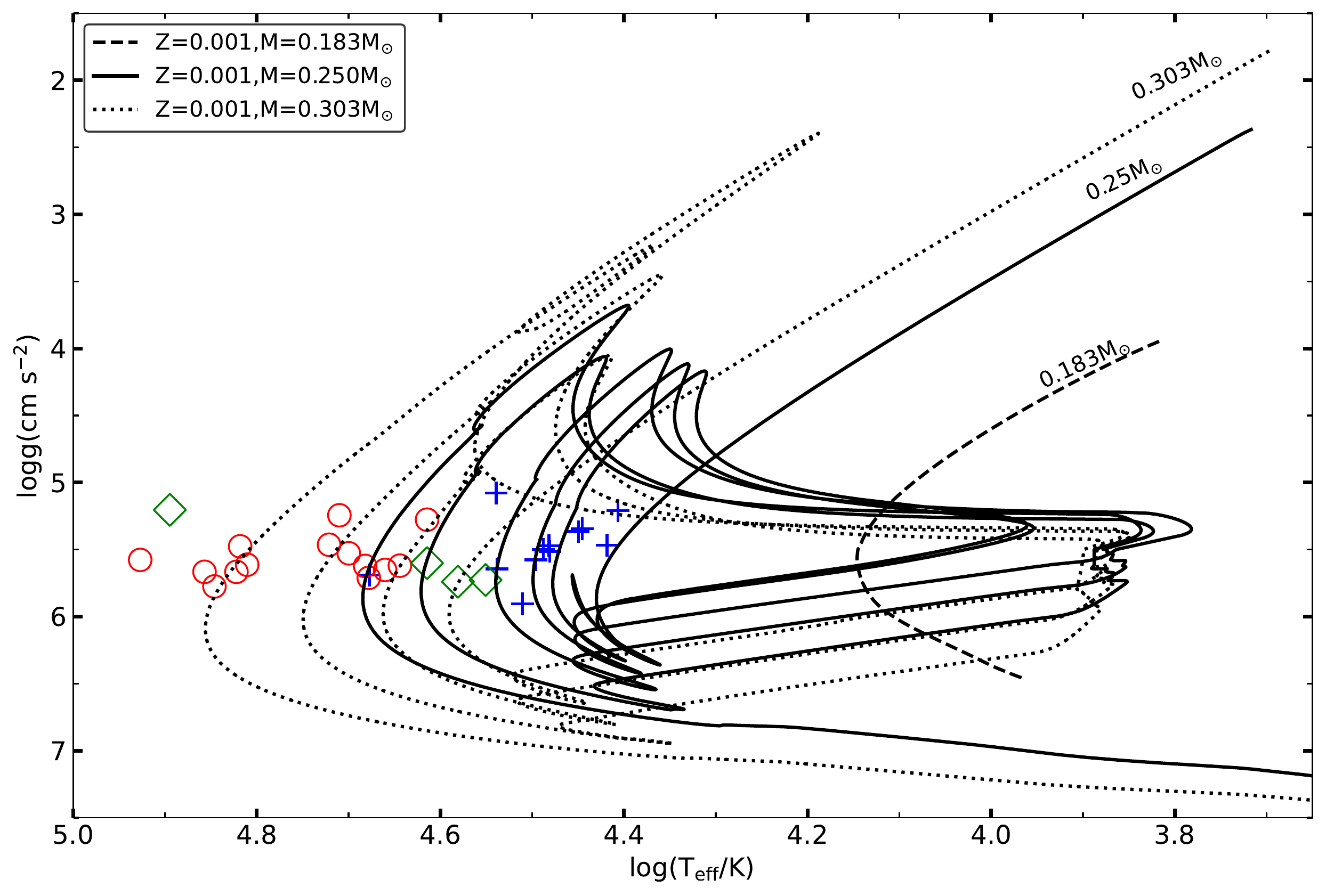}
    \caption{Left panel: $T_{\rm eff}$ -- luminosity plane for low-mass stars (i.e., less than 0.3 $\rm M_{\odot}$) in Group 3. Right panel: $T_{\rm eff}$ -- $\log{g}$ plane for low-mass stars in Group 3. Black dashed, solid and dotted curves in the panel are  evolutionary tracks of pre-ELM WDs with Z = 0.001 from \citet{2016A&A...595A..35I} for M = 0.183, 0.25 and 0.303 $\rm M_{\odot}$ respectively. }
    \label{fig13}
\end{figure} 

The minimum core mass for He ignition predicted by models is about 0.3 ${\rm M}_{\odot}$ \citep{2002MNRAS.336..449H, 2012ApJ...746..186C}, which could be a little different depending on metallicity and initial mass at zero-age main sequence (ZAMS). 
Therefore, hot subdwarf stars less massive than 0.3 ${\rm M}_{\odot}$ found in this study are potential post-RGB stars (e.g., low-mass WDs or pre-ELM WDs) that failed to ignite He burning in their cores.

It is believed that ELM WDs cannot be formed by the evolution of single stars since the evolutionary timescale of such low-mass stars would exceed the Hubble time. 
\citet{2016A&A...595A..35I} studied the effects of element diffusion and rotational mixing on the evolution of pre-ELM WDs with various metallicities. 
They found that element diffusion plays a significant role in the evolution of pre-ELM WDs which experience hydrogen shell flashes. 
Moreover, rotational mixing plays a key role in determining their surface chemical abundances, but it does not influence significantly the number of hydrogen shell flashes and the hydrogen envelope mass at the beginning of the cooling track. 
They calculated a large number of evolutionary tracks of pre-ELM WDs within a mass range of 0.16 - 0.45 ${\rm M}_{\odot}$, which provides great convenience to analyze the nature of low-mass hot subdwarf stars in our sample.

Fig \ref{fig13} shows the $T_{\rm eff}$ -- luminosity (left panel) and $T_{\rm eff}$ -- $\log{g}$ planes (right panel) for the low-mass hot subdwarfs (e.g., M $\leq 0.3 {\rm M}_{\odot}$) in Group 3, together with three evolutionary tracks of pre-ELM WDs at Z= 0.001 from \citet{2016A&A...595A..35I}. 
As can be seen in the figure, the positions of all low-mass sdB/sdOB stars (blue pluses), most of low-mass sdO stars (red circles), and He-rich stars (green squares) can be covered by the evolutionary tracks of pre-ELM WDs with higher stellar masses, e.g., 0.25 ${\rm M}_{\odot}$ (black solid curve) and 0.303 ${\rm M}_{\odot}$ (black dotted curve) both in $T_{\rm eff}$ -- luminosity plane and $T_{\rm eff}$ -- $\log{g}$ plane. 
These models experience several hydrogen shell flashes before entering the final cooling curve. 
On the other hand, our low-mass stars cannot be covered by the less massive evolutionary track of M = 0.183 ${\rm M}_{\odot}$, which did not experience any hydrogen shell flashes due to the thick shell (see Section 2.1 in \citealt{2016A&A...595A..35I} for detailed discussion). 
Furthermore, we checked the mass distribution of low-mass hot subdwarfs in Group 3 and found that most of the low-mass sdB/sdOB stars and He-rich stars have masses in the range of 0.25 - 0.30 ${\rm M}_{\odot}$, while most of the low-mass sdO stars have masses around 0.18 ${\rm M}_{\odot}$ as described in Table 1.

The results described above indicate that low-mass sdB/sdOB stars and He-rich stars should be post-RGB stars, which would lose too much envelope mass at the RGB stage to ignite He burning in their cores, and now are on the way towards the ELM WD cooling curves. \citet{2003A&A...411L.477H} discovered a pre-ELM WD in a close binary system, named HD\,188112, with a mass of 0.24 ${\rm M}_{\odot}$. 
The primary in that system presents $T_{\rm eff}$ = 21\,500 $\pm$ 500 K and $\log{g}$ = 5.66 $\pm$ 0.05, which is located at a similar region as our low-mass sdB/sdOB stars in the $T_{\rm eff}$ -- $\log{g}$ plane. 
However, pre-ELM models seem unable to explain our low-mass sdO stars. 
Because the evolutionary tracks with the same mass (e.g., about 0.18 ${\rm M}_{\odot}$, black dashed curve in Fig \ref{fig13}) predict much lower effective temperatures than our low-mass sdO stars (red circles in Fig \ref{fig13}, also see \citealt{2022ApJ...933...94B}). 
Thus, this result demonstrates that some other unknown mechanisms could play roles in the formation and evolution of these low-mass sdO stars, which need to be studied in the future.  

\section{Summary}
In this study, we obtained the radii, luminosities, and masses for 664 hot subdwarf stars identified in LAMOST by comparing synthetic fluxes from theoretical spectral energy distributions with observed fluxes from virtual observatory services. 
The relationship between stellar masses and atmospheric parameters was explored, interpreted, and shown. 
To study the mass distribution in our sample, three groups of hot subdwarf stars were selected by using their parallax precision. 
A wide mass distribution of hot subdwarf stars from 0.1 to 1.0 ${\rm M}_{\odot}$ is presented, and a sharp peak was found at 0.46 ${\rm M}_{\odot}$. 
The mass distribution of sdB/sdOB stars is consistent with the ones from model predictions. 
However, canonical binary models seem to predict fewer sdB stars with masses larger than 0.48 ${\rm M}_{\odot}$ when compared with the mass distribution obtained in this study.  
He-poor sdO and He-rich hot subdwarfs present a much flatter mass distribution and have a peak mass at 0.18 and 0.42 ${\rm M}_{\odot}$, respectively. 
sdB/sdOB and sdO stars have similar close binary fractions in our sample (e.g., about 40\%), and the similar mass - $\Delta RV_{\rm max}$ distribution between them supports that sdO stars are the subsequent evolution stage of sdB/sdOB stars. 
While, He-rich stars present a much lower close binary fraction than He-poor stars, which supports the binary merger formation channel for these stars. 
However, He-rich hot subdwarf stars in our sample present a peak mass at 0.42 ${\rm M}_{\odot}$, which is less than the wide mass peak of 0.5 - 0.6 ${\rm M}_{\odot}$ from the predictions of the binary merger channel, but consistent with the SN Ia explosion channel. 
Our results indicate that binary merger could not be the only main formation channel for He-rich hot subdwarf stars, other channels such as the SN Ia explosion or some other unknown channels must also play an important role in producing this special population, especially for He-rich hot subdwarf stars less massive than 0.44 ${\rm M}_{\odot}$. 

\begin{acknowledgments}
We thank the anonymous referee for his/her valuable suggestions and comments which helped improve the manuscript greatly. 
This work acknowledges support from the National Natural Science Foundation of China (Nos. 12073020 and 12273055, 12273055, 11973048), Scientific Research Fund of Hunan Provincial Education Department 
 grant No. 20K124, Cultivation Project for LAMOST Scientific Payoff and Research Achievement of CAMS-CAS, the science research grants from the China Manned Space Project with No. CMS-CSST-2021-B05. P.N. acknowledges support from the Grant Agency of the Czech Republic (GA\v{C}R 22-34467S) and from the Polish National Science Centre under projects UMO-2017/26/E/ST9/00703 and UMO-2017/25/B/ST9/02218. 
 This research has made use of the Spanish Virtual Observatory (\url{https://svo.cab.inta-csic.es}) project funded by MCIN/AEI/10.13039/501100011033/ through grant PID2020-112949GB-I00. Guoshoujing Telescope (the Large Sky Area Multi-Object Fiber Spectroscopic Telescope LAMOST) is a National Major Scientific Project built by the Chinese Academy of Sciences. Funding for the project has been provided by the National Development and Reform Commission. LAMOST is operated and managed by the National Astronomical Observatories, Chinese Academy of Sciences.
 This research has used the services of \mbox{\url{www.Astroserver.org}} under references: X074VU, KS5NVO, XD8O2C, D879YE and D880YE.
\end{acknowledgments}
%


\software{TOPCAT \citep{2005ASPC..347...29T}
          }
          


\bibliography{mass_estimation}{}

\begin{thebibliography}{}
\expandafter\ifx\csname natexlab\endcsname\relax\def\natexlab#1{#1}\fi
\providecommand{\url}[1]{\href{#1}{#1}}
\providecommand{\dodoi}[1]{doi:~\href{http://doi.org/#1}{\nolinkurl{#1}}}
\providecommand{\doeprint}[1]{\href{http://ascl.net/#1}{\nolinkurl{http://ascl.net/#1}}}
\providecommand{\doarXiv}[1]{\href{https://arxiv.org/abs/#1}{\nolinkurl{https://arxiv.org/abs/#1}}}

\bibitem[{{Astraatmadja} \& {Bailer-Jones}(2016)}]{2016ApJ...832..137A}
{Astraatmadja}, T.~L., \& {Bailer-Jones}, C. A.~L. 2016, \apj, 832, 137,
  \dodoi{10.3847/0004-637X/832/2/137}

\bibitem[{{Bailer-Jones}(2015)}]{2015PASP..127..994B}
{Bailer-Jones}, C. A.~L. 2015, \pasp, 127, 994, \dodoi{10.1086/683116}

\bibitem[{{Bailer-Jones} {et~al.}(2021){Bailer-Jones}, {Rybizki}, {Fouesneau},
  {Demleitner}, \& {Andrae}}]{2021AJ....161..147B}
{Bailer-Jones}, C.~A.~L., {Rybizki}, J., {Fouesneau}, M., {Demleitner}, M., \&
  {Andrae}, R. 2021, \aj, 161, 147, \dodoi{10.3847/1538-3881/abd806}

\bibitem[{{Baran} {et~al.}(2021){Baran}, {{\O}stensen}, {Heber}, {Irrgang},
  {Sanjayan}, {Telting}, {Reed}, \& {Ostrowski}}]{2021MNRAS.503.2157B}
{Baran}, A.~S., {{\O}stensen}, R.~H., {Heber}, U., {et~al.} 2021, \mnras, 503,
  2157, \dodoi{10.1093/mnras/stab620}

\bibitem[{{Battich} {et~al.}(2018){Battich}, {Miller Bertolami}, {C{\'o}rsico},
  \& {Althaus}}]{2018A&A...614A.136B}
{Battich}, T., {Miller Bertolami}, M.~M., {C{\'o}rsico}, A.~H., \& {Althaus},
  L.~G. 2018, \aap, 614, A136, \dodoi{10.1051/0004-6361/201731463}

\bibitem[{{Bayo} {et~al.}(2008){Bayo}, {Rodrigo}, {Barrado Y Navascu{\'e}s},
  {Solano}, {Guti{\'e}rrez}, {Morales-Calder{\'o}n}, \&
  {Allard}}]{2008A&A...492..277B}
{Bayo}, A., {Rodrigo}, C., {Barrado Y Navascu{\'e}s}, D., {et~al.} 2008, \aap,
  492, 277, \dodoi{10.1051/0004-6361:200810395}

\bibitem[{{Brown} {et~al.}(2010){Brown}, {Sweigart}, {Lanz}, {Smith},
  {Landsman}, \& {Hubeny}}]{2010ApJ...718.1332B}
{Brown}, T.~M., {Sweigart}, A.~V., {Lanz}, T., {et~al.} 2010, \apj, 718, 1332,
  \dodoi{10.1088/0004-637X/718/2/1332}

\bibitem[{{Brown} {et~al.}(2016){Brown}, {Cassisi}, {D'Antona}, {Salaris},
  {Milone}, {Dalessandro}, {Piotto}, {Renzini}, {Sweigart}, {Bellini},
  {Ortolani}, {Sarajedini}, {Aparicio}, {Bedin}, {Anderson}, {Pietrinferni}, \&
  {Nardiello}}]{2016ApJ...822...44B}
{Brown}, T.~M., {Cassisi}, S., {D'Antona}, F., {et~al.} 2016, \apj, 822, 44,
  \dodoi{10.3847/0004-637X/822/1/44}

\bibitem[{{Brown} {et~al.}(2022){Brown}, {Kilic}, {Kosakowski}, \&
  {Gianninas}}]{2022ApJ...933...94B}
{Brown}, W.~R., {Kilic}, M., {Kosakowski}, A., \& {Gianninas}, A. 2022, \apj,
  933, 94, \dodoi{10.3847/1538-4357/ac72ac}

\bibitem[{{Castellani} \& {Castellani}(1993)}]{1993ApJ...407..649C}
{Castellani}, M., \& {Castellani}, V. 1993, \apj, 407, 649,
  \dodoi{10.1086/172547}

\bibitem[{{Chen} {et~al.}(2013){Chen}, {Han}, {Deca}, \&
  {Podsiadlowski}}]{2013MNRAS.434..186C}
{Chen}, X., {Han}, Z., {Deca}, J., \& {Podsiadlowski}, P. 2013, \mnras, 434,
  186, \dodoi{10.1093/mnras/stt992}

\bibitem[{{Clausen} {et~al.}(2012){Clausen}, {Wade}, {Kopparapu}, \&
  {O'Shaughnessy}}]{2012ApJ...746..186C}
{Clausen}, D., {Wade}, R.~A., {Kopparapu}, R.~K., \& {O'Shaughnessy}, R. 2012,
  \apj, 746, 186, \dodoi{10.1088/0004-637X/746/2/186}

\bibitem[{{Copperwheat} {et~al.}(2011){Copperwheat}, {Morales-Rueda}, {Marsh},
  {Maxted}, \& {Heber}}]{2011MNRAS.415.1381C}
{Copperwheat}, C.~M., {Morales-Rueda}, L., {Marsh}, T.~R., {Maxted}, P.~F.~L.,
  \& {Heber}, U. 2011, \mnras, 415, 1381,
  \dodoi{10.1111/j.1365-2966.2011.18786.x}

\bibitem[{{Culpan} {et~al.}(2022){Culpan}, {Geier}, {Reindl}, {Pelisoli},
  {Gentile Fusillo}, \& {Vorontseva}}]{2022A&A...662A..40C}
{Culpan}, R., {Geier}, S., {Reindl}, N., {et~al.} 2022, \aap, 662, A40,
  \dodoi{10.1051/0004-6361/202243337}

\bibitem[{{Dorman} {et~al.}(1993){Dorman}, {Rood}, \&
  {O'Connell}}]{1993ApJ...419..596D}
{Dorman}, B., {Rood}, R.~T., \& {O'Connell}, R.~W. 1993, \apj, 419, 596,
  \dodoi{10.1086/173511}

\bibitem[{{Drilling} {et~al.}(2013){Drilling}, {Jeffery}, {Heber}, {Moehler},
  \& {Napiwotzki}}]{2013A&A...551A..31D}
{Drilling}, J.~S., {Jeffery}, C.~S., {Heber}, U., {Moehler}, S., \&
  {Napiwotzki}, R. 2013, \aap, 551, A31, \dodoi{10.1051/0004-6361/201219433}

\bibitem[{{Edelmann} {et~al.}(2003){Edelmann}, {Heber}, {Hagen}, {Lemke},
  {Dreizler}, {Napiwotzki}, \& {Engels}}]{2003A&A...400..939E}
{Edelmann}, H., {Heber}, U., {Hagen}, H.~J., {et~al.} 2003, \aap, 400, 939,
  \dodoi{10.1051/0004-6361:20030135}

\bibitem[{{Fontaine} {et~al.}(2012){Fontaine}, {Brassard}, {Charpinet},
  {Green}, {Randall}, \& {Van Grootel}}]{2012A&A...539A..12F}
{Fontaine}, G., {Brassard}, P., {Charpinet}, S., {et~al.} 2012, \aap, 539, A12,
  \dodoi{10.1051/0004-6361/201118220}

\bibitem[{{Fontaine} {et~al.}(2014){Fontaine}, {Green}, {Brassard}, {Latour},
  \& {Chayer}}]{2014ASPC..481...83F}
{Fontaine}, G., {Green}, E., {Brassard}, P., {Latour}, M., \& {Chayer}, P.
  2014, in Astronomical Society of the Pacific Conference Series, Vol. 481, 6th
  Meeting on Hot Subdwarf Stars and Related Objects, ed. V.~{van Grootel},
  E.~{Green}, G.~{Fontaine}, \& S.~{Charpinet}, 83

\bibitem[{{Gaia Collaboration} {et~al.}(2016){Gaia Collaboration}, {Prusti},
  {de Bruijne}, {Brown}, {Vallenari}, {Babusiaux}, {Bailer-Jones}, {Bastian},
  {Biermann}, {Evans}, {Eyer}, {Jansen}, {Jordi}, {Klioner}, {Lammers},
  {Lindegren}, {Luri}, {Mignard}, {Milligan}, {Panem}, {Poinsignon},
  {Pourbaix}, {Randich}, {Sarri}, {Sartoretti}, {Siddiqui}, {Soubiran},
  {Valette}, {van Leeuwen}, {Walton}, {Aerts}, {Arenou}, {Cropper}, {Drimmel},
  {H{\o}g}, {Katz}, {Lattanzi}, {O'Mullane}, {Grebel}, {Holland}, {Huc},
  {Passot}, {Bramante}, {Cacciari}, {Casta{\~n}eda}, {Chaoul}, {Cheek}, {De
  Angeli}, {Fabricius}, {Guerra}, {Hern{\'a}ndez}, {Jean-Antoine-Piccolo},
  {Masana}, {Messineo}, {Mowlavi}, {Nienartowicz}, {Ord{\'o}{\~n}ez-Blanco},
  {Panuzzo}, {Portell}, {Richards}, {Riello}, {Seabroke}, {Tanga},
  {Th{\'e}venin}, {Torra}, {Els}, {Gracia-Abril}, {Comoretto},
  {Garcia-Reinaldos}, {Lock}, {Mercier}, {Altmann}, {Andrae}, {Astraatmadja},
  {Bellas-Velidis}, {Benson}, {Berthier}, {Blomme}, {Busso}, {Carry},
  {Cellino}, {Clementini}, {Cowell}, {Creevey}, {Cuypers}, {Davidson}, {De
  Ridder}, {de Torres}, {Delchambre}, {Dell'Oro}, {Ducourant}, {Fr{\'e}mat},
  {Garc{\'\i}a-Torres}, {Gosset}, {Halbwachs}, {Hambly}, {Harrison}, {Hauser},
  {Hestroffer}, {Hodgkin}, {Huckle}, {Hutton}, {Jasniewicz}, {Jordan},
  {Kontizas}, {Korn}, {Lanzafame}, {Manteiga}, {Moitinho}, {Muinonen},
  {Osinde}, {Pancino}, {Pauwels}, {Petit}, {Recio-Blanco}, {Robin}, {Sarro},
  {Siopis}, {Smith}, {Smith}, {Sozzetti}, {Thuillot}, {van Reeven}, {Viala},
  {Abbas}, {Abreu Aramburu}, {Accart}, {Aguado}, {Allan}, {Allasia},
  {Altavilla}, {{\'A}lvarez}, {Alves}, {Anderson}, {Andrei}, {Anglada Varela},
  {Antiche}, {Antoja}, {Ant{\'o}n}, {Arcay}, {Atzei}, {Ayache}, {Bach},
  {Baker}, {Balaguer-N{\'u}{\~n}ez}, {Barache}, {Barata}, {Barbier}, {Barblan},
  {Baroni}, {Barrado y Navascu{\'e}s}, {Barros}, {Barstow}, {Becciani},
  {Bellazzini}, {Bellei}, {Bello Garc{\'\i}a}, {Belokurov}, {Bendjoya},
  {Berihuete}, {Bianchi}, {Bienaym{\'e}}, {Billebaud}, {Blagorodnova},
  {Blanco-Cuaresma}, {Boch}, {Bombrun}, {Borrachero}, {Bouquillon}, {Bourda},
  {Bouy}, {Bragaglia}, {Breddels}, {Brouillet}, {Br{\"u}semeister},
  {Bucciarelli}, {Budnik}, {Burgess}, {Burgon}, {Burlacu}, {Busonero}, {Buzzi},
  {Caffau}, {Cambras}, {Campbell}, {Cancelliere}, {Cantat-Gaudin}, {Carlucci},
  {Carrasco}, {Castellani}, {Charlot}, {Charnas}, {Charvet}, {Chassat},
  {Chiavassa}, {Clotet}, {Cocozza}, {Collins}, {Collins}, {Costigan}, {Crifo},
  {Cross}, {Crosta}, {Crowley}, {Dafonte}, {Damerdji}, {Dapergolas}, {David},
  {David}, {De Cat}, {de Felice}, {de Laverny}, {De Luise}, {De March}, {de
  Martino}, {de Souza}, {Debosscher}, {del Pozo}, {Delbo}, {Delgado},
  {Delgado}, {di Marco}, {Di Matteo}, {Diakite}, {Distefano}, {Dolding}, {Dos
  Anjos}, {Drazinos}, {Dur{\'a}n}, {Dzigan}, {Ecale}, {Edvardsson}, {Enke},
  {Erdmann}, {Escolar}, {Espina}, {Evans}, {Eynard Bontemps}, {Fabre},
  {Fabrizio}, {Faigler}, {Falc{\~a}o}, {Farr{\`a}s Casas}, {Faye}, {Federici},
  {Fedorets}, {Fern{\'a}ndez-Hern{\'a}ndez}, {Fernique}, {Fienga}, {Figueras},
  {Filippi}, {Findeisen}, {Fonti}, {Fouesneau}, {Fraile}, {Fraser}, {Fuchs},
  {Furnell}, {Gai}, {Galleti}, {Galluccio}, {Garabato}, {Garc{\'\i}a-Sedano},
  {Gar{\'e}}, {Garofalo}, {Garralda}, {Gavras}, {Gerssen}, {Geyer}, {Gilmore},
  {Girona}, {Giuffrida}, {Gomes}, {Gonz{\'a}lez-Marcos},
  {Gonz{\'a}lez-N{\'u}{\~n}ez}, {Gonz{\'a}lez-Vidal}, {Granvik}, {Guerrier},
  {Guillout}, {Guiraud}, {G{\'u}rpide}, {Guti{\'e}rrez-S{\'a}nchez}, {Guy},
  {Haigron}, {Hatzidimitriou}, {Haywood}, {Heiter}, {Helmi}, {Hobbs},
  {Hofmann}, {Holl}, {Holland}, {Hunt}, {Hypki}, {Icardi}, {Irwin}, {Jevardat
  de Fombelle}, {Jofr{\'e}}, {Jonker}, {Jorissen}, {Julbe}, {Karampelas},
  {Kochoska}, {Kohley}, {Kolenberg}, {Kontizas}, {Koposov}, {Kordopatis},
  {Koubsky}, {Kowalczyk}, {Krone-Martins}, {Kudryashova}, {Kull}, {Bachchan},
  {Lacoste-Seris}, {Lanza}, {Lavigne}, {Le Poncin-Lafitte}, {Lebreton},
  {Lebzelter}, {Leccia}, {Leclerc}, {Lecoeur-Taibi}, {Lemaitre}, {Lenhardt},
  {Leroux}, {Liao}, {Licata}, {Lindstr{\o}m}, {Lister}, {Livanou}, {Lobel},
  {L{\"o}ffler}, {L{\'o}pez}, {Lopez-Lozano}, {Lorenz}, {Loureiro},
  {MacDonald}, {Magalh{\~a}es Fernandes}, {Managau}, {Mann}, {Mantelet},
  {Marchal}, {Marchant}, {Marconi}, {Marie}, {Marinoni}, {Marrese},
  {Marschalk{\'o}}, {Marshall}, {Mart{\'\i}n-Fleitas}, {Martino}, {Mary},
  {Matijevi{\v{c}}}, {Mazeh}, {McMillan}, {Messina}, {Mestre}, {Michalik},
  {Millar}, {Miranda}, {Molina}, {Molinaro}, {Molinaro}, {Moln{\'a}r},
  {Moniez}, {Montegriffo}, {Monteiro}, {Mor}, {Mora}, {Morbidelli}, {Morel},
  {Morgenthaler}, {Morley}, {Morris}, {Mulone}, {Muraveva}, {Musella},
  {Narbonne}, {Nelemans}, {Nicastro}, {Noval}, {Ord{\'e}novic},
  {Ordieres-Mer{\'e}}, {Osborne}, {Pagani}, {Pagano}, {Pailler}, {Palacin},
  {Palaversa}, {Parsons}, {Paulsen}, {Pecoraro}, {Pedrosa}, {Pentik{\"a}inen},
  {Pereira}, {Pichon}, {Piersimoni}, {Pineau}, {Plachy}, {Plum}, {Poujoulet},
  {Pr{\v{s}}a}, {Pulone}, {Ragaini}, {Rago}, {Rambaux}, {Ramos-Lerate},
  {Ranalli}, {Rauw}, {Read}, {Regibo}, {Renk}, {Reyl{\'e}}, {Ribeiro},
  {Rimoldini}, {Ripepi}, {Riva}, {Rixon}, {Roelens}, {Romero-G{\'o}mez},
  {Rowell}, {Royer}, {Rudolph}, {Ruiz-Dern}, {Sadowski}, {Sagrist{\`a}
  Sell{\'e}s}, {Sahlmann}, {Salgado}, {Salguero}, {Sarasso}, {Savietto},
  {Schnorhk}, {Schultheis}, {Sciacca}, {Segol}, {Segovia}, {Segransan},
  {Serpell}, {Shih}, {Smareglia}, {Smart}, {Smith}, {Solano}, {Solitro},
  {Sordo}, {Soria Nieto}, {Souchay}, {Spagna}, {Spoto}, {Stampa}, {Steele},
  {Steidelm{\"u}ller}, {Stephenson}, {Stoev}, {Suess}, {S{\"u}veges}, {Surdej},
  {Szabados}, {Szegedi-Elek}, {Tapiador}, {Taris}, {Tauran}, {Taylor},
  {Teixeira}, {Terrett}, {Tingley}, {Trager}, {Turon}, {Ulla}, {Utrilla},
  {Valentini}, {van Elteren}, {Van Hemelryck}, {van Leeuwen}, {Varadi},
  {Vecchiato}, {Veljanoski}, {Via}, {Vicente}, {Vogt}, {Voss}, {Votruba},
  {Voutsinas}, {Walmsley}, {Weiler}, {Weingrill}, {Werner}, {Wevers},
  {Whitehead}, {Wyrzykowski}, {Yoldas}, {{\v{Z}}erjal}, {Zucker}, {Zurbach},
  {Zwitter}, {Alecu}, {Allen}, {Allende Prieto}, {Amorim},
  {Anglada-Escud{\'e}}, {Arsenijevic}, {Azaz}, {Balm}, {Beck}, {Bernstein},
  {Bigot}, {Bijaoui}, {Blasco}, {Bonfigli}, {Bono}, {Boudreault}, {Bressan},
  {Brown}, {Brunet}, {Bunclark}, {Buonanno}, {Butkevich}, {Carret}, {Carrion},
  {Chemin}, {Ch{\'e}reau}, {Corcione}, {Darmigny}, {de Boer}, {de Teodoro}, {de
  Zeeuw}, {Delle Luche}, {Domingues}, {Dubath}, {Fodor}, {Fr{\'e}zouls},
  {Fries}, {Fustes}, {Fyfe}, {Gallardo}, {Gallegos}, {Gardiol}, {Gebran},
  {Gomboc}, {G{\'o}mez}, {Grux}, {Gueguen}, {Heyrovsky}, {Hoar}, {Iannicola},
  {Isasi Parache}, {Janotto}, {Joliet}, {Jonckheere}, {Keil}, {Kim},
  {Klagyivik}, {Klar}, {Knude}, {Kochukhov}, {Kolka}, {Kos}, {Kutka}, {Lainey},
  {LeBouquin}, {Liu}, {Loreggia}, {Makarov}, {Marseille}, {Martayan},
  {Martinez-Rubi}, {Massart}, {Meynadier}, {Mignot}, {Munari}, {Nguyen},
  {Nordlander}, {Ocvirk}, {O'Flaherty}, {Olias Sanz}, {Ortiz}, {Osorio},
  {Oszkiewicz}, {Ouzounis}, {Palmer}, {Park}, {Pasquato}, {Peltzer}, {Peralta},
  {P{\'e}turaud}, {Pieniluoma}, {Pigozzi}, {Poels}, {Prat}, {Prod'homme},
  {Raison}, {Rebordao}, {Risquez}, {Rocca-Volmerange}, {Rosen}, {Ruiz-Fuertes},
  {Russo}, {Sembay}, {Serraller Vizcaino}, {Short}, {Siebert}, {Silva},
  {Sinachopoulos}, {Slezak}, {Soffel}, {Sosnowska}, {Strai{\v{z}}ys}, {ter
  Linden}, {Terrell}, {Theil}, {Tiede}, {Troisi}, {Tsalmantza}, {Tur},
  {Vaccari}, {Vachier}, {Valles}, {Van Hamme}, {Veltz}, {Virtanen}, {Wallut},
  {Wichmann}, {Wilkinson}, {Ziaeepour}, \& {Zschocke}}]{2016A&A...595A...1G}
{Gaia Collaboration}, {Prusti}, T., {de Bruijne}, J.~H.~J., {et~al.} 2016,
  \aap, 595, A1, \dodoi{10.1051/0004-6361/201629272}

\bibitem[{{Gaia Collaboration} {et~al.}(2018){Gaia Collaboration}, {Brown},
  {Vallenari}, {Prusti}, {de Bruijne}, {Babusiaux}, {Bailer-Jones}, {Biermann},
  {Evans}, {Eyer}, {Jansen}, {Jordi}, {Klioner}, {Lammers}, {Lindegren},
  {Luri}, {Mignard}, {Panem}, {Pourbaix}, {Randich}, {Sartoretti}, {Siddiqui},
  {Soubiran}, {van Leeuwen}, {Walton}, {Arenou}, {Bastian}, {Cropper},
  {Drimmel}, {Katz}, {Lattanzi}, {Bakker}, {Cacciari}, {Casta{\~n}eda},
  {Chaoul}, {Cheek}, {De Angeli}, {Fabricius}, {Guerra}, {Holl}, {Masana},
  {Messineo}, {Mowlavi}, {Nienartowicz}, {Panuzzo}, {Portell}, {Riello},
  {Seabroke}, {Tanga}, {Th{\'e}venin}, {Gracia-Abril}, {Comoretto},
  {Garcia-Reinaldos}, {Teyssier}, {Altmann}, {Andrae}, {Audard},
  {Bellas-Velidis}, {Benson}, {Berthier}, {Blomme}, {Burgess}, {Busso},
  {Carry}, {Cellino}, {Clementini}, {Clotet}, {Creevey}, {Davidson}, {De
  Ridder}, {Delchambre}, {Dell'Oro}, {Ducourant},
  {Fern{\'a}ndez-Hern{\'a}ndez}, {Fouesneau}, {Fr{\'e}mat}, {Galluccio},
  {Garc{\'\i}a-Torres}, {Gonz{\'a}lez-N{\'u}{\~n}ez}, {Gonz{\'a}lez-Vidal},
  {Gosset}, {Guy}, {Halbwachs}, {Hambly}, {Harrison}, {Hern{\'a}ndez},
  {Hestroffer}, {Hodgkin}, {Hutton}, {Jasniewicz}, {Jean-Antoine-Piccolo},
  {Jordan}, {Korn}, {Krone-Martins}, {Lanzafame}, {Lebzelter}, {L{\"o}ffler},
  {Manteiga}, {Marrese}, {Mart{\'\i}n-Fleitas}, {Moitinho}, {Mora}, {Muinonen},
  {Osinde}, {Pancino}, {Pauwels}, {Petit}, {Recio-Blanco}, {Richards},
  {Rimoldini}, {Robin}, {Sarro}, {Siopis}, {Smith}, {Sozzetti}, {S{\"u}veges},
  {Torra}, {van Reeven}, {Abbas}, {Abreu Aramburu}, {Accart}, {Aerts},
  {Altavilla}, {{\'A}lvarez}, {Alvarez}, {Alves}, {Anderson}, {Andrei},
  {Anglada Varela}, {Antiche}, {Antoja}, {Arcay}, {Astraatmadja}, {Bach},
  {Baker}, {Balaguer-N{\'u}{\~n}ez}, {Balm}, {Barache}, {Barata}, {Barbato},
  {Barblan}, {Barklem}, {Barrado}, {Barros}, {Barstow}, {Bartholom{\'e}
  Mu{\~n}oz}, {Bassilana}, {Becciani}, {Bellazzini}, {Berihuete}, {Bertone},
  {Bianchi}, {Bienaym{\'e}}, {Blanco-Cuaresma}, {Boch}, {Boeche}, {Bombrun},
  {Borrachero}, {Bossini}, {Bouquillon}, {Bourda}, {Bragaglia}, {Bramante},
  {Breddels}, {Bressan}, {Brouillet}, {Br{\"u}semeister}, {Brugaletta},
  {Bucciarelli}, {Burlacu}, {Busonero}, {Butkevich}, {Buzzi}, {Caffau},
  {Cancelliere}, {Cannizzaro}, {Cantat-Gaudin}, {Carballo}, {Carlucci},
  {Carrasco}, {Casamiquela}, {Castellani}, {Castro-Ginard}, {Charlot},
  {Chemin}, {Chiavassa}, {Cocozza}, {Costigan}, {Cowell}, {Crifo}, {Crosta},
  {Crowley}, {Cuypers}, {Dafonte}, {Damerdji}, {Dapergolas}, {David}, {David},
  {de Laverny}, {De Luise}, {De March}, {de Martino}, {de Souza}, {de Torres},
  {Debosscher}, {del Pozo}, {Delbo}, {Delgado}, {Delgado}, {Di Matteo},
  {Diakite}, {Diener}, {Distefano}, {Dolding}, {Drazinos}, {Dur{\'a}n},
  {Edvardsson}, {Enke}, {Eriksson}, {Esquej}, {Eynard Bontemps}, {Fabre},
  {Fabrizio}, {Faigler}, {Falc{\~a}o}, {Farr{\`a}s Casas}, {Federici},
  {Fedorets}, {Fernique}, {Figueras}, {Filippi}, {Findeisen}, {Fonti},
  {Fraile}, {Fraser}, {Fr{\'e}zouls}, {Gai}, {Galleti}, {Garabato},
  {Garc{\'\i}a-Sedano}, {Garofalo}, {Garralda}, {Gavel}, {Gavras}, {Gerssen},
  {Geyer}, {Giacobbe}, {Gilmore}, {Girona}, {Giuffrida}, {Glass}, {Gomes},
  {Granvik}, {Gueguen}, {Guerrier}, {Guiraud}, {Guti{\'e}rrez-S{\'a}nchez},
  {Haigron}, {Hatzidimitriou}, {Hauser}, {Haywood}, {Heiter}, {Helmi}, {Heu},
  {Hilger}, {Hobbs}, {Hofmann}, {Holland}, {Huckle}, {Hypki}, {Icardi},
  {Jan{\ss}en}, {Jevardat de Fombelle}, {Jonker}, {Juh{\'a}sz}, {Julbe},
  {Karampelas}, {Kewley}, {Klar}, {Kochoska}, {Kohley}, {Kolenberg},
  {Kontizas}, {Kontizas}, {Koposov}, {Kordopatis}, {Kostrzewa-Rutkowska},
  {Koubsky}, {Lambert}, {Lanza}, {Lasne}, {Lavigne}, {Le Fustec}, {Le
  Poncin-Lafitte}, {Lebreton}, {Leccia}, {Leclerc}, {Lecoeur-Taibi},
  {Lenhardt}, {Leroux}, {Liao}, {Licata}, {Lindstr{\o}m}, {Lister}, {Livanou},
  {Lobel}, {L{\'o}pez}, {Managau}, {Mann}, {Mantelet}, {Marchal}, {Marchant},
  {Marconi}, {Marinoni}, {Marschalk{\'o}}, {Marshall}, {Martino}, {Marton},
  {Mary}, {Massari}, {Matijevi{\v{c}}}, {Mazeh}, {McMillan}, {Messina},
  {Michalik}, {Millar}, {Molina}, {Molinaro}, {Moln{\'a}r}, {Montegriffo},
  {Mor}, {Morbidelli}, {Morel}, {Morris}, {Mulone}, {Muraveva}, {Musella},
  {Nelemans}, {Nicastro}, {Noval}, {O'Mullane}, {Ord{\'e}novic},
  {Ord{\'o}{\~n}ez-Blanco}, {Osborne}, {Pagani}, {Pagano}, {Pailler},
  {Palacin}, {Palaversa}, {Panahi}, {Pawlak}, {Piersimoni}, {Pineau}, {Plachy},
  {Plum}, {Poggio}, {Poujoulet}, {Pr{\v{s}}a}, {Pulone}, {Racero}, {Ragaini},
  {Rambaux}, {Ramos-Lerate}, {Regibo}, {Reyl{\'e}}, {Riclet}, {Ripepi}, {Riva},
  {Rivard}, {Rixon}, {Roegiers}, {Roelens}, {Romero-G{\'o}mez}, {Rowell},
  {Royer}, {Ruiz-Dern}, {Sadowski}, {Sagrist{\`a} Sell{\'e}s}, {Sahlmann},
  {Salgado}, {Salguero}, {Sanna}, {Santana-Ros}, {Sarasso}, {Savietto},
  {Schultheis}, {Sciacca}, {Segol}, {Segovia}, {S{\'e}gransan}, {Shih},
  {Siltala}, {Silva}, {Smart}, {Smith}, {Solano}, {Solitro}, {Sordo}, {Soria
  Nieto}, {Souchay}, {Spagna}, {Spoto}, {Stampa}, {Steele},
  {Steidelm{\"u}ller}, {Stephenson}, {Stoev}, {Suess}, {Surdej}, {Szabados},
  {Szegedi-Elek}, {Tapiador}, {Taris}, {Tauran}, {Taylor}, {Teixeira},
  {Terrett}, {Teyssandier}, {Thuillot}, {Titarenko}, {Torra Clotet}, {Turon},
  {Ulla}, {Utrilla}, {Uzzi}, {Vaillant}, {Valentini}, {Valette}, {van Elteren},
  {Van Hemelryck}, {van Leeuwen}, {Vaschetto}, {Vecchiato}, {Veljanoski},
  {Viala}, {Vicente}, {Vogt}, {von Essen}, {Voss}, {Votruba}, {Voutsinas},
  {Walmsley}, {Weiler}, {Wertz}, {Wevers}, {Wyrzykowski}, {Yoldas},
  {{\v{Z}}erjal}, {Ziaeepour}, {Zorec}, {Zschocke}, {Zucker}, {Zurbach}, \&
  {Zwitter}}]{2018A&A...616A...1G}
{Gaia Collaboration}, {Brown}, A.~G.~A., {Vallenari}, A., {et~al.} 2018, \aap,
  616, A1, \dodoi{10.1051/0004-6361/201833051}

\bibitem[{{Gaia Collaboration} {et~al.}(2021){Gaia Collaboration}, {Brown},
  {Vallenari}, {Prusti}, {de Bruijne}, {Babusiaux}, {Biermann}, {Creevey},
  {Evans}, {Eyer}, {Hutton}, {Jansen}, {Jordi}, {Klioner}, {Lammers},
  {Lindegren}, {Luri}, {Mignard}, {Panem}, {Pourbaix}, {Randich}, {Sartoretti},
  {Soubiran}, {Walton}, {Arenou}, {Bailer-Jones}, {Bastian}, {Cropper},
  {Drimmel}, {Katz}, {Lattanzi}, {van Leeuwen}, {Bakker}, {Cacciari},
  {Casta{\~n}eda}, {De Angeli}, {Ducourant}, {Fabricius}, {Fouesneau},
  {Fr{\'e}mat}, {Guerra}, {Guerrier}, {Guiraud}, {Jean-Antoine Piccolo},
  {Masana}, {Messineo}, {Mowlavi}, {Nicolas}, {Nienartowicz}, {Pailler},
  {Panuzzo}, {Riclet}, {Roux}, {Seabroke}, {Sordo}, {Tanga}, {Th{\'e}venin},
  {Gracia-Abril}, {Portell}, {Teyssier}, {Altmann}, {Andrae}, {Bellas-Velidis},
  {Benson}, {Berthier}, {Blomme}, {Brugaletta}, {Burgess}, {Busso}, {Carry},
  {Cellino}, {Cheek}, {Clementini}, {Damerdji}, {Davidson}, {Delchambre},
  {Dell'Oro}, {Fern{\'a}ndez-Hern{\'a}ndez}, {Galluccio}, {Garc{\'\i}a-Lario},
  {Garcia-Reinaldos}, {Gonz{\'a}lez-N{\'u}{\~n}ez}, {Gosset}, {Haigron},
  {Halbwachs}, {Hambly}, {Harrison}, {Hatzidimitriou}, {Heiter},
  {Hern{\'a}ndez}, {Hestroffer}, {Hodgkin}, {Holl}, {Jan{\ss}en}, {Jevardat de
  Fombelle}, {Jordan}, {Krone-Martins}, {Lanzafame}, {L{\"o}ffler}, {Lorca},
  {Manteiga}, {Marchal}, {Marrese}, {Moitinho}, {Mora}, {Muinonen}, {Osborne},
  {Pancino}, {Pauwels}, {Petit}, {Recio-Blanco}, {Richards}, {Riello},
  {Rimoldini}, {Robin}, {Roegiers}, {Rybizki}, {Sarro}, {Siopis}, {Smith},
  {Sozzetti}, {Ulla}, {Utrilla}, {van Leeuwen}, {van Reeven}, {Abbas}, {Abreu
  Aramburu}, {Accart}, {Aerts}, {Aguado}, {Ajaj}, {Altavilla}, {{\'A}lvarez},
  {{\'A}lvarez Cid-Fuentes}, {Alves}, {Anderson}, {Anglada Varela}, {Antoja},
  {Audard}, {Baines}, {Baker}, {Balaguer-N{\'u}{\~n}ez}, {Balbinot}, {Balog},
  {Barache}, {Barbato}, {Barros}, {Barstow}, {Bartolom{\'e}}, {Bassilana},
  {Bauchet}, {Baudesson-Stella}, {Becciani}, {Bellazzini}, {Bernet}, {Bertone},
  {Bianchi}, {Blanco-Cuaresma}, {Boch}, {Bombrun}, {Bossini}, {Bouquillon},
  {Bragaglia}, {Bramante}, {Breedt}, {Bressan}, {Brouillet}, {Bucciarelli},
  {Burlacu}, {Busonero}, {Butkevich}, {Buzzi}, {Caffau}, {Cancelliere},
  {C{\'a}novas}, {Cantat-Gaudin}, {Carballo}, {Carlucci}, {Carnerero},
  {Carrasco}, {Casamiquela}, {Castellani}, {Castro-Ginard}, {Castro Sampol},
  {Chaoul}, {Charlot}, {Chemin}, {Chiavassa}, {Cioni}, {Comoretto}, {Cooper},
  {Cornez}, {Cowell}, {Crifo}, {Crosta}, {Crowley}, {Dafonte}, {Dapergolas},
  {David}, {David}, {de Laverny}, {De Luise}, {De March}, {De Ridder}, {de
  Souza}, {de Teodoro}, {de Torres}, {del Peloso}, {del Pozo}, {Delbo},
  {Delgado}, {Delgado}, {Delisle}, {Di Matteo}, {Diakite}, {Diener},
  {Distefano}, {Dolding}, {Eappachen}, {Edvardsson}, {Enke}, {Esquej}, {Fabre},
  {Fabrizio}, {Faigler}, {Fedorets}, {Fernique}, {Fienga}, {Figueras},
  {Fouron}, {Fragkoudi}, {Fraile}, {Franke}, {Gai}, {Garabato},
  {Garcia-Gutierrez}, {Garc{\'\i}a-Torres}, {Garofalo}, {Gavras}, {Gerlach},
  {Geyer}, {Giacobbe}, {Gilmore}, {Girona}, {Giuffrida}, {Gomel}, {Gomez},
  {Gonzalez-Santamaria}, {Gonz{\'a}lez-Vidal}, {Granvik},
  {Guti{\'e}rrez-S{\'a}nchez}, {Guy}, {Hauser}, {Haywood}, {Helmi}, {Hidalgo},
  {Hilger}, {H{\l}adczuk}, {Hobbs}, {Holland}, {Huckle}, {Jasniewicz},
  {Jonker}, {Juaristi Campillo}, {Julbe}, {Karbevska}, {Kervella}, {Khanna},
  {Kochoska}, {Kontizas}, {Kordopatis}, {Korn}, {Kostrzewa-Rutkowska},
  {Kruszy{\'n}ska}, {Lambert}, {Lanza}, {Lasne}, {Le Campion}, {Le Fustec},
  {Lebreton}, {Lebzelter}, {Leccia}, {Leclerc}, {Lecoeur-Taibi}, {Liao},
  {Licata}, {Lindstr{\o}m}, {Lister}, {Livanou}, {Lobel}, {Madrero Pardo},
  {Managau}, {Mann}, {Marchant}, {Marconi}, {Marcos Santos}, {Marinoni},
  {Marocco}, {Marshall}, {Martin Polo}, {Mart{\'\i}n-Fleitas}, {Masip},
  {Massari}, {Mastrobuono-Battisti}, {Mazeh}, {McMillan}, {Messina},
  {Michalik}, {Millar}, {Mints}, {Molina}, {Molinaro}, {Moln{\'a}r},
  {Montegriffo}, {Mor}, {Morbidelli}, {Morel}, {Morris}, {Mulone}, {Munoz},
  {Muraveva}, {Murphy}, {Musella}, {Noval}, {Ord{\'e}novic}, {Orr{\`u}},
  {Osinde}, {Pagani}, {Pagano}, {Palaversa}, {Palicio}, {Panahi}, {Pawlak},
  {Pe{\~n}alosa Esteller}, {Penttil{\"a}}, {Piersimoni}, {Pineau}, {Plachy},
  {Plum}, {Poggio}, {Poretti}, {Poujoulet}, {Pr{\v{s}}a}, {Pulone}, {Racero},
  {Ragaini}, {Rainer}, {Raiteri}, {Rambaux}, {Ramos}, {Ramos-Lerate}, {Re
  Fiorentin}, {Regibo}, {Reyl{\'e}}, {Ripepi}, {Riva}, {Rixon}, {Robichon},
  {Robin}, {Roelens}, {Rohrbasser}, {Romero-G{\'o}mez}, {Rowell}, {Royer},
  {Rybicki}, {Sadowski}, {Sagrist{\`a} Sell{\'e}s}, {Sahlmann}, {Salgado},
  {Salguero}, {Samaras}, {Sanchez Gimenez}, {Sanna}, {Santove{\~n}a},
  {Sarasso}, {Schultheis}, {Sciacca}, {Segol}, {Segovia}, {S{\'e}gransan},
  {Semeux}, {Shahaf}, {Siddiqui}, {Siebert}, {Siltala}, {Slezak}, {Smart},
  {Solano}, {Solitro}, {Souami}, {Souchay}, {Spagna}, {Spoto}, {Steele},
  {Steidelm{\"u}ller}, {Stephenson}, {S{\"u}veges}, {Szabados}, {Szegedi-Elek},
  {Taris}, {Tauran}, {Taylor}, {Teixeira}, {Thuillot}, {Tonello}, {Torra},
  {Torra}, {Turon}, {Unger}, {Vaillant}, {van Dillen}, {Vanel}, {Vecchiato},
  {Viala}, {Vicente}, {Voutsinas}, {Weiler}, {Wevers}, {Wyrzykowski}, {Yoldas},
  {Yvard}, {Zhao}, {Zorec}, {Zucker}, {Zurbach}, \&
  {Zwitter}}]{2021A&A...649A...1G}
---. 2021, \aap, 649, A1, \dodoi{10.1051/0004-6361/202039657}

\bibitem[{{Geier}(2020)}]{2020A&A...635A.193G}
{Geier}, S. 2020, \aap, 635, A193, \dodoi{10.1051/0004-6361/202037526}

\bibitem[{{Geier} {et~al.}(2022){Geier}, {Dorsch}, {Pelisoli}, {Reindl},
  {Heber}, \& {Irrgang}}]{2022A&A...661A.113G}
{Geier}, S., {Dorsch}, M., {Pelisoli}, I., {et~al.} 2022, \aap, 661, A113,
  \dodoi{10.1051/0004-6361/202143022}

\bibitem[{{Geier} {et~al.}(2017){Geier}, {{\O}stensen}, {Nemeth}, {Gentile
  Fusillo}, {G{\"a}nsicke}, {Telting}, {Green}, \&
  {Schaffenroth}}]{2017A&A...600A..50G}
{Geier}, S., {{\O}stensen}, R.~H., {Nemeth}, P., {et~al.} 2017, \aap, 600, A50,
  \dodoi{10.1051/0004-6361/201630135}

\bibitem[{{Geier} {et~al.}(2015){Geier}, {Kupfer}, {Heber}, {Schaffenroth},
  {Barlow}, {{\O}stensen}, {O'Toole}, {Ziegerer}, {Heuser}, {Maxted},
  {G{\"a}nsicke}, {Marsh}, {Napiwotzki}, {Br{\"u}nner}, {Schindewolf}, \&
  {Niederhofer}}]{2015A&A...577A..26G}
{Geier}, S., {Kupfer}, T., {Heber}, U., {et~al.} 2015, \aap, 577, A26,
  \dodoi{10.1051/0004-6361/201525666}

\bibitem[{{Han} {et~al.}(2003){Han}, {Podsiadlowski}, {Maxted}, \&
  {Marsh}}]{2003MNRAS.341..669H}
{Han}, Z., {Podsiadlowski}, P., {Maxted}, P.~F.~L., \& {Marsh}, T.~R. 2003,
  \mnras, 341, 669, \dodoi{10.1046/j.1365-8711.2003.06451.x}

\bibitem[{{Han} {et~al.}(2002){Han}, {Podsiadlowski}, {Maxted}, {Marsh}, \&
  {Ivanova}}]{2002MNRAS.336..449H}
{Han}, Z., {Podsiadlowski}, P., {Maxted}, P.~F.~L., {Marsh}, T.~R., \&
  {Ivanova}, N. 2002, \mnras, 336, 449,
  \dodoi{10.1046/j.1365-8711.2002.05752.x}

\bibitem[{{Heber}(2009)}]{2009ARA&A..47..211H}
{Heber}, U. 2009, \araa, 47, 211, \dodoi{10.1146/annurev-astro-082708-101836}

\bibitem[{{Heber}(2016)}]{2016PASP..128h2001H}
---. 2016, \pasp, 128, 082001, \dodoi{10.1088/1538-3873/128/966/082001}

\bibitem[{{Heber} {et~al.}(2003){Heber}, {Edelmann}, {Lisker}, \&
  {Napiwotzki}}]{2003A&A...411L.477H}
{Heber}, U., {Edelmann}, H., {Lisker}, T., \& {Napiwotzki}, R. 2003, \aap, 411,
  L477, \dodoi{10.1051/0004-6361:20031553}

\bibitem[{{Heber} {et~al.}(2018){Heber}, {Irrgang}, \&
  {Schaffenroth}}]{2018OAst...27...35H}
{Heber}, U., {Irrgang}, A., \& {Schaffenroth}, J. 2018, Open Astronomy, 27, 35,
  \dodoi{10.1515/astro-2018-0008}

\bibitem[{{Heber} {et~al.}(2000){Heber}, {Reid}, \&
  {Werner}}]{2000A&A...363..198H}
{Heber}, U., {Reid}, I.~N., \& {Werner}, K. 2000, \aap, 363, 198,
  \dodoi{10.48550/arXiv.astro-ph/0009159}

\bibitem[{{Howell} {et~al.}(2014){Howell}, {Sobeck}, {Haas}, {Still},
  {Barclay}, {Mullally}, {Troeltzsch}, {Aigrain}, {Bryson}, {Caldwell},
  {Chaplin}, {Cochran}, {Huber}, {Marcy}, {Miglio}, {Najita}, {Smith},
  {Twicken}, \& {Fortney}}]{2014PASP..126..398H}
{Howell}, S.~B., {Sobeck}, C., {Haas}, M., {et~al.} 2014, \pasp, 126, 398,
  \dodoi{10.1086/676406}

\bibitem[{{Hubeny} \& {Lanz}(2017)}]{2017arXiv170601859H}
{Hubeny}, I., \& {Lanz}, T. 2017, arXiv e-prints, arXiv:1706.01859.
\newblock \doarXiv{1706.01859}

\bibitem[{{Irrgang} {et~al.}(2021){Irrgang}, {Geier}, {Heber}, {Kupfer},
  {El-Badry}, \& {Bloemen}}]{2021A&A...650A.102I}
{Irrgang}, A., {Geier}, S., {Heber}, U., {et~al.} 2021, \aap, 650, A102,
  \dodoi{10.1051/0004-6361/202038757}

\bibitem[{{Istrate} {et~al.}(2016){Istrate}, {Marchant}, {Tauris}, {Langer},
  {Stancliffe}, \& {Grassitelli}}]{2016A&A...595A..35I}
{Istrate}, A.~G., {Marchant}, P., {Tauris}, T.~M., {et~al.} 2016, \aap, 595,
  A35, \dodoi{10.1051/0004-6361/201628874}

\bibitem[{{Jeffery} {et~al.}(2021){Jeffery}, {Miszalski}, \&
  {Snowdon}}]{2021MNRAS.501..623J}
{Jeffery}, C.~S., {Miszalski}, B., \& {Snowdon}, E. 2021, \mnras, 501, 623,
  \dodoi{10.1093/mnras/staa3648}

\bibitem[{{Kepler} {et~al.}(2015){Kepler}, {Pelisoli}, {Koester}, {Ourique},
  {Kleinman}, {Romero}, {Nitta}, {Eisenstein}, {Costa}, {K{\"u}lebi}, {Jordan},
  {Dufour}, {Giommi}, \& {Rebassa-Mansergas}}]{2015MNRAS.446.4078K}
{Kepler}, S.~O., {Pelisoli}, I., {Koester}, D., {et~al.} 2015, \mnras, 446,
  4078, \dodoi{10.1093/mnras/stu2388}

\bibitem[{{Kepler} {et~al.}(2016){Kepler}, {Pelisoli}, {Koester}, {Ourique},
  {Romero}, {Reindl}, {Kleinman}, {Eisenstein}, {Valois}, \&
  {Amaral}}]{2016MNRAS.455.3413K}
---. 2016, \mnras, 455, 3413, \dodoi{10.1093/mnras/stv2526}

\bibitem[{{Kepler} {et~al.}(2019){Kepler}, {Pelisoli}, {Koester}, {Reindl},
  {Geier}, {Romero}, {Ourique}, {Oliveira}, \& {Amaral}}]{2019MNRAS.486.2169K}
---. 2019, \mnras, 486, 2169, \dodoi{10.1093/mnras/stz960}

\bibitem[{{Krzesinski} {et~al.}(2022){Krzesinski}, {{\c{S}}ener}, {Zola}, \&
  {Siwak}}]{2022MNRAS.516.1509K}
{Krzesinski}, J., {{\c{S}}ener}, H.~T., {Zola}, S., \& {Siwak}, M. 2022,
  \mnras, 516, 1509, \dodoi{10.1093/mnras/stac2088}

\bibitem[{{Lanz} \& {Hubeny}(2007)}]{2007ApJS..169...83L}
{Lanz}, T., \& {Hubeny}, I. 2007, \apjs, 169, 83, \dodoi{10.1086/511270}

\bibitem[{{Lei} {et~al.}(2019{\natexlab{a}}){Lei}, {Bu}, {Zhao}, {N{\'e}meth},
  \& {Zhao}}]{2019PASJ...71...41L}
{Lei}, Z., {Bu}, Y., {Zhao}, J., {N{\'e}meth}, P., \& {Zhao}, G.
  2019{\natexlab{a}}, \pasj, 71, 41, \dodoi{10.1093/pasj/psz006}

\bibitem[{{Lei} {et~al.}(2018){Lei}, {Zhao}, {N{\'e}meth}, \&
  {Zhao}}]{2018ApJ...868...70L}
{Lei}, Z., {Zhao}, J., {N{\'e}meth}, P., \& {Zhao}, G. 2018, \apj, 868, 70,
  \dodoi{10.3847/1538-4357/aae82b}

\bibitem[{{Lei} {et~al.}(2019{\natexlab{b}}){Lei}, {Zhao}, {N{\'e}meth}, \&
  {Zhao}}]{2019ApJ...881..135L}
---. 2019{\natexlab{b}}, \apj, 881, 135, \dodoi{10.3847/1538-4357/ab2edc}

\bibitem[{{Lei} {et~al.}(2020){Lei}, {Zhao}, {N{\'e}meth}, \&
  {Zhao}}]{2020ApJ...889..117L}
---. 2020, \apj, 889, 117, \dodoi{10.3847/1538-4357/ab660a}

\bibitem[{{Lei} {et~al.}(2023){Lei}, {He}, {N{\'e}meth}, {Vos}, {Zou}, {Hu},
  {Xiao}, {Yan}, \& {Zhao}}]{2023ApJ...942..109L}
{Lei}, Z., {He}, R., {N{\'e}meth}, P., {et~al.} 2023, \apj, 942, 109,
  \dodoi{10.3847/1538-4357/aca542}

\bibitem[{{Lindegren} {et~al.}(2021{\natexlab{a}}){Lindegren}, {Klioner},
  {Hern{\'a}ndez}, {Bombrun}, {Ramos-Lerate}, {Steidelm{\"u}ller}, {Bastian},
  {Biermann}, {de Torres}, {Gerlach}, {Geyer}, {Hilger}, {Hobbs}, {Lammers},
  {McMillan}, {Stephenson}, {Casta{\~n}eda}, {Davidson}, {Fabricius},
  {Gracia-Abril}, {Portell}, {Rowell}, {Teyssier}, {Torra}, {Bartolom{\'e}},
  {Clotet}, {Garralda}, {Gonz{\'a}lez-Vidal}, {Torra}, {Abbas}, {Altmann},
  {Anglada Varela}, {Balaguer-N{\'u}{\~n}ez}, {Balog}, {Barache}, {Becciani},
  {Bernet}, {Bertone}, {Bianchi}, {Bouquillon}, {Brown}, {Bucciarelli},
  {Busonero}, {Butkevich}, {Buzzi}, {Cancelliere}, {Carlucci}, {Charlot},
  {Cioni}, {Crosta}, {Crowley}, {del Peloso}, {del Pozo}, {Drimmel}, {Esquej},
  {Fienga}, {Fraile}, {Gai}, {Garcia-Reinaldos}, {Guerra}, {Hambly}, {Hauser},
  {Jan{\ss}en}, {Jordan}, {Kostrzewa-Rutkowska}, {Lattanzi}, {Liao}, {Licata},
  {Lister}, {L{\"o}ffler}, {Marchant}, {Masip}, {Mignard}, {Mints}, {Molina},
  {Mora}, {Morbidelli}, {Murphy}, {Pagani}, {Panuzzo}, {Pe{\~n}alosa Esteller},
  {Poggio}, {Re Fiorentin}, {Riva}, {Sagrist{\`a} Sell{\'e}s}, {Sanchez
  Gimenez}, {Sarasso}, {Sciacca}, {Siddiqui}, {Smart}, {Souami}, {Spagna},
  {Steele}, {Taris}, {Utrilla}, {van Reeven}, \&
  {Vecchiato}}]{2021A&A...649A...2L}
{Lindegren}, L., {Klioner}, S.~A., {Hern{\'a}ndez}, J., {et~al.}
  2021{\natexlab{a}}, \aap, 649, A2, \dodoi{10.1051/0004-6361/202039709}

\bibitem[{{Lindegren} {et~al.}(2021{\natexlab{b}}){Lindegren}, {Bastian},
  {Biermann}, {Bombrun}, {de Torres}, {Gerlach}, {Geyer}, {Hern{\'a}ndez},
  {Hilger}, {Hobbs}, {Klioner}, {Lammers}, {McMillan}, {Ramos-Lerate},
  {Steidelm{\"u}ller}, {Stephenson}, \& {van Leeuwen}}]{2021A&A...649A...4L}
{Lindegren}, L., {Bastian}, U., {Biermann}, M., {et~al.} 2021{\natexlab{b}},
  \aap, 649, A4, \dodoi{10.1051/0004-6361/202039653}

\bibitem[{{Lisker} {et~al.}(2005){Lisker}, {Heber}, {Napiwotzki}, {Christlieb},
  {Han}, {Homeier}, \& {Reimers}}]{2005A&A...430..223L}
{Lisker}, T., {Heber}, U., {Napiwotzki}, R., {et~al.} 2005, \aap, 430, 223,
  \dodoi{10.1051/0004-6361:20040232}

\bibitem[{{Luo} {et~al.}(2019){Luo}, {N{\'e}meth}, {Deng}, \&
  {Han}}]{2019ApJ...881....7L}
{Luo}, Y., {N{\'e}meth}, P., {Deng}, L., \& {Han}, Z. 2019, \apj, 881, 7,
  \dodoi{10.3847/1538-4357/ab298d}

\bibitem[{{Luo} {et~al.}(2021){Luo}, {N{\'e}meth}, {Wang}, {Wang}, \&
  {Han}}]{2021ApJS..256...28L}
{Luo}, Y., {N{\'e}meth}, P., {Wang}, K., {Wang}, X., \& {Han}, Z. 2021, \apjs,
  256, 28, \dodoi{10.3847/1538-4365/ac11f6}

\bibitem[{{Luo} {et~al.}(2016){Luo}, {N{\'e}meth}, {Liu}, {Deng}, \&
  {Han}}]{2016ApJ...818..202L}
{Luo}, Y.-P., {N{\'e}meth}, P., {Liu}, C., {Deng}, L.-C., \& {Han}, Z.-W. 2016,
  \apj, 818, 202, \dodoi{10.3847/0004-637X/818/2/202}

\bibitem[{{Martin} {et~al.}(2005){Martin}, {Fanson}, {Schiminovich},
  {Morrissey}, {Friedman}, {Barlow}, {Conrow}, {Grange}, {Jelinsky},
  {Milliard}, {Siegmund}, {Bianchi}, {Byun}, {Donas}, {Forster}, {Heckman},
  {Lee}, {Madore}, {Malina}, {Neff}, {Rich}, {Small}, {Surber}, {Szalay},
  {Welsh}, \& {Wyder}}]{2005ApJ...619L...1M}
{Martin}, D.~C., {Fanson}, J., {Schiminovich}, D., {et~al.} 2005, \apjl, 619,
  L1, \dodoi{10.1086/426387}

\bibitem[{{Maxted} {et~al.}(2001){Maxted}, {Heber}, {Marsh}, \&
  {North}}]{2001MNRAS.326.1391M}
{Maxted}, P.~F.~L., {Heber}, U., {Marsh}, T.~R., \& {North}, R.~C. 2001,
  \mnras, 326, 1391, \dodoi{10.1111/j.1365-2966.2001.04714.x}

\bibitem[{{Meng} \& {Luo}(2021)}]{2021MNRAS.507.4603M}
{Meng}, X.-C., \& {Luo}, Y.-P. 2021, \mnras, 507, 4603,
  \dodoi{10.1093/mnras/stab2369}

\bibitem[{{Mengel} {et~al.}(1976){Mengel}, {Norris}, \&
  {Gross}}]{1976ApJ...204..488M}
{Mengel}, J.~G., {Norris}, J., \& {Gross}, P.~G. 1976, \apj, 204, 488,
  \dodoi{10.1086/154193}

\bibitem[{{Miller Bertolami} {et~al.}(2008){Miller Bertolami}, {Althaus},
  {Unglaub}, \& {Weiss}}]{2008A&A...491..253M}
{Miller Bertolami}, M.~M., {Althaus}, L.~G., {Unglaub}, K., \& {Weiss}, A.
  2008, \aap, 491, 253, \dodoi{10.1051/0004-6361:200810373}

\bibitem[{{Miller Bertolami} {et~al.}(2022){Miller Bertolami}, {Battich},
  {C{\'o}rsico}, {Althaus}, \& {Wachlin}}]{2022MNRAS.511L..60M}
{Miller Bertolami}, M.~M., {Battich}, T., {C{\'o}rsico}, A.~H., {Althaus},
  L.~G., \& {Wachlin}, F.~C. 2022, \mnras, 511, L60,
  \dodoi{10.1093/mnrasl/slab134}

\bibitem[{{Moehler} {et~al.}(1990){Moehler}, {Richtler}, {de Boer}, {Dettmar},
  \& {Heber}}]{1990A&AS...86...53M}
{Moehler}, S., {Richtler}, T., {de Boer}, K.~S., {Dettmar}, R.~J., \& {Heber},
  U. 1990, \aaps, 86, 53

\bibitem[{{Napiwotzki} {et~al.}(2004){Napiwotzki}, {Karl}, {Lisker}, {Heber},
  {Christlieb}, {Reimers}, {Nelemans}, \& {Homeier}}]{2004Ap&SS.291..321N}
{Napiwotzki}, R., {Karl}, C.~A., {Lisker}, T., {et~al.} 2004, \apss, 291, 321,
  \dodoi{10.1023/B:ASTR.0000044362.07416.6c}

\bibitem[{{N{\'e}meth} {et~al.}(2012){N{\'e}meth}, {Kawka}, \&
  {Vennes}}]{2012MNRAS.427.2180N}
{N{\'e}meth}, P., {Kawka}, A., \& {Vennes}, S. 2012, \mnras, 427, 2180,
  \dodoi{10.1111/j.1365-2966.2012.22009.x}

\bibitem[{{Pelisoli} {et~al.}(2020){Pelisoli}, {Vos}, {Geier}, {Schaffenroth},
  \& {Baran}}]{2020A&A...642A.180P}
{Pelisoli}, I., {Vos}, J., {Geier}, S., {Schaffenroth}, V., \& {Baran}, A.~S.
  2020, \aap, 642, A180, \dodoi{10.1051/0004-6361/202038473}

\bibitem[{{Raddi} {et~al.}(2022){Raddi}, {Torres}, {Rebassa-Mansergas},
  {Maldonado}, {Camisassa}, {Koester}, {Gentile Fusillo}, {Tremblay}, {Dimpel},
  {Heber}, {Cunningham}, \& {Ren}}]{2022A&A...658A..22R}
{Raddi}, R., {Torres}, S., {Rebassa-Mansergas}, A., {et~al.} 2022, \aap, 658,
  A22, \dodoi{10.1051/0004-6361/202141837}

\bibitem[{{Ricker} {et~al.}(2015){Ricker}, {Winn}, {Vanderspek}, {Latham},
  {Bakos}, {Bean}, {Berta-Thompson}, {Brown}, {Buchhave}, {Butler}, {Butler},
  {Chaplin}, {Charbonneau}, {Christensen-Dalsgaard}, {Clampin}, {Deming},
  {Doty}, {De Lee}, {Dressing}, {Dunham}, {Endl}, {Fressin}, {Ge}, {Henning},
  {Holman}, {Howard}, {Ida}, {Jenkins}, {Jernigan}, {Johnson}, {Kaltenegger},
  {Kawai}, {Kjeldsen}, {Laughlin}, {Levine}, {Lin}, {Lissauer}, {MacQueen},
  {Marcy}, {McCullough}, {Morton}, {Narita}, {Paegert}, {Palle}, {Pepe},
  {Pepper}, {Quirrenbach}, {Rinehart}, {Sasselov}, {Sato}, {Seager},
  {Sozzetti}, {Stassun}, {Sullivan}, {Szentgyorgyi}, {Torres}, {Udry}, \&
  {Villasenor}}]{2015JATIS...1a4003R}
{Ricker}, G.~R., {Winn}, J.~N., {Vanderspek}, R., {et~al.} 2015, Journal of
  Astronomical Telescopes, Instruments, and Systems, 1, 014003,
  \dodoi{10.1117/1.JATIS.1.1.014003}

\bibitem[{{Rodrigo} \& {Solano}(2020)}]{2020sea..confE.182R}
{Rodrigo}, C., \& {Solano}, E. 2020, in XIV.0 Scientific Meeting (virtual) of
  the Spanish Astronomical Society, 182

\bibitem[{{Rodrigo} {et~al.}(2012){Rodrigo}, {Solano}, \&
  {Bayo}}]{2012ivoa.rept.1015R}
{Rodrigo}, C., {Solano}, E., \& {Bayo}, A. 2012, {SVO Filter Profile Service
  Version 1.0}, IVOA Working Draft 15 October 2012,
  \dodoi{10.5479/ADS/bib/2012ivoa.rept.1015R}

\bibitem[{{Schaffenroth} {et~al.}(2022){Schaffenroth}, {Pelisoli}, {Barlow},
  {Geier}, \& {Kupfer}}]{2022A&A...666A.182S}
{Schaffenroth}, V., {Pelisoli}, I., {Barlow}, B.~N., {Geier}, S., \& {Kupfer},
  T. 2022, \aap, 666, A182, \dodoi{10.1051/0004-6361/202244214}

\bibitem[{{Schlafly} \& {Finkbeiner}(2011)}]{2011ApJ...737..103S}
{Schlafly}, E.~F., \& {Finkbeiner}, D.~P. 2011, \apj, 737, 103,
  \dodoi{10.1088/0004-637X/737/2/103}

\bibitem[{{Schlegel} {et~al.}(1998){Schlegel}, {Finkbeiner}, \&
  {Davis}}]{1998ApJ...500..525S}
{Schlegel}, D.~J., {Finkbeiner}, D.~P., \& {Davis}, M. 1998, \apj, 500, 525,
  \dodoi{10.1086/305772}

\bibitem[{{Skrutskie} {et~al.}(2006){Skrutskie}, {Cutri}, {Stiening},
  {Weinberg}, {Schneider}, {Carpenter}, {Beichman}, {Capps}, {Chester},
  {Elias}, {Huchra}, {Liebert}, {Lonsdale}, {Monet}, {Price}, {Seitzer},
  {Jarrett}, {Kirkpatrick}, {Gizis}, {Howard}, {Evans}, {Fowler}, {Fullmer},
  {Hurt}, {Light}, {Kopan}, {Marsh}, {McCallon}, {Tam}, {Van Dyk}, \&
  {Wheelock}}]{2006AJ....131.1163S}
{Skrutskie}, M.~F., {Cutri}, R.~M., {Stiening}, R., {et~al.} 2006, \aj, 131,
  1163, \dodoi{10.1086/498708}

\bibitem[{{Stroeer} {et~al.}(2007){Stroeer}, {Heber}, {Lisker}, {Napiwotzki},
  {Dreizler}, {Christlieb}, \& {Reimers}}]{2007A&A...462..269S}
{Stroeer}, A., {Heber}, U., {Lisker}, T., {et~al.} 2007, \aap, 462, 269,
  \dodoi{10.1051/0004-6361:20065564}

\bibitem[{{Taylor}(2005)}]{2005ASPC..347...29T}
{Taylor}, M.~B. 2005, in Astronomical Society of the Pacific Conference Series,
  Vol. 347, Astronomical Data Analysis Software and Systems XIV, ed.
  P.~{Shopbell}, M.~{Britton}, \& R.~{Ebert}, 29

\bibitem[{{Vennes} {et~al.}(2011){Vennes}, {Kawka}, \&
  {N{\'e}meth}}]{2011MNRAS.410.2095V}
{Vennes}, S., {Kawka}, A., \& {N{\'e}meth}, P. 2011, \mnras, 410, 2095,
  \dodoi{10.1111/j.1365-2966.2010.17584.x}

\bibitem[{{Webbink}(1984)}]{1984ApJ...277..355W}
{Webbink}, R.~F. 1984, \apj, 277, 355, \dodoi{10.1086/161701}

\bibitem[{{Werner} {et~al.}(2022){Werner}, {Reindl}, {Geier}, \&
  {Pritzkuleit}}]{2022MNRAS.511L..66W}
{Werner}, K., {Reindl}, N., {Geier}, S., \& {Pritzkuleit}, M. 2022, \mnras,
  511, L66, \dodoi{10.1093/mnrasl/slac005}

\bibitem[{{Wright} {et~al.}(2010){Wright}, {Eisenhardt}, {Mainzer}, {Ressler},
  {Cutri}, {Jarrett}, {Kirkpatrick}, {Padgett}, {McMillan}, {Skrutskie},
  {Stanford}, {Cohen}, {Walker}, {Mather}, {Leisawitz}, {Gautier}, {McLean},
  {Benford}, {Lonsdale}, {Blain}, {Mendez}, {Irace}, {Duval}, {Liu}, {Royer},
  {Heinrichsen}, {Howard}, {Shannon}, {Kendall}, {Walsh}, {Larsen}, {Cardon},
  {Schick}, {Schwalm}, {Abid}, {Fabinsky}, {Naes}, \&
  {Tsai}}]{2010AJ....140.1868W}
{Wright}, E.~L., {Eisenhardt}, P. R.~M., {Mainzer}, A.~K., {et~al.} 2010, \aj,
  140, 1868, \dodoi{10.1088/0004-6256/140/6/1868}

\bibitem[{{Zhang} {et~al.}(2017){Zhang}, {Hall}, {Jeffery}, \&
  {Bi}}]{2017ApJ...835..242Z}
{Zhang}, X., {Hall}, P.~D., {Jeffery}, C.~S., \& {Bi}, S. 2017, \apj, 835, 242,
  \dodoi{10.3847/1538-4357/835/2/242}

\bibitem[{{Zhang} \& {Jeffery}(2012)}]{2012MNRAS.419..452Z}
{Zhang}, X., \& {Jeffery}, C.~S. 2012, \mnras, 419, 452,
  \dodoi{10.1111/j.1365-2966.2011.19711.x}

\end{thebibliography}
\bibliographystyle{aasjournal}



\end{document}